\documentclass[12pt]{article}
\usepackage{epsfig}
\usepackage{a4}

\def\to{\rightarrow}
\def\ra{\rightarrow}

\newcommand\nue{{\nu_e}}
\newcommand\anue{\bar{\nu}_e}
\def\sq2{sin^2(2\Theta)}

\def\dms2{\Delta m^2}
\def\Dm32{$\Delta m^2_{32}$}
\def\dm31{\Delta m^2_{31}}
\def\dm21{\Delta m^2_{21}}
\def\t12{\theta_{12}}
\def\th13{$\theta_{13}$}
\def\t23{\theta_{23}}
\def\s2t13{$\sin^2\theta_{13}$}
\def\nuall{\nu^{^{\hbox{\hspace*{-3mm}{\tiny (---)}}}}}

\def\etal{{\it et\ al.}}

\def\NPA{{\em Nucl. Phys.} A}

\def\PLB{{\em Phys. Lett.}  B}
\def\PRL{{\em Phys. Rev. Lett.}}

\def\PRD{{\em Phys. Rev.} D}

\begin{document}
\thispagestyle{empty}
\begin{flushright}
{\tt ICARUS-TM/03-03}\\ 
July 18, 2003\\
Revised \today
\end{flushright}
\vspace*{1cm}
\begin{center}
{\Large{\bf Oscillation effects on supernova neutrino rates and spectra
and  detection
of the shock breakout 
in a liquid Argon TPC}}\\

\vspace{1cm}
{\large I. Gil-Botella}\footnote{Ines.Gil.Botella@cern.ch},
{\large A. Rubbia}\footnote{Andre.Rubbia@cern.ch}

Institut f\"{u}r Teilchenphysik, ETHZ, CH-8093 Z\"{u}rich,
Switzerland
\end{center}
\vspace{2.cm}
\begin{abstract}
\noindent
A liquid Argon TPC (ICARUS-like) has the ability to detect clean
neutrino bursts from type-II supernova collapses. In this paper, we consider
for the first time the four possible detectable
channels, namely, the elastic scattering on electrons from all
neutrino species, $\nu_e$ charged current absorption on $Ar$ with
production of excited $K$, $\bar\nu_e$ charged current absorption on
$Ar$ with production of excited $Cl$ and neutral current interactions
on $Ar$ from all neutrino flavors. 
We compute the total rates and energy spectra of supernova
neutrino events including the effects of the three--flavor neutrino
oscillation with matter effects in the propagation in the supernova. 
Results show a dramatic dependence on
the oscillation parameters and in the energy
spectrum, especially for charged-current events. 
The shock breakout phase has also been investigated
using recent simulations of the core collapse supernova. 
We stress the importance of the neutral current signal to decouple
supernova from neutrino oscillation physics.
\end{abstract}

\newpage
\pagestyle{plain} 
\setcounter{page}{1}
\setcounter{footnote}{0}

\section{Introduction}
\label{sec:intro}
Core collapse supernovae are a huge source of all flavor
neutrinos. Neutrino astrophysics entered a new phase with the
detection of neutrinos from the supernova SN1987A in the Large
Magellanic Cloud by the Kamiokande and IMB detectors \cite{SN1987}. In
spite of these fundamental neutrino observations, the 19 events
observed are not statistically significant enough to obtain
precise quantitative information on the neutrino spectrum. 
Currently running neutrino detectors like Superkamiokande
or SNO have the capabilities to provide high
statistics information about supernova and neutrino properties if a
supernova collapse were to take place in the near future.

The flavor composition, energy spectrum and time structure of
the neutrino burst  from a galactic supernova can give information
about the explosion mechanism
and the mechanisms of proto neutron star cooling. In addition, because
neutrinos arrive before any other signal from a supernova, it is
possible to provide an early alert to the astronomical community
combining the observations of several experiments. This will allow an
early observation of the first stages of the supernova explosion.

The neutrino signal from a galactic supernova can also give information
about the intrinsic properties of the neutrino such as flavor
oscillations. Although new data from solar, atmospheric, reactor and
accelerator neutrinos \cite{nuobserv} have contributed to the
understanding of the neutrino properties, still the neutrino mixing
angle \th13 and the nature of the mass hierarchy, i.e., normal or
inverted, remain unknown. These parameters can be probed by the
observation of supernovae neutrino bursts, since neutrinos will travel
long distances before reaching the Earth and will, as they travel through
the exploding star, in addition traverse
regions of different matter densities where matter enhanced oscillations
will take place. In the case of the  \th13 angle, matter enhancement has 
the striking feature that  very small mixing angles, beyond any
value detectable by next generation accelerators, could in fact
alter significantly the neutrino spectrum. Hence, supernova neutrinos
provide indeed a complementary tool to study  the  \th13 angle.

The main question of course is to understand to which extend can the
supernova and the neutrino physics be decoupled in the observation
of a single supernova? On one hand, the understanding of the supernova
explosion mechanism is still plagued by uncertainties which have an impact
on the precision with which one can predict time, energy and flavor-dependent
neutrino fluxes. On the other hand, the intrinsic neutrino properties are
not fully known, since the type of mass hierarchy and the value of
the  \th13 angle are unknown, and in fact large uncertainty still exist on
the prediction of the actual effect of neutrino oscillation.

There have been various studies on future supernova
neutrino detection and their interplay with neutrino oscillations physics.
Dighe and Smirnov \cite{Dighe} estimated qualitatively
the effects of neutrino oscillations in a collapse-driven supernova
on the neutronization peak, the distortion of energy spectra
and the Earth matter effect. Dutta et al \cite{Dutta:1999ir}
and also Takahashi et al \cite{Takahashi:2001ep}
showed numerically that the total number of events for Superkamiokande and SNO
detectors increases dramatically when there is neutrino mixing.

Detailed supernova simulations (see e.g. \cite{Takahashi:2002yj})
indicate that the average neutrino
energy is flavor dependent and hierarchical (E$_{\nu_e}$ $<$ E$_{\bar\nu_e}$ $<$
E$_{\nu_{\mu,\tau},\bar\nu_{\mu,\tau}}$). Great uncertainties still exist in the spectral
and temporal evolution of the neutrino fluxes. Although the hierarchy of the
average energies of the $\nu_e$, $\bar\nu_e$ and $\nu_x$, $\bar\nu_x$ is believed
to hold, quantitatively their specific spectra remain a matter of detailed calculations (see
\cite{Raffelt:2003en}, and \cite{Takahashi:2003rn} and references therein).
Under the assumption of hierarchical neutrinos,
neutrino flavor oscillations should make
the spectra of $\nu_e$ and $\bar\nu_e$ harder (see 
\cite{Lunardini:2003eh} and references therein). 
This will lead to a
change in the neutrino observations in the detector because the
cross sections depend strongly on the neutrino energy and they are different
for different flavors. 

It is hence well known that uncertainties in the supernova explosion
mechanism and the effect of neutrino oscillations can change
dramatically the number of observed events in a given
experiment. Indeed, this dependence can be used to determine
the supernova and mixing parameters, once the next supernova
has been detected.
Most detailed studies in the literature however concern water
or heavy-water targets (see e.g. 
\cite{Totani:1997vj,Minakata:2001cd,Dutta:2001nf,Takahashi:2002cm}). 
In addition, some authors advocate the use of specific variables
in order to disentangle supernova from neutrino oscillation physics (see e.g. \cite{Lunardini:2003eh}).

In this work we study the capabilities of a liquid Argon TPC (ICARUS-like) 
to detect neutrinos
from supernova collapses and its possible contributions to this field.
Our study is motivated by the
proposed 3 kton ICARUS detector \cite{icarus3000} that should become
operational at the Gran Sasso Underground Laboratory and
by ideas for a very large liquid Argon TPC with
a mass in the range of 100 kton's (see e.g. \cite{Cline:2001pt}).

We think that a liquid Argon detector has excellent potentialities 
for supernova physics, namely because of its high granularity
and bubble-chamber like features, all channels, namely elastic 
scattering off atomic electrons, as well as charged and neutral nuclear scattering processes
are detectable and identifiable. The simultaneous observation of nuclear charged
and neutral currents provide high-statistics and golden channels
to decouple supernova from neutrino
oscillation physics. 

It is known that precise elastic scattering can also yield
information on the flavor content of the flux (see e.g.  \cite{nuobserv} in the
context of the solar neutrino puzzle), because of different cross-sections
depending on flavor and helicity. However, this requires good statistical
precision that is more difficult to achieve with target electrons than 
with nuclear target. Hence, nuclear processes are expected to contribute
most.

A Liquid Argon TPC can also separate the various channels by a classification
of the associated photons from the $K$, $Cl$ or $Ar$ de-excitation,
which exhibit specific spectral lines, or by the absence of associated photons in case
of the elastic scattering. It is hence possible to detect $\nu_e$, $\bar\nu_e$
and other flavors independently.

In section~\ref{sec:framework}, we briefly recall the general framework we used
in this study. In particular, we identify five scenario for the cooling
of the supernova. Neutrino oscillations in vacuum
and in the supernova matter are discussed in section~\ref{sec:matter}.  

In section~\ref{sec:signal}, 
we discuss the relevant neutrino cross-section on Argon and identify
the experimentally detectable classes of events. New
calculations on the cross section for the absorption processes based
in the random phase approximation (RPA) were used allowing computing
for the first time the contribution of the $\bar\nu_e ~^{40}Ar$
absorption reaction and of neutral current reactions. 

In section~\ref{sec:burst}, a particular interest is taken on detection of the 
early phase neutrino signal from
supernova in liquid Argon. We investigate the expected events coming from the first
hundreds of milliseconds after the supernova bounce using the calculations
on the shock breakout process in core collapse supernovae and taking
neutrino oscillations into account. 

In section~\ref{sec:cool}, we compute the expected neutrino rates
in liquid Argon, for different primary supernova neutrino fluxes
scenarios. Moreover, the energy spectra modified by oscillations are
plotted for different conditions on the oscillation parameters.
From supernova neutrino observations it is possible to extract
information about the character of the mass hierarchy and the \th13 
mixing angle. 

\section{General framework and supernova explosion scenarios}
\label{sec:framework}

When the iron core of a massive star (M $\geq$ 8 M$_\odot$) overcomes its
hydrodynamical stability limit (the Chandrasekhar mass), the core
collapses raising its density up to many times the nuclear
density. During this infall stage of the collapse, a first $\nue$ burst is
emitted (``{\it infall burst}''), since the high density of matter enhances
the electron capture by protons. The total energy radiated during this
phase is roughly 10$^{51}$ ergs.

This emission does not continue indefinitely, since at a density
$\sim$ 10$^{12}$ g cm$^{-3}$ neutrinos are trapped in the stellar core
and go to equilibrium with matter. 
The anomalous density produces an elastic bounce of the core, which
results in a shock wave. After the {\it core bounce}, also neutrinos of
other flavors begin to be produced. The neutrinos are trapped in a
region (``neutrinosphere'') whose size is different for different
neutrino flavors. A deeper neutrinosphere corresponds to a higher temperature. 

The wave propagates through the star and loses energy in dissociating
nucleons. When the shock crosses the $\nue$ neutrinosphere, an intense
burst of $\nue$ (``{\it shock breakout}'' or ``{\it neutronization
burst}'') is produced  by electron capture on the large number of protons 
liberated by the shock. The characteristic time of the breakout burst
is 3--10~ms and the total energy radiated in $\nue$ neutrinos during
breakout is $\sim$ 3 $\times$ 10$^{51}$ ergs. 

After the breakout stage, the $\nue$ luminosity rapidly
decreases, while the luminosities of other flavor neutrinos
increase. At the end of the neutrino diffusion inside the mantle the
energy is practically equally distributed between the various neutrino
flavors. 
The shock wave stalls without driving off the stellar mantle and envelope.
$\nue$ and $\anue$ are absorbed on the nucleons liberated by the
shock. Such processes supply new energy to the wave, which is revived
$\sim$ 500 ms after the bounce. This energy transfer is known as
``neutrino heating''. Continued mass accretion and convection below
the shock wave also deposit energy ans thus contribute to the shock
revival. The reinforced shock can propagate within the
stellar matter and expel the external layers into the space.

After the explosion the star loses energy, mainly by neutrino emission, and cools
down, forming a neutron star or a black hole. This final stage (``{\it
cooling}'') requires $\sim$ 10 s and takes away $>$ 99 \% of the
gravitational energy of the star ($\sim$ 2--4 $\times$ 10$^{53}$ ergs).  

In order to obtain quantitative estimates, the following general considerations are taken into 
account for the present analysis: 
\begin{itemize}
\item
SN type-II at a distance of 10 kpc (galactic supernovae)
\item
Gravitational binding energy of 3 $\times$ 10$^{53}$ ergs (radiated away as neutrinos).
\end{itemize}
In our analysis of the event rates and energy spectra, we simplify the supernova neutrino emission
process and assume that 
it can be split into two distinct phases: the
early $\nu_e$ pulse expected in the first 40 ms of the collapse ({\it shock
breakout}) and the rest beyond the first 40 ms which we treat as a {\it cooling stage},
which could in reality last up to $\sim$ 10~s.
\begin{itemize}
\item For the shock breakout we use the time-dependent energy spectra
from Burrows et al. The
calculation is based on dynamical models of core collapse supernovae
in one spatial dimension, employing a newly-developed Boltzmann
neutrino radiation transport algorithm, coupled to Newtonian
Lagrangean hydrodynamics and a consistent high-density nuclear
equation of state. Full details of the simulation can be found in
\cite{Burrows}. 
Recent studies on the initial mass progenitor dependence seem to indicate
that the major features of the early neutrino burst should be almost
independent of the initial mass of the star \cite{Takahashi:2003rn}. 
\item
For the cooling stage we assume that the energy spectra of neutrinos
is a Fermi-Dirac distribution. In order to take into account the uncertainties
in the modelling of the cooling phase, we consider five different scenarios
which correspond to simulations found in the literature. The various scenarios
can be parameterized by different assumptions on the average energies
and relative luminosities of the different neutrino flavors.
Table \ref{tab:sncoolscenario} summarizes the different possibilities
for the primary supernova neutrino flux parameters and neutrino
average energies considered in this work. 

\end{itemize}
\begin{table}[htbp]
\centering
\begin{tabular}{cccccc} \hline
Cooling & $\langle E_{\nu_e} \rangle$ & $\langle E_{\bar\nu_e} \rangle$  & $\langle
E_{\nu_{\mu,\tau}} \rangle$ = $\langle E_{\bar\nu_{\mu,\tau}} \rangle$ & Luminosity & Reference \\
scenario & MeV & MeV & MeV &  & \\
\hline
\\
{\bf I} & 11 & 16 & 25 &  L$_{\nu_e}$ = L$_{\bar\nu_e}$ = 
L$_{\nu_x}$ & \cite{Langanke} \\
\\
{\bf II} & 13 & 16 & 23 &  L$_{\nu_e}$ = L$_{\bar\nu_e}$ = 
L$_{\nu_x}$ & \cite{Totani:1997vj}\\
\\
{\bf III} & 13 & 16 & 17.6 &  L$_{\nu_e}$ = L$_{\bar\nu_e}$ = 
L$_{\nu_x}$ & \cite{Raffelt:2003en,Raffelt:2002}\\
\\
{\bf IV} & 13 & 16 & 17.6 &  L$_{\nu_e}$ = L$_{\bar\nu_e}$ = 
2L$_{\nu_x}$ & \cite{Raffelt:2003en,Raffelt:2002}\\
\\
{\bf V} & 13 & 16 & 17.6 &  L$_{\nu_e}$ = L$_{\bar\nu_e}$ = 
0.5 L$_{\nu_x}$ & \cite{Raffelt:2003en,Raffelt:2002}\\
\\
\hline
\end{tabular}
\caption{Supernova cooling phase scenarios.}   
\label{tab:sncoolscenario}
\end{table}

We assume the following neutrino oscillation situations:
\begin{itemize}
\item
Normal (m$_1^2$ $<$ m$_2^2$ $<<$ m$_3^2$) and inverted (m$_3^2$ $<<$ m$_1^2$
$<$ m$_2^2$) mass hierarchies are investigated in the analysis. 
\item
Matter effects inside the supernova are included.
\item
The following neutrino oscillations parameters are used in the analysis,
obtained from other neutrino observations \cite{nuobserv}:
\begin{eqnarray}
\begin{tabular}{ll}
$\sin^22\theta_{23} = 1.$ & $\Delta m^2_{32} \approx \Delta m^2_{31} = 3 \times 10^{-3} eV^2$\\
$\sin^2 \theta_{12} = 0.3$ & $\Delta m^2_{21} = 7 \times 10^{-5} eV^2$\\
$\sin^2 \theta_{13} < 0.02$ & \\
\end{tabular}
\end{eqnarray}
\end{itemize}

\section{Neutrino oscillations and matter effects}
\label{sec:matter}

Neutrinos produced in the high density region of the iron core interact
with matter before emerging from the supernova. Due to the non-zero
masses and non-zero vacuum mixing angles among various neutrino
flavors, flavor conversions can occur in supernovae. 

The transitions are produced mainly in the so-called resonance layers. The
resonance density can be computed as \cite{Dighe}:

\begin{equation}
\rho_{res} \approx \frac{1}{2\sqrt{2}G_F}\frac{\Delta
m^2}{E}\frac{m_N}{Y_e}cos2\theta \\
\end{equation}

\noindent where $G_F$ is the Fermi constant, $\Delta m^2$ is the mass squared
difference, $\theta$ is the mixing angle, $E$ is the neutrino energy,
$m_N$ is the nucleon mass and $Y_e$ is the number of electrons per
nucleon. 

\begin{figure}[tbp]
\centering
\begin{tabular}{cc}
\epsfig{file=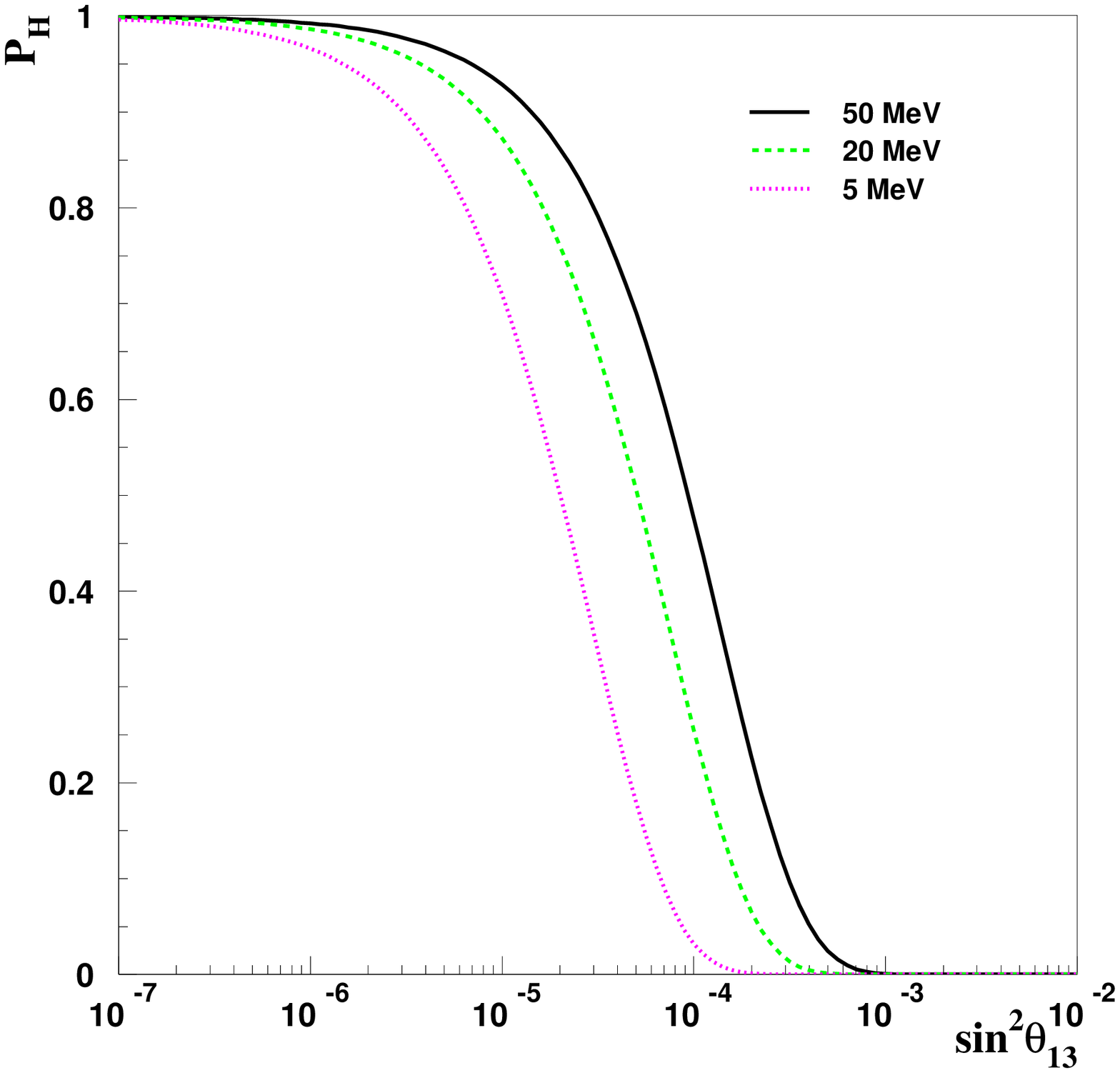,width=8.5cm}
&
\hspace{-1cm}
\epsfig{file=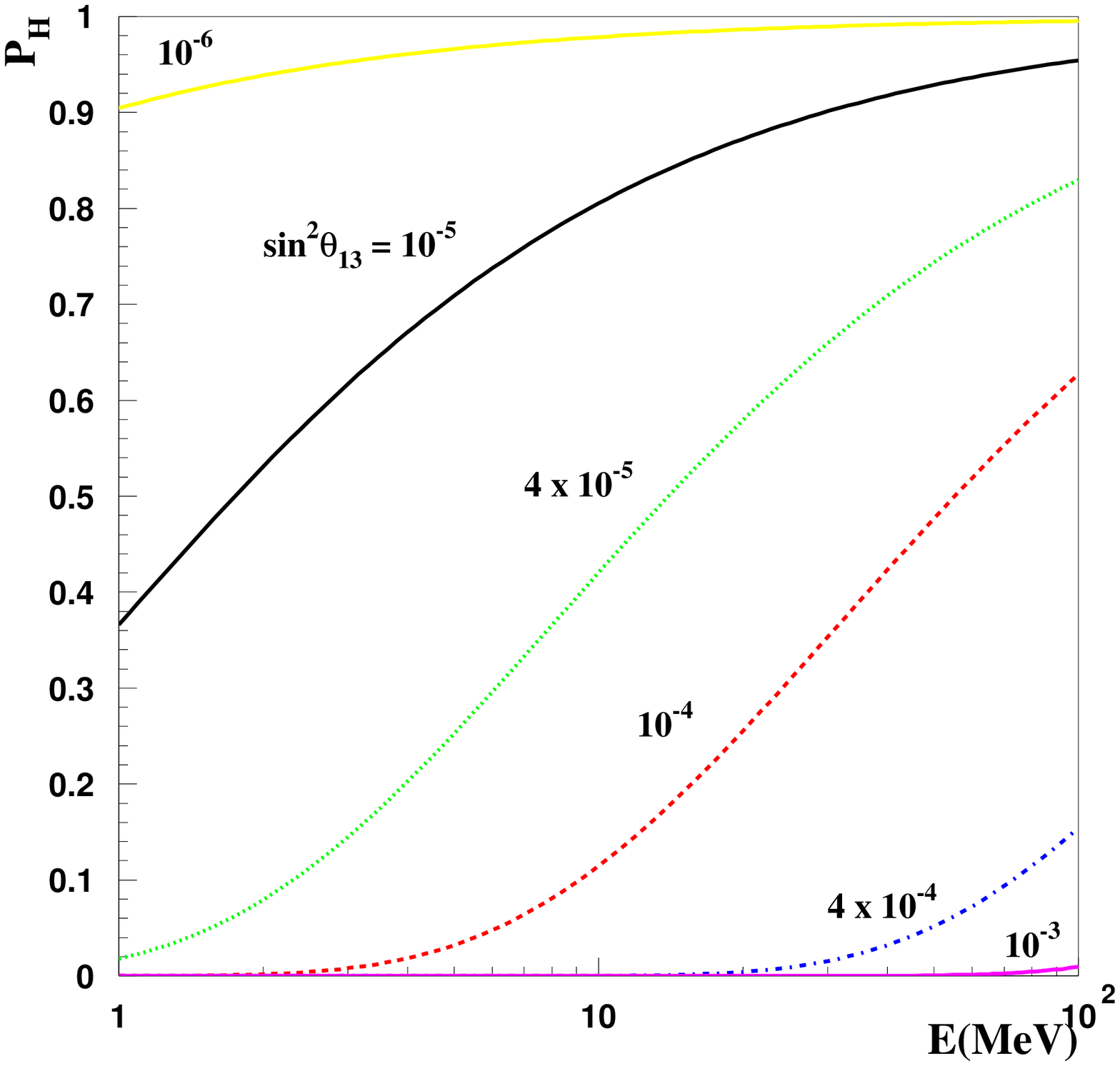,width=8.5cm}
\end{tabular}
\caption{Jump probability in the H-resonance, P$_H$, as a function of
$\sin^2\theta_{13}$ for three values of the neutrino energy (left), and
as a function of the neutrino energy for different values of
$\sin^2\theta_{13}$ (right).}
\label{fig:ph}
\end{figure}

Since the inner supernova core is too dense to allow resonance
conversion, we can consider two resonance points in the outer
supernova envelope: one at high density $\rho\approx $  10$^3$--10$^4$
g cm$^{-3}$ ({\bf H--resonance}) which is 
governed by the atmospheric parameters ($\Delta m^2_{31}$ and
$\theta_{13}$) and the other one at low density $\rho\approx$ 
10--30 g cm$^{-3}$ ({\bf L--resonance}),
characterized by the solar parameters ($\Delta m^2_{21}$ and
$\theta_{12}$).  

The transitions in the two resonance layers can be considered
independently and each transition is reduced to a two neutrino oscillation
problem. 
The H--resonance lies in the neutrino channel for normal mass
hierarchy and in the antineutrino channel for the inverted
hierarchy. The L--resonance lies in the neutrino channel for both the
hierarchies \cite{Dighe}.   

The probability of transition between one neutrino eigenstate to
another in the resonance layer (L or H) is called {\it jump
probability} (P$_L$ or P$_H$). 

The propagation through the low density region is always adiabatic for
the LMA solar parameters. It means that $P_L$ and $\overline{P}_L$ are both
equal to 0 and the neutrino mass eigenstates remain unchanged.

Considering a matter density profile of the star $\rho$(r) $\propto$
r$^{-3}$, the jump probability in the H-resonance can be written as \cite{Dighe}:
\begin{equation}
P_H \propto \exp\left[-const ~ \sin^2\theta_{13} \left(\frac{\Delta
m^2_{31}}{E}\right)^{2/3}\right]
\label{eq:ph}
\end{equation} 
Indeed, the neutrino conversion in the high density region depends on the
mixing angle \th13 and the mass squared difference between the
involved flavors, as shown in figure \ref{fig:ph}. The variation of
the jump probability with the 13-mixing angle and the neutrino energy
is plotted. We see that the crossing probability increases with
energy. Three regions can be distinguished: 

\begin{enumerate}
\item
{\it Non adiabatic region} (\s2t13 $<$ 2 $\times$ 10$^{-6}$): For these
values of $\theta_{13}$ the jump probability is almost equal to 1,
independently of the neutrino energy. 
\item
{\it Intermediate region} (2 $\times$ 10$^{-6}$ $<$ \s2t13 $<$ 3 $\times$
10$^{-4}$): In this region the value of P$_H$, between 0 and 1,
depends on $\theta_{13}$ and neutrino energy as expressed in the
equation \ref{eq:ph}.  
\item
{\it Adiabatic region} (\s2t13 $>$ 3 $\times$ 10$^{-4}$): The jump
probability is equal to 0 for any value of the neutrino energy. Then,
there is no conversion between eigenstates in the resonance
layer.
\end{enumerate}

\begin{figure}[tbp]
\centering
\vspace{-1cm}
\epsfig{file=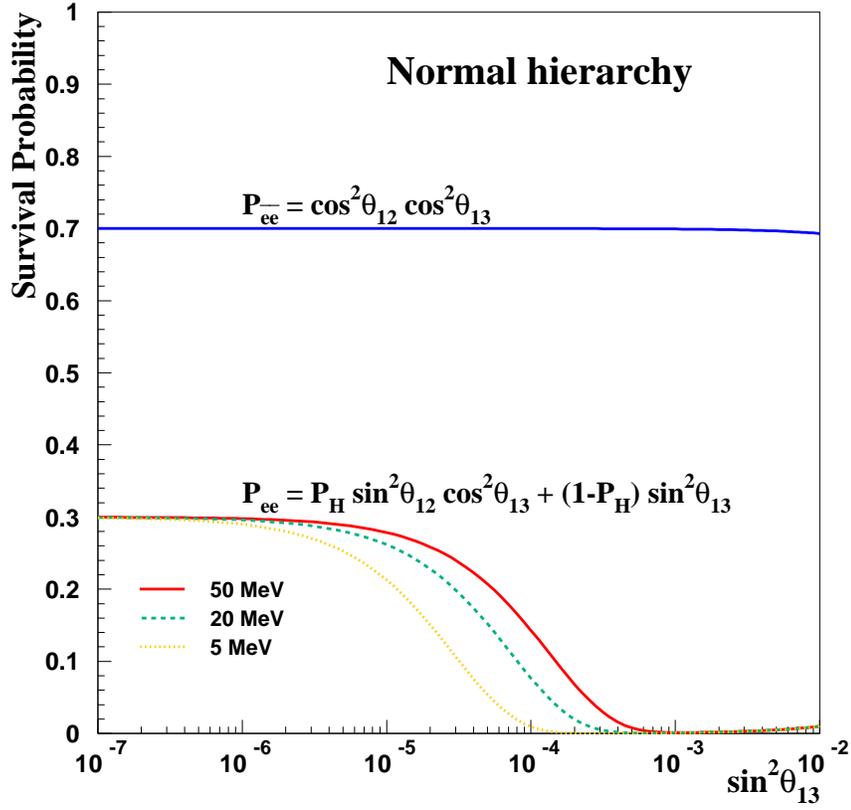,width=12cm} \\
\vspace{-1cm}
\epsfig{file=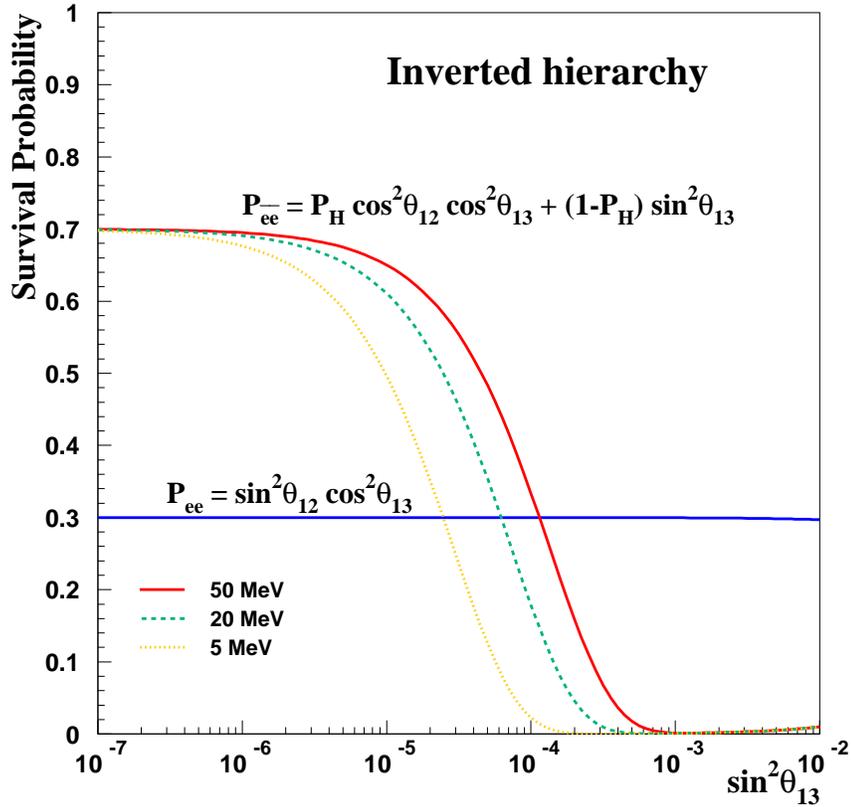,width=12cm}
\caption{Survival probabilities for normal (top) and inverted (bottom)
hierarchies as a function of \s2t13 and different values of neutrino
energies.} 
\label{fig:survprob}
\end{figure}

The effect of oscillations in the neutrino fluxes is summarized in
equation \ref{eq:flux}, where the expected fluxes at the detector
($\phi_{\nu}$) are computed in terms of the original fluxes at the SN
core ($\phi^o_{\nu}$) and the survival probabilities P$_{ee}$ =
P($\nu_e$ $\to$ $\nu_e$)  and $\overline{P}_{ee}$ = P($\bar\nu_e$
$\to$ $\bar\nu_e$)\footnote{We neglect the possible effect of the propagation through the mass of the Earth.}:
\begin{eqnarray}
\begin{tabular}{l}
$\phi_{\nu_e} = \phi^o_{\nu_e} P_{ee} + \phi^o_{\nu_x} (1-P_{ee})$ \\
$\phi_{\bar\nu_e} = \phi^o_{\bar\nu_e} \overline{P}_{ee} + \phi^o_{\bar\nu_x}
(1-\overline{P}_{ee})$ \\
$\phi_{\nu_\mu} + \phi_{\nu_\tau} = \phi^o_{\nu_e} (1-P_{ee}) +
\phi^o_{\nu_x} (1+P_{ee})$ \\
$\phi_{\bar\nu_\mu} + \phi_{\bar\nu_\tau} = \phi^o_{\bar\nu_e} (1-\overline{P}_{ee}) +
\phi^o_{\bar\nu_x} (1+\overline{P}_{ee})$
\end{tabular}
\label{eq:flux}
\end{eqnarray}

\noindent with $\phi^o_{\nu_x}$ = $\phi^o_{\nu_\mu}$ =
$\phi^o_{\nu_\tau}$ = $\phi^o_{\bar\nu_x}$ = $\phi^o_{\bar\nu_\mu}$ =
$\phi^o_{\bar\nu_\tau}$.
If the original fluxes were equal between flavors, any oscillation
effect could be observed.

The survival probabilities can be written in terms of jump
probabilities as \cite{Dighe}:
\begin{eqnarray}
\begin{tabular}{l}
$P_{ee} = P_H P_L |U_{e1}|^2 + P_H (1-P_L) |U_{e2}|^2 + (1-P_H)
|U_{e3}|^2$ \\
$\overline{P}_{ee} = (1-\overline{P}_L) |U_{e1}|^2 + \overline{P}_L |U_{e2}|^2$
\end{tabular}
\label{eq:peenh}
\end{eqnarray}
\noindent for normal hierarchy (\Dm32 $>$ 0) and
\begin{eqnarray}
\begin{tabular}{l}
$P_{ee} = P_L |U_{e1}|^2 + (1-P_L) |U_{e2}|^2$ \\
$\overline{P}_{ee} = P_H (1-\overline{P}_L) |U_{e1}|^2 +
\overline{P}_L P_H |U_{e2}|^2 + (1-P_H) |U_{e3}|^2$
\end{tabular}
\label{eq:peeih}
\end{eqnarray}
\noindent for inverted hierarchy (\Dm32 $<$ 0). 

Considering the values of the jump probabilities and using the standard
parametrization of the mixing matrix, the survival probabilities can
be computed as shown in figure \ref{fig:survprob}. The evolution of
the probability P$_{ee}$ ($\overline{P}_{ee}$) that a neutrino emitted
as $\nu_e$ ($\bar\nu_e$) at the neutrinosphere remains as $\nu_e$
($\bar\nu_e$) at Earth is plotted as a function of the \th13 mixing
angle for different values of neutrino energy.

The upper plot corresponds to the normal hierarchy case (n.h.) and the lower
one to the inverted hierarchy (i.h.). For n.h., the survival probability in
the antineutrino channel does not depend on the neutrino energy and
only very weakly on \th13. However, the neutrino channel is very
sensitive to \th13 via the $P_H$ parameter. 
For inverted hierarchy the neutrino channel is independent on \th13
while the antineutrino channel depends on the 13-mixing angle and
neutrino energy. 

According to the adiabaticity conditions in the H resonance and the
value of the \th13 angle, three regions can be distinguished. 
If energy spectra are different, then \th13 and the mass hierarchy can
be probed as follows:

\begin{itemize}
\item
For {\it small mixing angle} (\s2t13 $<$ 2 $\times$ 10$^{-6}$), there are no
effects on \th13 and both hierarchies give similar results. Therefore,
in this region we can not distinguish among hierarchies and only an
upper bound on \s2t13 can be set.
\item
For {\it intermediate \th13} (2 $\times$ 10$^{-6}$ $<$ \s2t13 $<$ 3
$\times$ 10$^{-4}$) maximal sensitivity to the angle is achieved and
measurements of the angle are possible in this region. 
\item
For {\it large mixing angle} (\s2t13 $>$ 3 $\times$ 10$^{-4}$) maximal
conversions occur. The effect of the mixing is strong in neutrino
(antineutrino) channel if the mass hierarchy is normal (inverted). We
are able to probe the mass hierarchy in this region but only a lower
bound on \th13 can be put.
\end{itemize}

\section{Cross-sections and expected supernova neutrino signal on
liquid Ar detectors}
\label{sec:signal}

At low energy, where nuclear processes dominate, three nuclei-scattering processes
can be studied ($\nu_e$ charged current
absorption on $Ar$ with production of excited $K$, $\bar\nu_e$
charged current absorption on $Ar$ with production of excited $Cl$ and
neutral current interactions on $Ar$ from all neutrino flavors) in addition
to elastic scattering on atomic electrons from all neutrino species. 
There is a possibility to separate the various channels by a classification
of the associated photons from the $K$, $Cl$ or $Ar$ de-excitation,
which exhibit specific spectral lines, or by the absence of associated photons in case
of the elastic scattering.

In a liquid Argon TPC, it is hence possible to identify the following events:
\begin{enumerate}
\item  {\bf elastic scattering} on electrons
\begin{equation}
\nuall ~e^- \to \nuall ~e^-
\label{eq:elas}
\end{equation}

\item {\bf charged-current (CC) interactions on argon}
\begin{equation}
\nue ~^{40}Ar \to e^- ~^{40}K^*
\label{eq:ccnue}
\end{equation}
\begin{equation}
\bar\nue ~^{40}Ar \to e^+ ~^{40}Cl^*
\label{eq:ccanue}
\end{equation}

\item {\bf neutral-current (NC) interactions on argon}
\begin{equation}
\nuall ~^{40}Ar \to \nuall ~^{40}Ar^*
\label{eq:ncnu}
\end{equation}
\end{enumerate}

\subsection{Elastic scattering processes}
As well known, the elastic neutrino scattering off target-electrons has a total cross
section that increases linearly with energy:
\begin{eqnarray}
\begin{tabular}{lccc}
$\sigma(\nu_e e^- \to \nu_e e^- )$ & = & $9.20 \times 10^{-45} E_{\nu_e} {\rm (MeV) \ \ \ \ cm}^2$ \\
$\sigma(\bar\nu_e e^-  \to \bar\nu_e e^- )$ & = & $3.83 \times 10^{-45} E_{\bar\nu_e} {\rm (MeV) \ \ \ \ cm}^2$ \\
$\sigma(\nu_{\mu ,\tau} e^-  \to \nu_{\mu ,\tau} e^- )$ &=& $1.57 \times 10^{-45} E_{\nu_{\mu ,\tau}} {\rm (MeV) \ \ cm}^2$ \\
$\sigma(\bar\nu_{\mu ,\tau} e^-  \to \bar\nu_{\mu ,\tau} e^- )$ &=& $1.29 \times 10^{-45} E_{\bar\nu_{\mu ,\tau}} {\rm (MeV) \ \ cm}^2$ \\
\end{tabular}
\end{eqnarray}

All neutrino species contribute to elastic scattering. The
experimental signature consists in a single recoil electron. Since,
the direction of this electron is highly correlated to the incoming
$\nu$ direction, these events have the potentiality of precisely
determining the location of the supernova source\cite{snmemo}.

Figure \ref{fig:allxsec} shows the cross
sections of the elastic processes as a function of neutrino
energy.  We recall that the reaction involving $\nu_e$'s is
the most probable since both CC and NC currents contribute.
Electron antineutrinos are suppressed by helicity. The cross-section
for muon and tau neutrinos is the smallest, since they proceed
through NC current only.

These features are important since while the nature of the neutrino
cannot be identified in elastic events, the total number of expected
elastic events for a given flux depends on their nature because of
its cross-section dependence. Hence, if the total
flux of neutrino is constrained, the number of elastic events is sensitive
to the nature of the neutrinos emitted by the supernova.

\subsection{Nuclear processes}
In order to compute the event rates, the relevant nuclear cross-sections
must be in principle known with relatively good precision from energies
ranging from a few MeV up to 100 MeV. 
For the nuclear processes, we use the calculations
computed for the first time using the RPA including all the
multipoles \cite{martinez}. 

\begin{figure}[htbp]
\centering
\epsfig{file=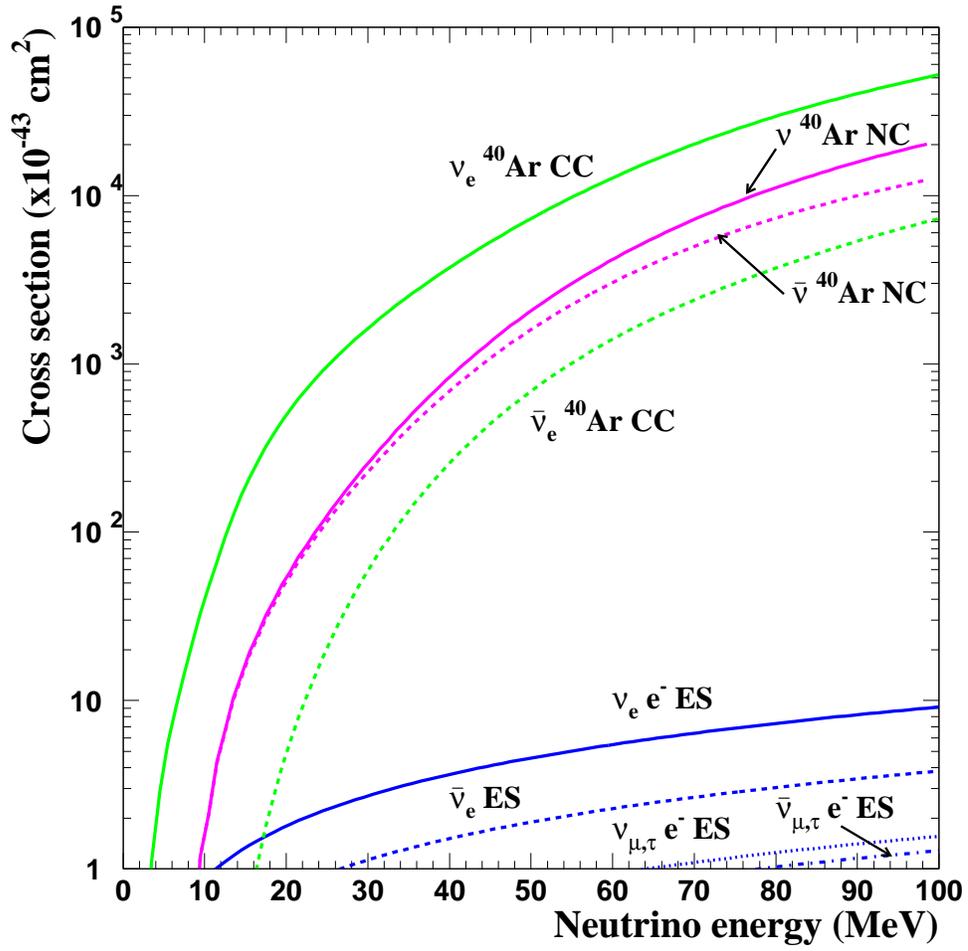,width=14.cm}
\caption{Neutrino cross sections relevant to the supernovae detection
with a liquid Argon TPC detector.} 
\label{fig:allxsec}
\end{figure}

Figure \ref{fig:allxsec} shows the cross
sections of all the processes as a function of neutrino
energy.  The NC processes present high cross sections of the same
order of magnitude as the ones from the CC reactions. They are even
bigger than those from the $\anue$CC events. 
This can be understood in terms of the energy threshold and nuclear effects
of the various reactions.
The $Q$--value of the $\nu_e$ CC is 1.50 MeV while for the $\bar\nu_e$ CC case is 7.48
MeV, hence electron antineutrinos are kinematically disfavored. In addition,
the 1$^+$ and 0$^+$ allowed transitions are suppressed by nuclear
structure reasons and the cross section is dominated by the other
multipoles. Hence, electron antineutrinos cross-section is smaller than for neutrinos.
The first excited state
of Argon has an energy of 1.46 MeV with respect to the ground state, so
the threshold for neutral current reaction is low.

\begin{figure}[htbp]
\centering
\epsfig{file=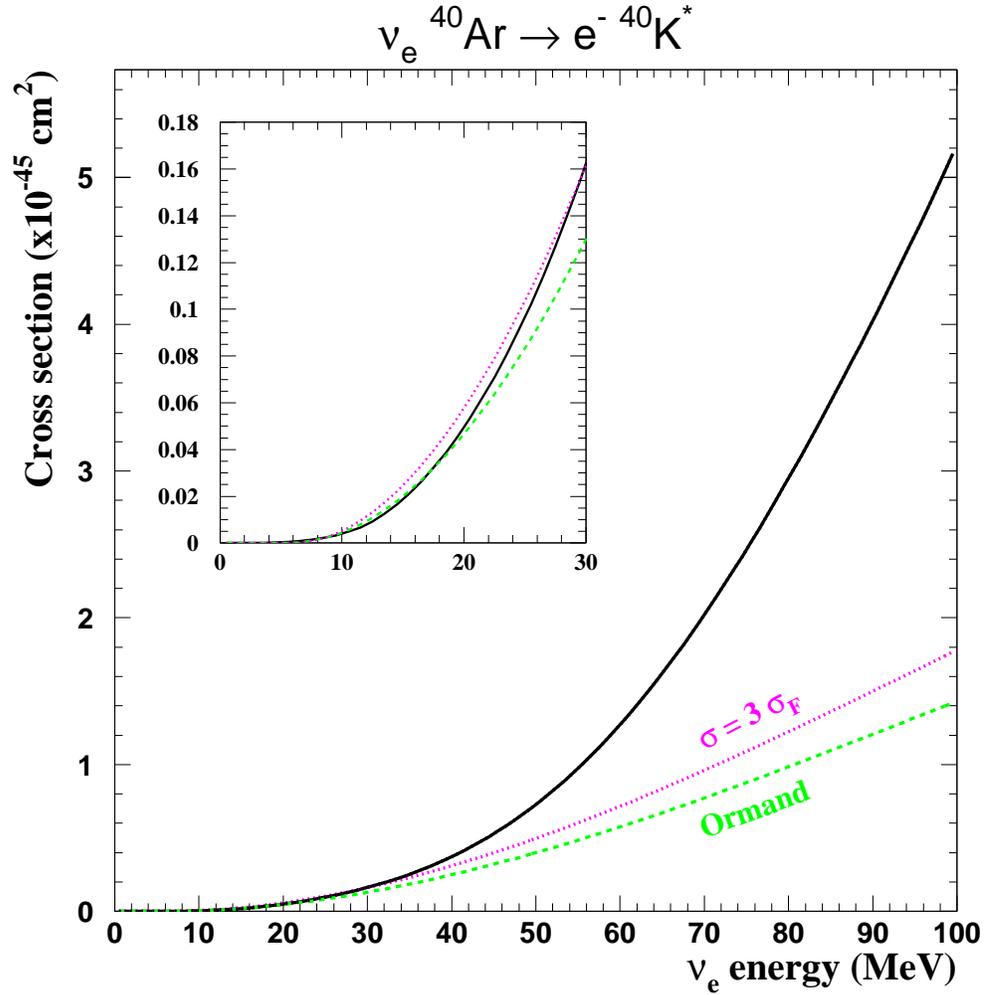,width=14.cm}
\caption{$\nu_e$ CC cross section as a function of the neutrino
energy. The dashed line corresponds to the Ormand \cite{Ormand} cross
section calculation, dotted line assumes that the total cross section of
the absorption interaction is 3 times the cross section of the Fermi
transition \cite{Burrows} and the solid line is the cross section used
in this analysis calculated from RPA including all the transitions
\cite{martinez}.}  
\label{fig:compxsec}
\end{figure}

In the case of the $\nu_e$ CC process, various estimates are
available in the literature (see e.g. \cite{Ormand} and \cite{Raghavan}).
It was assumed in the past, that the $\nue ^{40}Ar$ absorption rate proceeds 
through two main channels:
the superallowed Fermi transition to the 4.38 MeV excited isobaric
analog $^{40}$K$^*$ state, and other Gamow-Teller transitions to several excited $^{40}$K
states. This assumption is correct for neutrino energies relevant to solar neutrino
detection, however is not adequate for the energy range of supernova neutrinos,
which extends up to 100~MeV.
In the recent calculations that we use, the cross-section
has been computed by random phase approximation (RPA) for neutrino
energies up to 100 MeV, including all the multipoles
up to J=6 and both parities.
Figure \ref{fig:compxsec} compares this cross section from three
different sources: the dashed line corresponds to the shell model
calculations performed by Ormand \cite{Ormand} considering only Fermi
and Gamow--Teller transitions; the dotted line is the assumption made in
\cite{Burrows} where they take the total
$\nu_e$ absorption cross section as 3 times the cross section of the
pure Fermi transition computed in \cite{Raghavan}; finally the solid
line is the cross section used in this analysis \cite{martinez}.

We see that the assumption of $\sigma_{tot}$ = 3 $\sigma_F$
overestimates the cross section for energies smaller than 30 MeV,
which is the energy range of $\nu_e$ from breakout, and therefore affects the
expected number of events from the $\nue$ burst (discussed in section~\ref{sec:burst}).  

On
the other hand, both Ormand and $\sigma_{tot}$ = 3 $\sigma_F$ cross sections underestimate
the cross section for high energies up to 100 MeV, relevant for
supernova neutrinos from the cooling phase (discussed in section~\ref{sec:cool}).
The cross section including all the multipoles
grows faster than the others,  showing that indeed not only the Fermi and Gamow--Teller
transitions are important but also the rest of multipoles.

\section{Supernova collapse and neutrino breakout phase}
\label{sec:burst}

\subsection{Time and energy dependence without neutrino oscillations}

The neutrino shock breakout burst is the signal event in the supernova
core collapse evolution. Its detection and characterization could test
fundamental aspects of the current collapse supernova paradigm. The
prompt electron neutrino burst is in principle observable and represents a
diagnostic of the fundamental collapse supernova behavior. The
characteristic time of the breakout burst is of the order of tens of ms.

The advantage of the early phase analysis is that it should not be affected by
the time evolution of the density structure of the star nor whether the
remnant is a neutron star or a black hole (see e.g. \cite{Burrows}). 
The shock wave takes about
2 seconds to reach the H-resonance region. Therefore, the potential
time dependence of neutrino oscillations due to shock propagation can
be neglected in this phase. Clean detection of the early part of the supernova
signal is therefore a very important aspect of experimental supernova detection.



As already mentioned, a liquid Argon TPC has an excellent sensitivity to $\nu_e$ neutrinos mainly
through the CC process. This provide unique information
about the early breakout pulse. We will however show that neutral currents as well
as elastic scattering, although lesser in rate, can provide important information as well
on the shock breakout.

In order to study the detection of the shock breakout, 
we compute the differential number of neutrinos detected at a distance $D$ from the
supernova:
\begin{equation}
dN_{events} = \frac{N_{targets}}{4\pi D^2}\left(\frac{d^2N_{\nu}}{dE_{\nu}dt}\right)
\sigma(E_{\nu})dE_{\nu}dt
\end{equation}
\noindent where $N_{targets}$ is the number of targets, $D$ is the
supernova distance, $E_{\nu}$ is the
neutrino energy, $t$ is the detector time, ${dN_{\nu}}/{dE_{\nu}dt}$
is the time and energy spectra of the neutrinos and $\sigma(E_{\nu})$
is the cross section of a reaction process. 

We use the simulation of the early phase of the
neutrino burst performed by Burrows et al. \cite{Burrows} which yields
${d^2N_{\nu}}/{dE_{\nu}dt}$
in discrete steps of time.

\begin{figure}[htbp]
\vspace{-1cm}
\centering
\epsfig{file=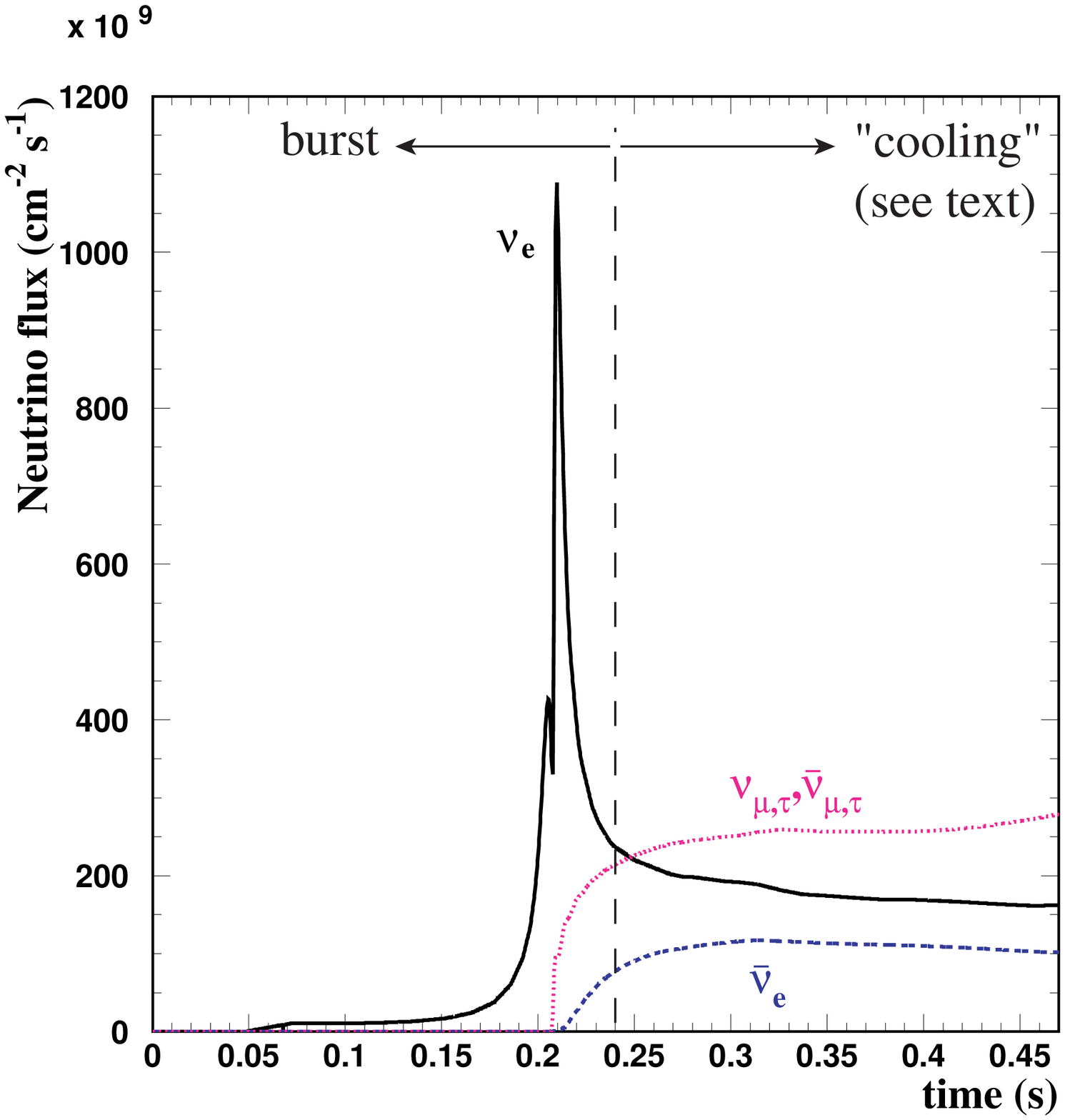,width=11cm}
\epsfig{file=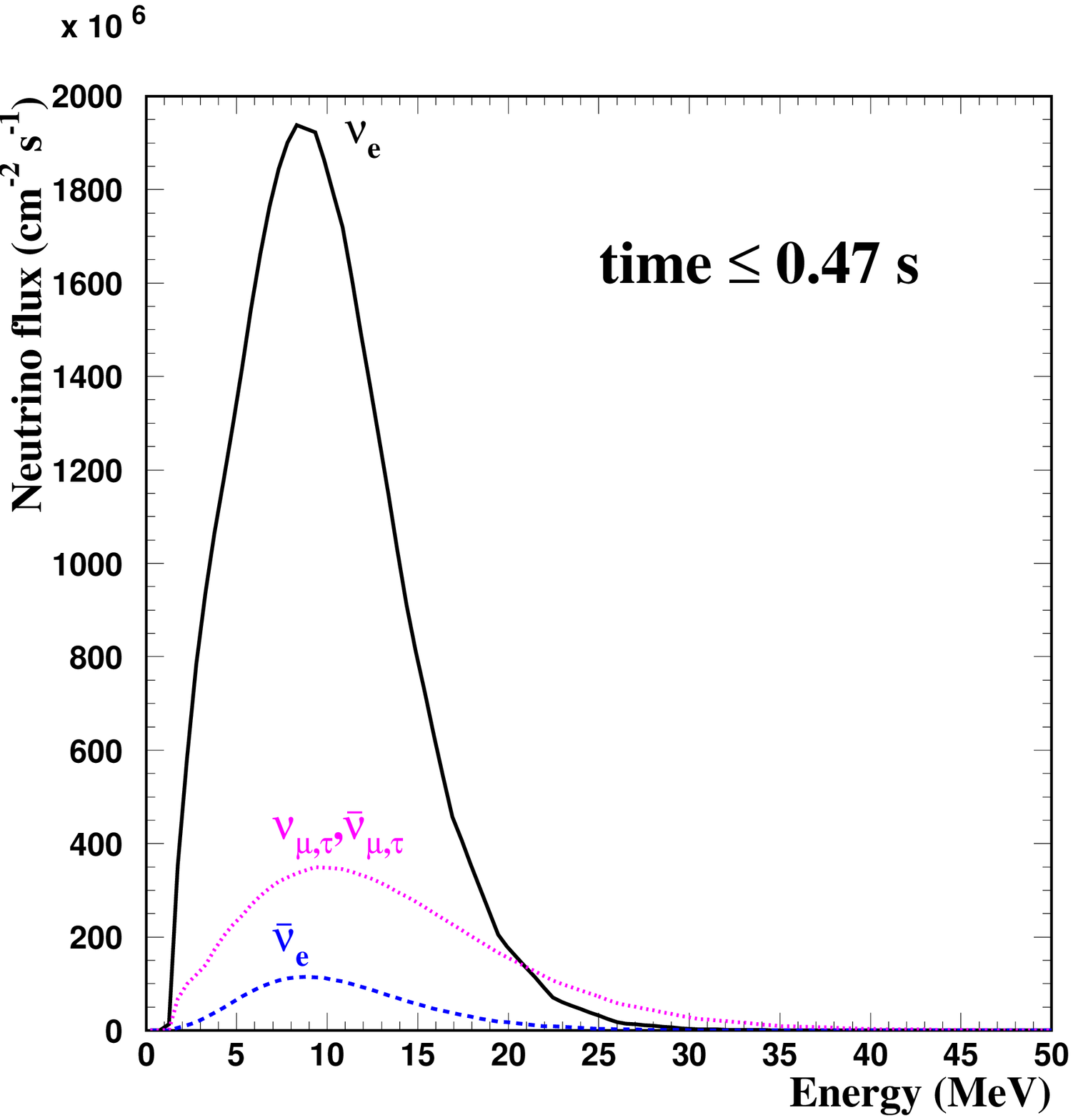,width=7.5cm}
\epsfig{file=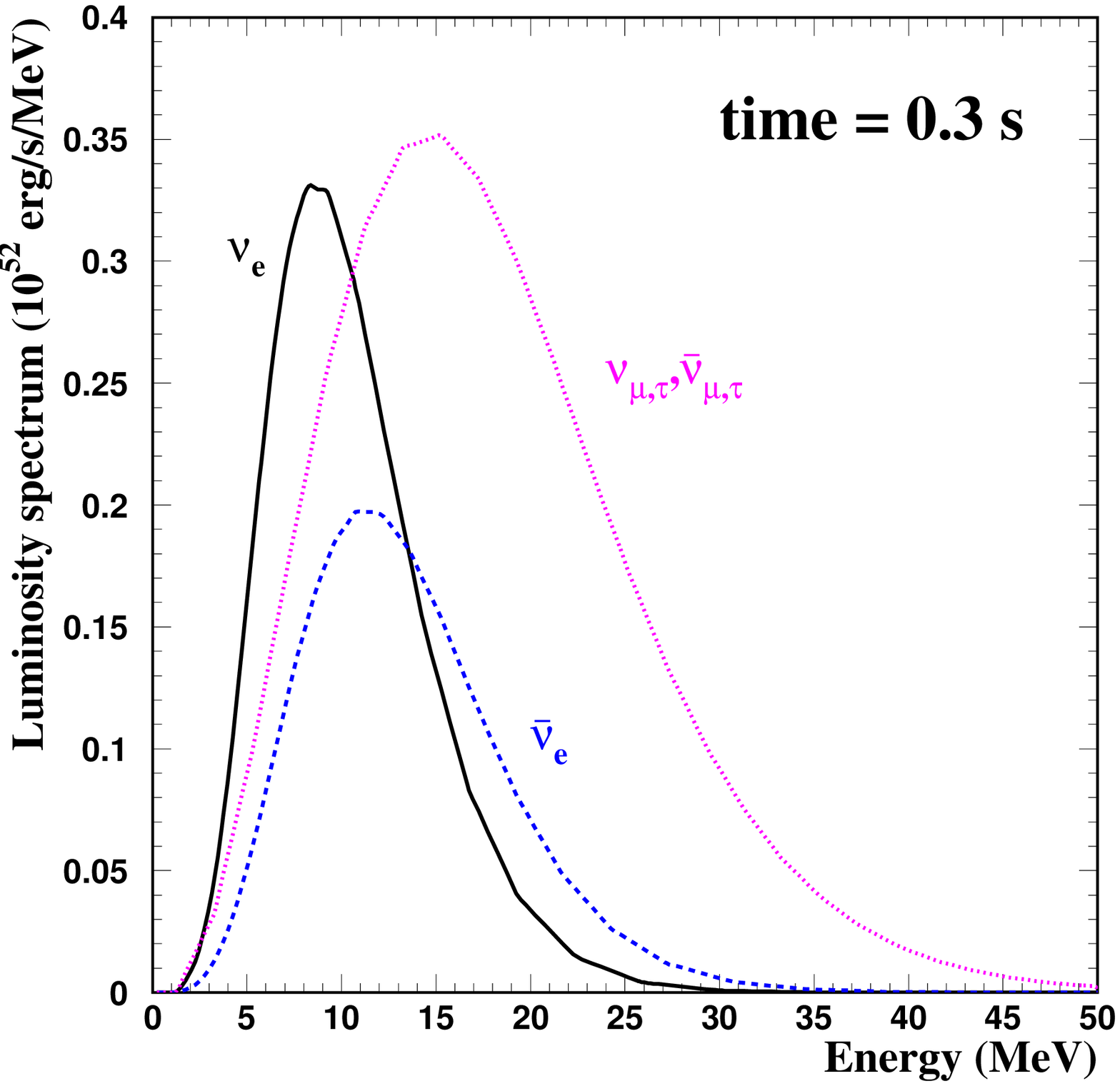,width=7.5cm}
\caption{Illustration of the prediction
of \cite{Burrows} without neutrino flavor oscillations:
prediction of the neutrino flux at Earth at a distance of 
10 kpc (top), time--integrated energy spectrum corresponding to
the first 270 ms after the bounce (bottom-left) and luminosity
spectrum of $\nu_e$ (solid line), $\bar\nu_e$ (dashed line) and
$\nu_{\mu,\tau}$, $\bar\nu_{\mu,\tau}$ (dotted line) neutrinos at 100
ms after the shock breakout (bottom-right). We treat the times before
t=240 ms as ``burst'' and after as ``cooling''.} 
\label{fig:flux}
\end{figure}

The prediction for the time (energy integrated) evolution of the neutrino fluxes
for the different flavors in absence of neutrino oscillations is shown in 
Figure~\ref{fig:flux}(top). The time origin is defined such that the supernova core
bounce time corresponds to $t=200\ \rm ms$. This time is taken as the
reference for the rest of the processes. The Figure~\ref{fig:flux} (bottom-left) shows
the time--integrated energy spectrum of the first 470 ms of the time, namely
up to 270 ms after the bounce. As expected, the flux is characterized
by the bright peak of the $\nu_e$ emission due to the supernova
shock breakout. 

For comparison, we show in Figure~\ref{fig:flux}(bottom-right) the particular
energy distribution at 100 ms after the bounce ($t=300\ \rm ms$). This picture
illustrates how the energy spectra of the different 
neutrino flavors is predicted to change with time and how the bright $\nu_e$
becomes superseded by the neutrinos of all other flavors and helicities.

For the sake of calculations, we define the shock breakout
burst by integrating up to $t=240\ \rm ms$, namely we define a duration
for the burst of 40 ms. We use these in the next section. In order
to simplify the analysis, we treat the times
after $t=240\ \rm ms$ as part of the cooling phase (see Section~\ref{sec:cool}).

\subsection{Effect of flavor oscillations and predicted event rates}

Neutrino oscillations modify
dramatically the flavor content of the neutrinos emitted from the supernova
during the burst. 
A reduction of the $\nu_e$ peak due to the effect of oscillations
can be readily understood from Figure~\ref{fig:survprob}, which
plots the survival probability of electron neutrinos for normal and 
inverted mass hierarchies. 

\subsubsection{Small and large mixing angle approximations}
We can first distinguish four limiting cases
depending on the type of mass
hierarchy and the value of the $\theta_{13}$ mixing angle (see equations~\ref{eq:peenh} and \ref{eq:peeih}
and Figure~\ref{fig:survprob}):
\begin{enumerate}
\item n.h.-L for
normal mass hierarchy and large $\theta_{13}$, where $P_{ee}\approx 0$
and $\overline{P}_{ee}\approx 0.7$,
\item n.h.-S for normal hierarchy
and small $\theta_{13}$, where $P_{ee}\approx 0.3$ and
$\overline{P}_{ee}\approx 0.7$, 
\item i.h.-L  for inverted mass hierarchy and large
$\theta_{13}$, where $P_{ee}\approx 0.3$ and $\overline{P}_{ee}\approx
0$,  
\item i.h.-S for inverted mass hierarchy and small $\theta_{13}$,
where $P_{ee}\approx 0.3$ and $\overline{P}_{ee}\approx 0.7$,
\end{enumerate}
and where ``small'' (resp. ``large'') mean $\sin^2\theta_{13}<2\times 10^{-6}$ 
(resp. $\sin^2\theta_{13}>3\times 10^{-4}$). In the intermediate mixing angle
region, there survival probability depends on the neutrino energy.
Note that the no oscillation case
corresponds to  $P_{ee}\equiv 1$ and $\overline{P}_{ee}\equiv 1$.

We can therefore summarize the situation as follows: the suppression of the $\nu_e$ 
flavor is maximal in the case
of a normal mass hierarchy and large \th13 mixing angle. This corresponds to the total
conversion of $\nu_e$'s into $\nu_{\mu,\tau}$'s as they traverse the supernova
matter. For the other cases, the oscillation reduces the expected $\nu_e$ flux by
a factor 3 compared to the non-oscillation case. 

Tables \ref{tab:ratesburst3kton} and
\ref{tab:ratesburst70kton} show the expected number
of events during the burst in a 3 kton and 70
kton liquid Argon TPC detector, respectively, for the four
oscillation scenarios and compares them to the non oscillation case.
For completeness we have separated the contributions from 
$\nu_e$'s, $\nu_{\mu,\tau}$'s and similarly for antineutrinos.
However, it should be recalled that four channels are in fact
accessible experimentally: the $\nu_e$ CC, the $\bar\nu_e$ CC,
the NC and the elastic scattering.

With the assumed distance of 10 kpc, the burst produces in a 3 kton
detector and in absence of neutrino oscillations 2 elastic events, 8 $\nu_e$ CC
and 1 NC event for a total of 11 events. In case of the 70 kton detector,
the rates are 37 elastic events, 192 $\nu_e$ CC
and 39 NC event for a total of 268 events.

\begin{table}[tb]
\centering
\begin{tabular}{clccccc} \hline
& &\multicolumn{5}{c}{\bf Burst (t $<$ 240 ms) -- 3 kton detector} \\
Reaction & & No & \multicolumn{2}{c}{Oscillation
(n.h.)} & \multicolumn{2}{c}{Oscillation (i.h.)} \\
 & & oscillation & Large $\theta_{13}$ & Small $\theta_{13}$ & Large
$\theta_{13}$ & Small $\theta_{13}$\\ \hline
{\bf Elastic} & & & & & & \\
& $\nu_e \, e^-$ & 2 & $<$1 & 1 & 1 & 1 \\
& $\bar\nu_e\, e^-$ & $<$1 & $<$1 & $<$1 &  $<$1& $<$1 \\
& $(\nu_\mu+\nu_\tau)\, e^-$ & $<$1 & $<$1 & $<$1 & $<$1 & $<$1 \\
& $(\bar\nu_\mu +\bar\nu_\tau)\, e^-$ & $<$1 & $<$1 & $<$1 & $<$1 & $<$1 \\
& total $\nu\, e^-$ & 2 & $<$1 & 1 & 1 & 1 \\
\hline
{\bf Absorption} & & & & & & \\
{\bf CC} & $\nu_e$ $^{40}$Ar       & 8 & 2 & 4 & 4 & 4 \\ 
& $\anue$ $^{40}$Ar                  & $<$1 & $<$1 & $<$1 & $<$1 & $<$1 \\
\hline
{\bf NC} & $\nu$ $^{40}$Ar       & 1 & 1 & 1 & 1 & 1 \\
          & $\bar{\nu}$ $^{40}$Ar & $<$1 & $<$1 & $<$1 & $<$1 & $<$1 \\ 
\hline
{\bf Total} & & 11 & 3 & 6 & 6 & 6 \\\hline
\end{tabular}
\caption{Expected neutrino events in a 3 kton detector
during the burst (see text for definition). Neutrino
oscillations with matter effects are included for
different mass hierarchies and \th13 values.}   
\label{tab:ratesburst3kton}
\end{table}

We consider first the elastic events. As already mentioned, 
oscillation has a strong effect on the $\nu_e$ burst. The $\nu_e$ will be
suppressed by a factor 3 for three of the four oscillation scenarios
we have considered, or can even be maximally suppressed in the case
of normal mass hierarchy and large $\theta_{13}$. Neutrino flavor oscillation
is of course a unitary process, hence, the oscillated $\nu_e$'s will be transformed
into other flavors. However, since the cross-section for the elastic $\nu_e$ process
is higher than for other processes, we expect a reduction of the number of elastic
events. Indeed, as visible in Tables \ref{tab:ratesburst3kton} and
\ref{tab:ratesburst70kton}, the elastic burst signal is highly suppressed for the
oscillation scenarios. With our definition of burst, with the assumption of
a 40 ms duration, there is in fact a smaller component of $\nu_\mu$'s,  $\nu_\tau$'s,
 $\bar\nu_\mu$'s,  $\bar\nu_\tau$'s and  $\bar\nu_e$'s (see Figure~\ref{fig:flux} ). 
Finally, the elastic events are suppressed by a factor roughly 4 for the case
of a normal mass hierarchy and large \th13 mixing angle, and roughly by a factor 2 for
the other cases.

We now turn to the nuclear charged current events. The $\nu_e$ CC reaction is in principle
a golden channel to study the burst. Because of the oscillation of $\nu_e$ into
other flavors (not completely compensated by oscillation of other flavors into $\nu_e$), 
the size of the peak is reduced compared
to the non-oscillation case: the suppression is almost a factor 5 for the case
of a normal mass hierarchy and large \th13 mixing angle, and roughly a factor 2.5 for
the other cases.

Finally, we can consider the neutral current events. Obviously they are not affected
by oscillations, since neutral currents are flavor symmetric. They hence constitute
an excellent probe for the supernova properties independent of the neutrino oscillation
intrinsic properties. We therefore stress the importance of this signal to decouple
supernova from oscillation physics. Unfortunately the rates in the 3 kton detector
are rather modest: at the level of one event (independent of the oscillation
scenario). For the case of a 70 kton detector,
we expect in total 39 events (independent of the oscillation scenario).

\begin{table}[tb]
\centering
\begin{tabular}{clccccc} \hline
& &\multicolumn{5}{c}{\bf Burst (t $<$ 240 ms) -- 70 kton detector} \\
Reaction & & No & \multicolumn{2}{c}{Oscillation
(n.h.)} & \multicolumn{2}{c}{Oscillation (i.h.)} \\
 & & oscillation & Large $\theta_{13}$ & Small $\theta_{13}$ & Large
$\theta_{13}$ & Small $\theta_{13}$\\ \hline
{\bf Elastic} & & & & & & \\
& $\nu_e \, e^-$ & 34 & 3 & 12 & 12 & 12 \\
& $\bar\nu_e\, e^-$ & 1 & 1 & 1 & 1 & 1 \\
& $(\nu_\mu+\nu_\tau)\, e^-$ & 1 & 6 & 5 & 5 & 5 \\
& $(\bar\nu_\mu +\bar\nu_\tau)\, e^-$ & 1 & 1 & 1 & 1 & 1 \\
& total $\nu\, e^-$ & 37 & 11 & 19 & 19 & 19 \\
\hline
{\bf Absorption} & & & & & & \\
{\bf CC} & $\nu_e$ $^{40}$Ar       & 192 & 41 & 86 & 86 & 86 \\ 
& $\anue$ $^{40}$Ar                  & $<$1 & $<$1 & $<$1 & 1 & $<$1 \\ 
\hline
{\bf NC} & $\nu$ $^{40}$Ar       & 28 & 28 & 28 & 28 & 28 \\
          & $\bar{\nu}$ $^{40}$Ar & 11 & 11 & 11 & 11 & 11 \\ 
\hline
{\bf Total} & & 268 & 91 & 144 & 145 & 144 \\\hline
\end{tabular}
\caption{Expected neutrino events in a 70 kton detector 
during the burst (see text for definition). Neutrino
oscillations with matter effects are included for
different mass hierarchies and \th13 values.}
\label{tab:ratesburst70kton}
\end{table}

The information about the shock breakout 
mechanism that can be
extracted from the 3 kton detector is hence marginal, in particular in the case of the reduced number of
expected events because of oscillation effects. A supernova could explode at
a closer distance, however, this is an unlikely event.
In contrast, a 70 kton
detector could provide hundreds of events, 
enough statistics to study the $\nue$ neutrinos from
the burst. 

\subsubsection{Study of energy spectra}
The importance of the statistics can be appreciated when one studies
the energy spectra of the events. Sufficient statistics would allow
to study the energy spectra of the events which provides an extra
signature to separate supernova from oscillation physics.

\begin{figure}[htbp]
\centering
\epsfig{file=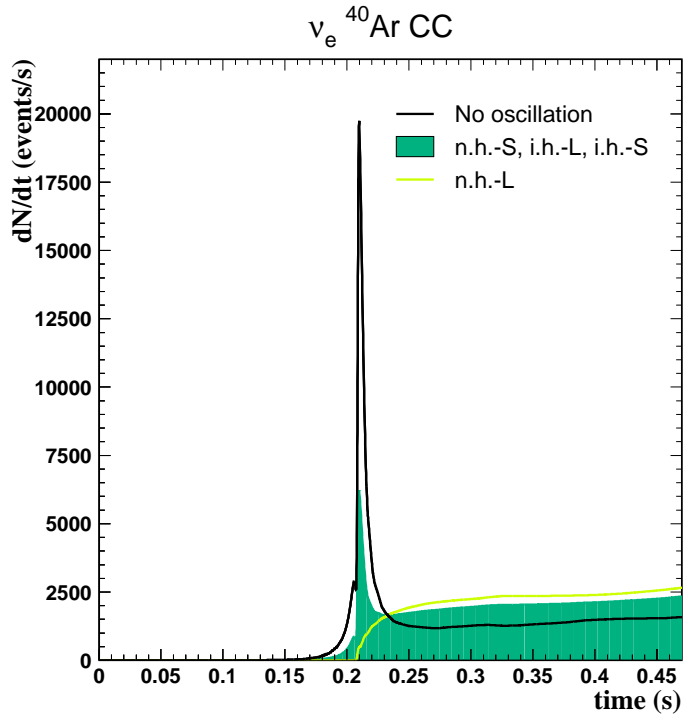,width=10cm}
\epsfig{file=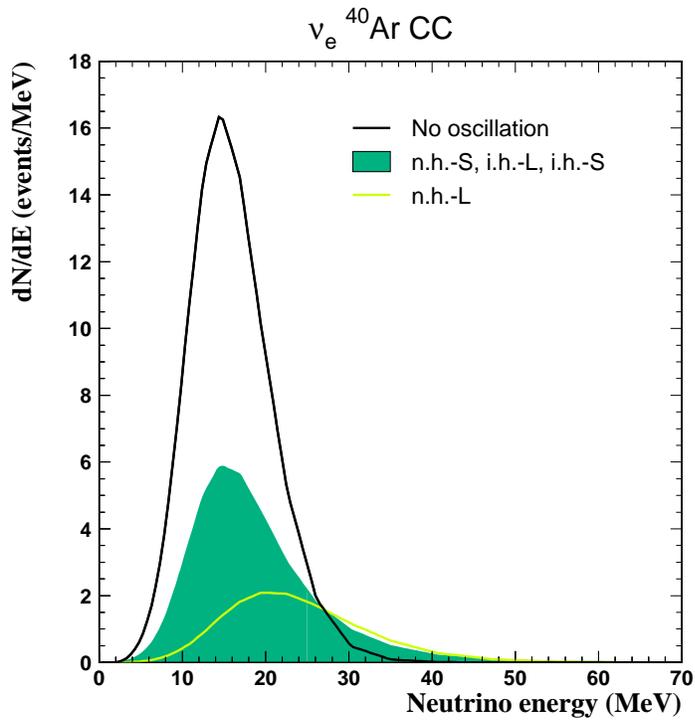,width=10cm}
\caption{SN burst: time evolution of the $\nu_e$ CC event rate  (top) and the corresponding time integrated event
spectra (bottom). The non oscillation (dark curve), n.h.--L (light
curve) and n.h.--S, i.h.--L and i.h.--S (dark region) cases are
considered.} 
\label{fig:nueccburst}
\end{figure}

\begin{figure}[htbp]
\centering
\epsfig{file=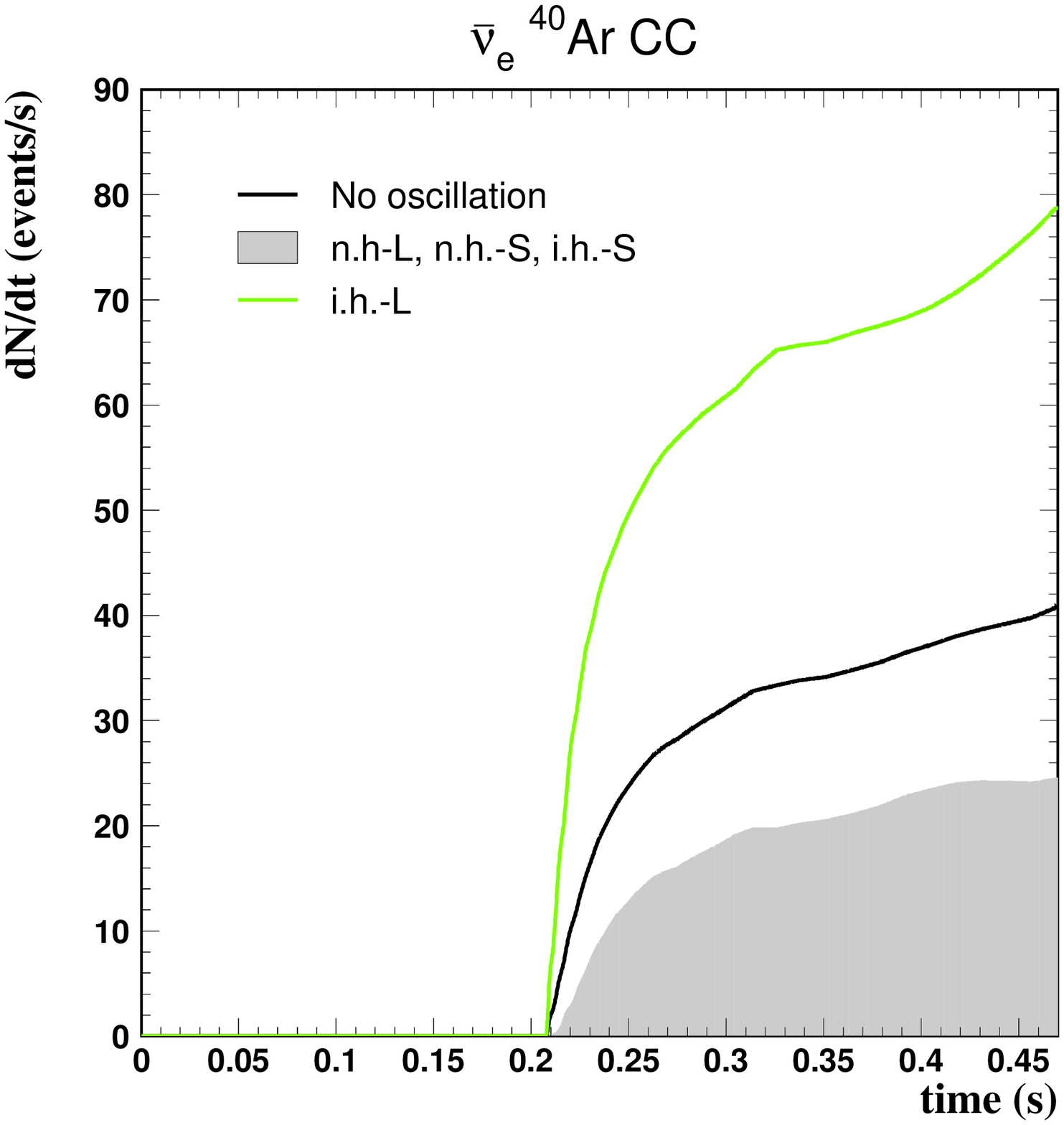,width=10cm}
\epsfig{file=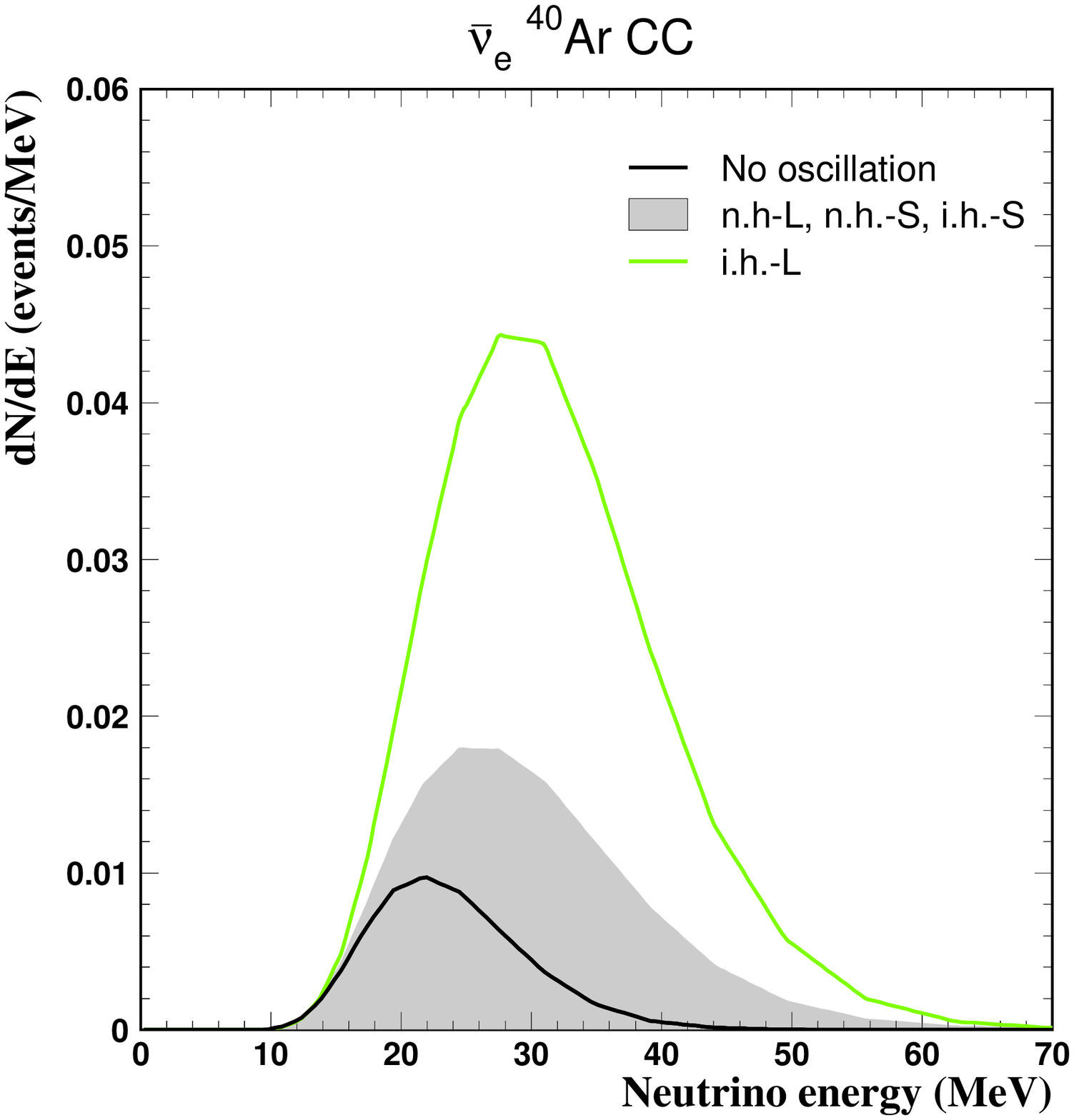,width=10cm}
\caption{SN burst: time evolution of the $\anue$ CC event rate (top) 
and the corresponding time integrated event
spectra (bottom). The non oscillation (dark curve), i.h.--L (light
curve) and n.h.--L, n.h.--S and i.h.--S (gray region) cases are
considered.} 
\label{fig:anueccburst}
\end{figure}

\begin{figure}[htbp]
\centering
\epsfig{file=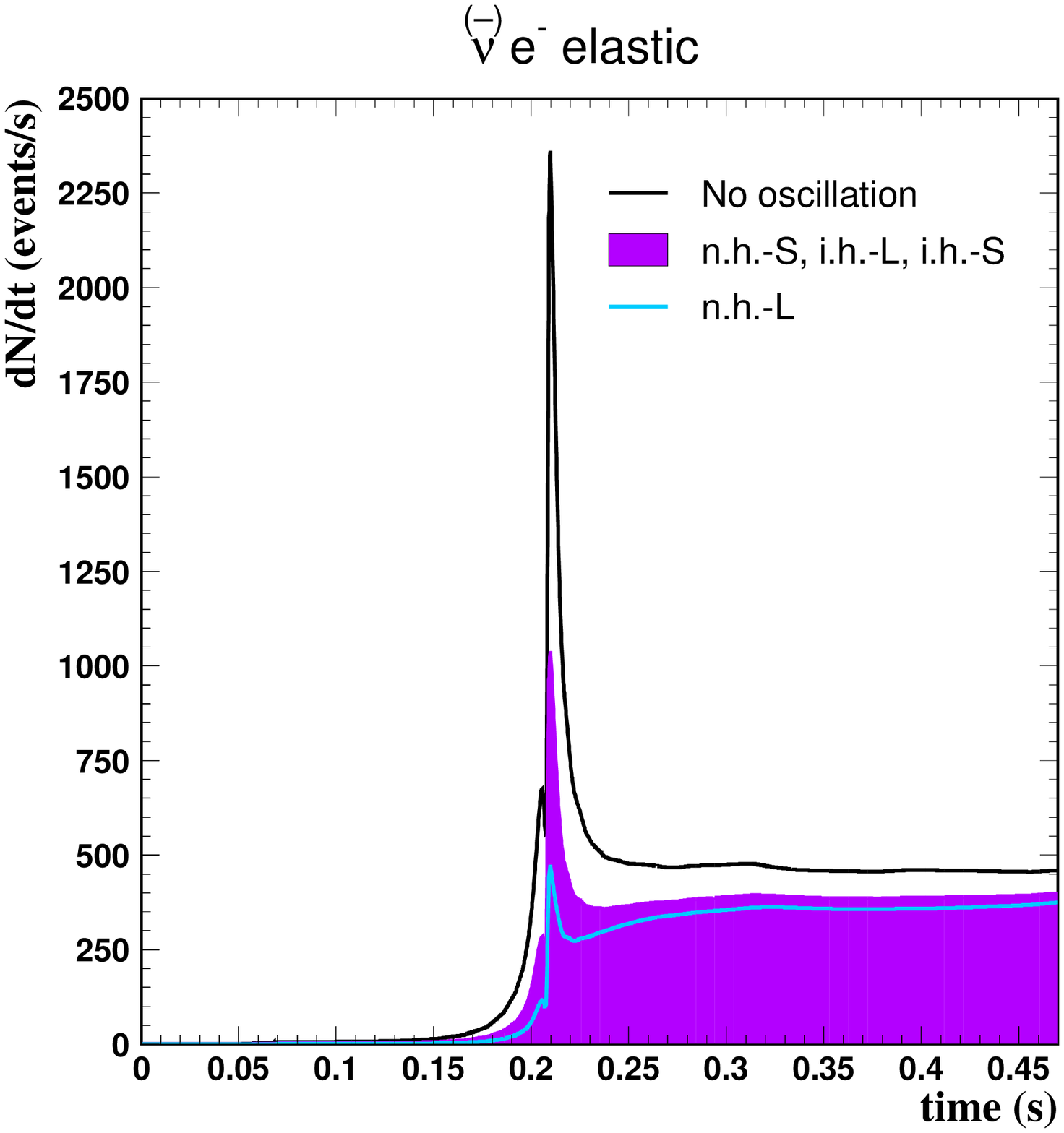,width=10cm}
\epsfig{file=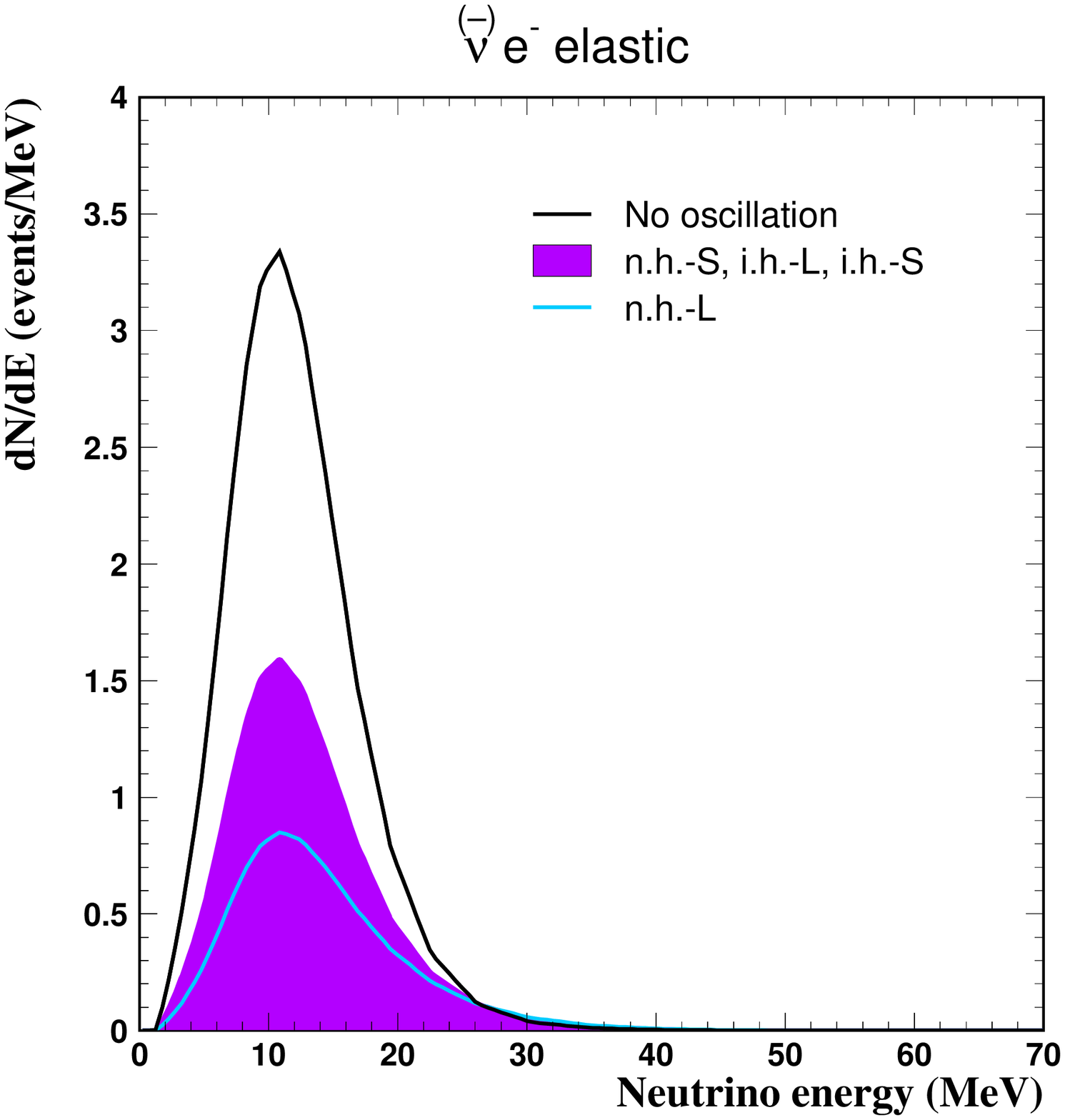,width=10cm}
\caption{SN burst: time evolution of the elastic event rate (top) and 
the corresponding time integrated event
spectra (bottom). The non oscillation (dark curve), n.h.--L (light
curve) and n.h.--S, i.h.--L and i.h.--S (dark region) cases are
considered.} 
\label{fig:elasburst}
\end{figure}

\begin{figure}[htbp]
\centering
\epsfig{file=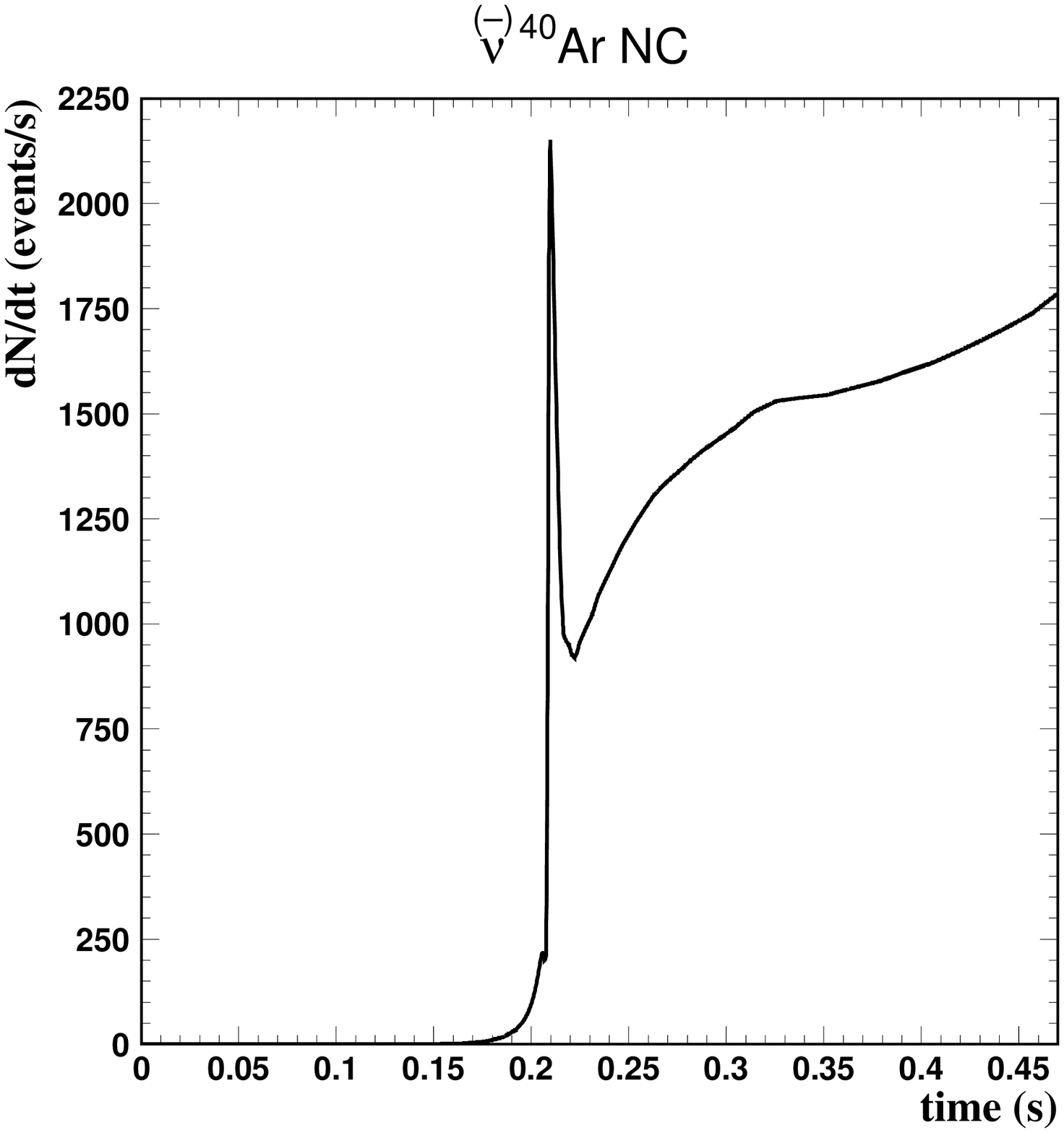,width=10cm}
\caption{SN burst: time evolution of the NC event rate.} 
\label{fig:ncburst}
\end{figure}

Figures \ref{fig:nueccburst}--\ref{fig:ncburst} contain the time
evolution of the expected events in the burst for the different detection channels, namely
$\nu_e$ CC, $\bar\nu_e$ CC, elastic and neutral currents, 
normalized to 70 kton detector and the corresponding time integrated event
spectra. The non oscillation and the four different oscillation cases
are compared in the plots.    For the $\nu_e$ CC, $\bar\nu_e$ CC and elastic
events we also show the event energy distributions. For the neutral current events,
the energy cannot be determined due to the neutrino in the final state.
  
The reduction of the $\nu_e$ peak due to the effect of oscillations
can be seen in figure \ref{fig:nueccburst}. The suppression is maximal
for normal hierarchy and large \th13 mixing angle due to the total
conversion of $\nu_e$ into $\nu_{\mu,\tau}$. At the same time, the
energy spectrum moves slightly to higher neutrino energy values.

Elastic events (figure \ref{fig:elasburst}) are also sensitive to the
oscillations with a reduction of the shock breakout peak. However, the
energy spectrum remains almost unchanged. 

The neutral current events
(figure \ref{fig:ncburst}) are not affected by neutrino oscillations
and the time evolution clearly shows the shock breakout peak coming
from $\nue$ events. 

Overall, we note that the energy distribution of the burst is marginally
affected by oscillations.

\subsubsection{Consequence of the neutrino burst detection}

We can hence conclude that the burst signal is only affected
in rate by the effect of neutrino oscillation and the effect on the shape
is marginal. This means that:
\begin{itemize}
\item a clear SN burst peak should be visible in the neutral current sample,
regardless of the neutrino oscillation physics. This signature should provide
a clear indication for the nature of the supernova explosion. The size of the
peak determines the burst absolute neutrino flux from the supernova.
\item the oscillation physics occurring during the early phase of the supernova
can be studied by detection of the $\nu_e$ and $\bar\nu_e$ CC channels.
\item the number of observed  $\nu_e$ CC observed in the burst can distinguish
between the non oscillation case, the n.h.-L oscillation and the three
other oscillation scenarios. The n.h.-S, i.h.-L and i.h.-S cases cannot
be distinguished by $\nu_e$  CC.
\item a roughly factor 4 enhancement of observed  $\bar\nu_e$  CC observed in the burst  
would be a clear indication for an oscillation matter enhancement only possible
in the inverted mass hierarchy with large mixing angle. Otherwise, a roughly factor 2 enhancement
is expected from the effect of oscillations (see Figure
\ref{fig:anueccburst}). It should be stressed that this effect is not
visible even for a 70 kton detector.
\item elastic events can help distinguish between the non oscillation
case, the n.h.-L oscillation case and the three other oscillation scenario. 
\end{itemize}

\subsubsection{General mixing case -- intermediate mixing angle}
For the intermediate mixing angle region defined approximately
by $\approx 10^{-5}<\sin^2\theta_{13}<2\times 10^{-4}$, the
number of expected events varies strongly with the mixing angle. 
In fact, in this region, we expect that the neutrino oscillation effect
depends on the neutrino energy within the range up to 100~MeV
relevant for supernova neutrinos.

\begin{figure}[p]
\centering
\begin{tabular}{cc}
\epsfig{file=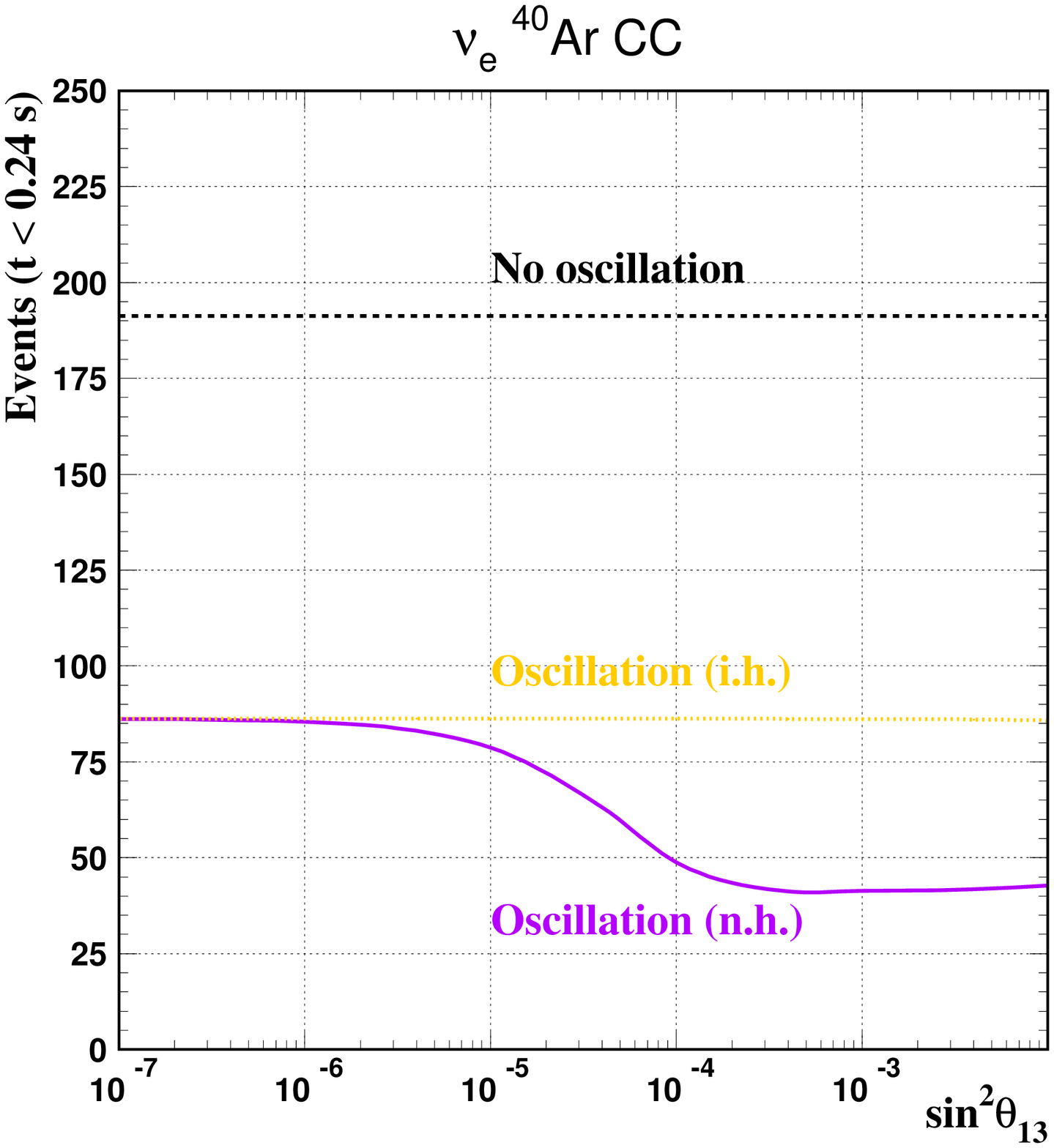,width=8.cm}
&
\hspace{-1cm}
\epsfig{file=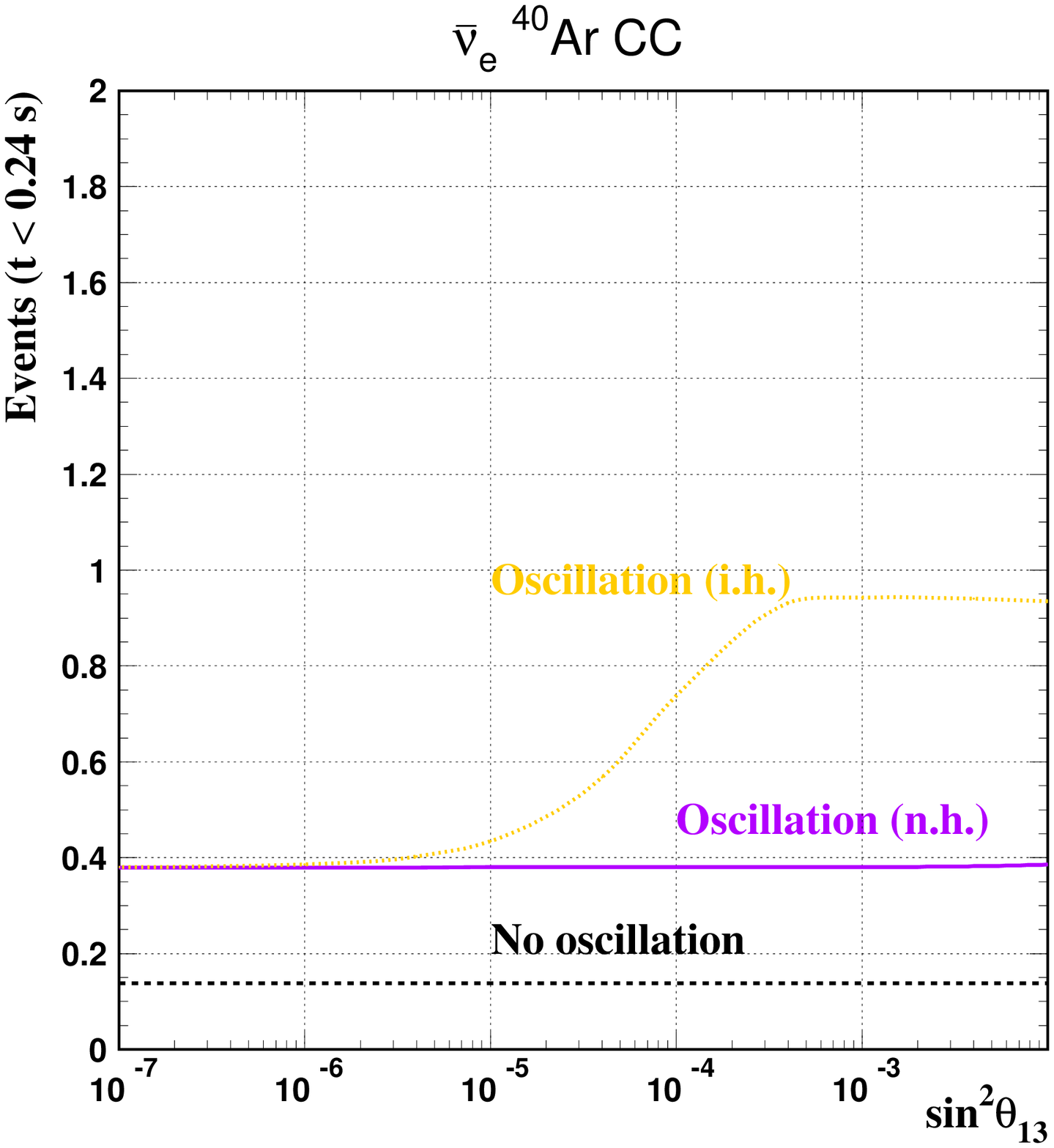,width=8.cm}
\end{tabular}
\centering
\begin{tabular}{cc}
\epsfig{file=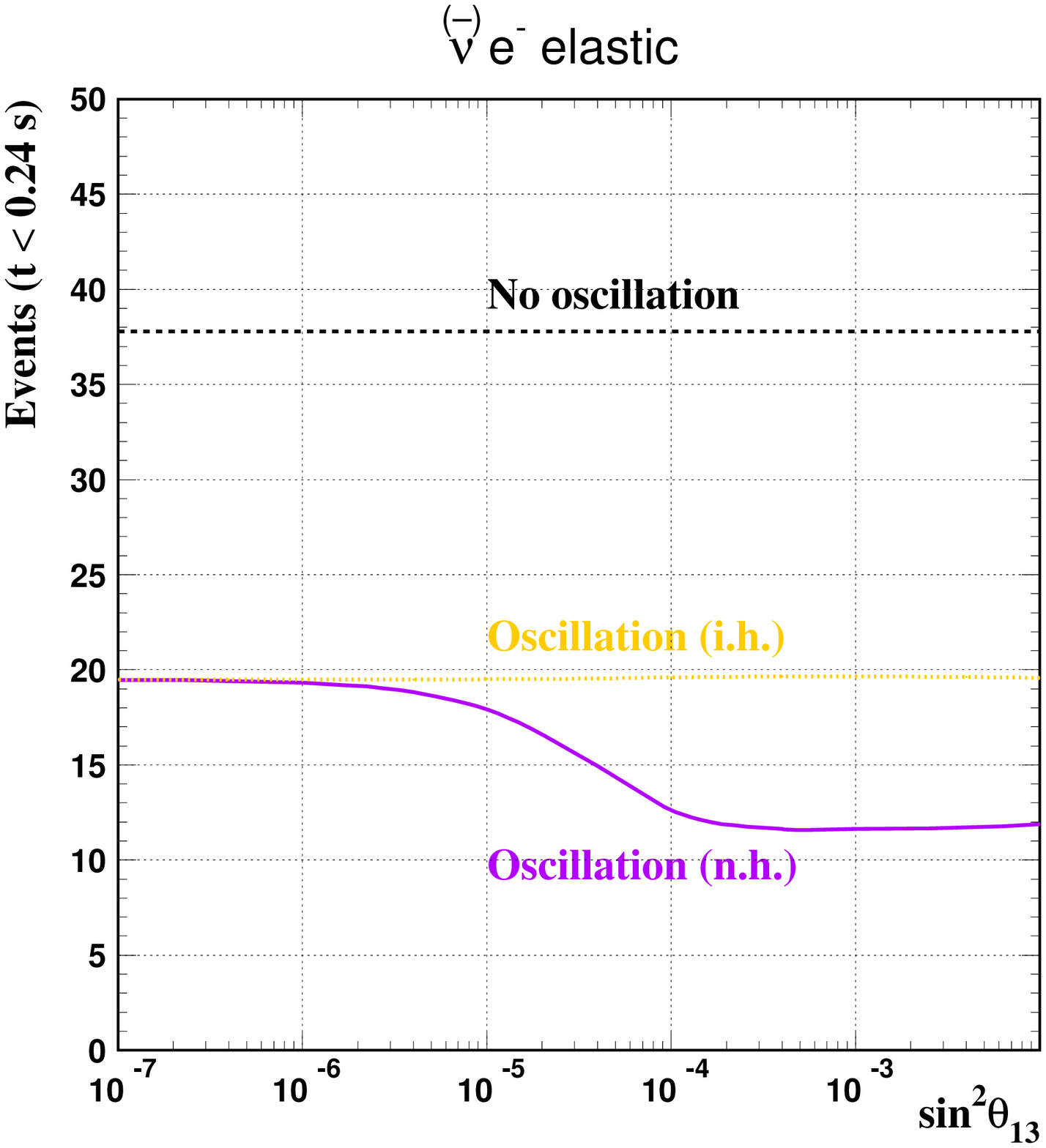,width=8.cm}
&
\hspace{-1cm}
\epsfig{file=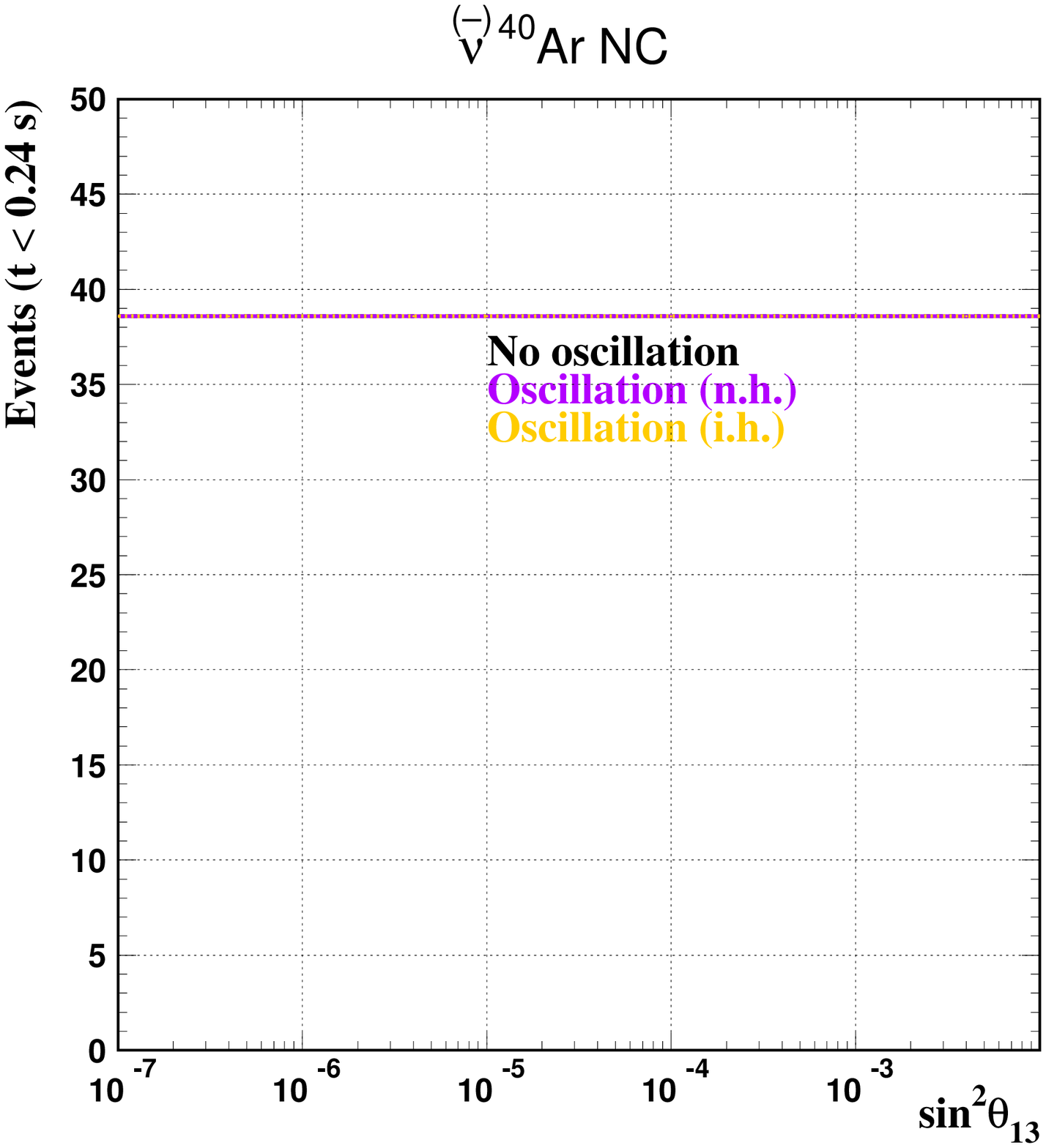,width=8.cm}
\end{tabular}
\caption{Expected number of events in a 70 kton detector during the burst (see text for definition) 
as a function of \s2t13. The different neutrino interaction processes 
are plotted separately. Solid lines correspond to the n.h. oscillation
case, dotted lines to the i.h. case and dashed lines to the non
oscillation case.}  
\label{fig:ratesbursts2t13}
\end{figure}

The actual variation of the number of events
as a function of the \th13 value are summarized in Figure
\ref{fig:ratesbursts2t13} for the four detection channels. Solid lines correspond to the
normal mass hierarchy case, dotted lines to the inverted mass hierarchy case and dashed lines
to the non oscillation case. We consider events in burst and the plots
are normalized to a 70 kton detector. 

This variation can be exploited to measure the mixing angle if it lies
in this intermediate region. Quantitative estimate of the precision
with which the angle can be determined will be described elsewhere~\cite{superfit}.

\subsection{Comparison with other experiments}
All experiments will be affected in the same way by the effect
of neutrino flavor oscillation in their expected rates of $\nu_e$ events.
In order to compare experiments, we hence restrict ourself to the
case where neutrino oscillations are neglected and compare
rates in Superkamiokande and SNO with our results on Argon.
Results can be readily scaled to include neutrino oscillation effects.

\begin{figure}[htbp]
\vspace{-1cm}
\centering
\epsfig{file=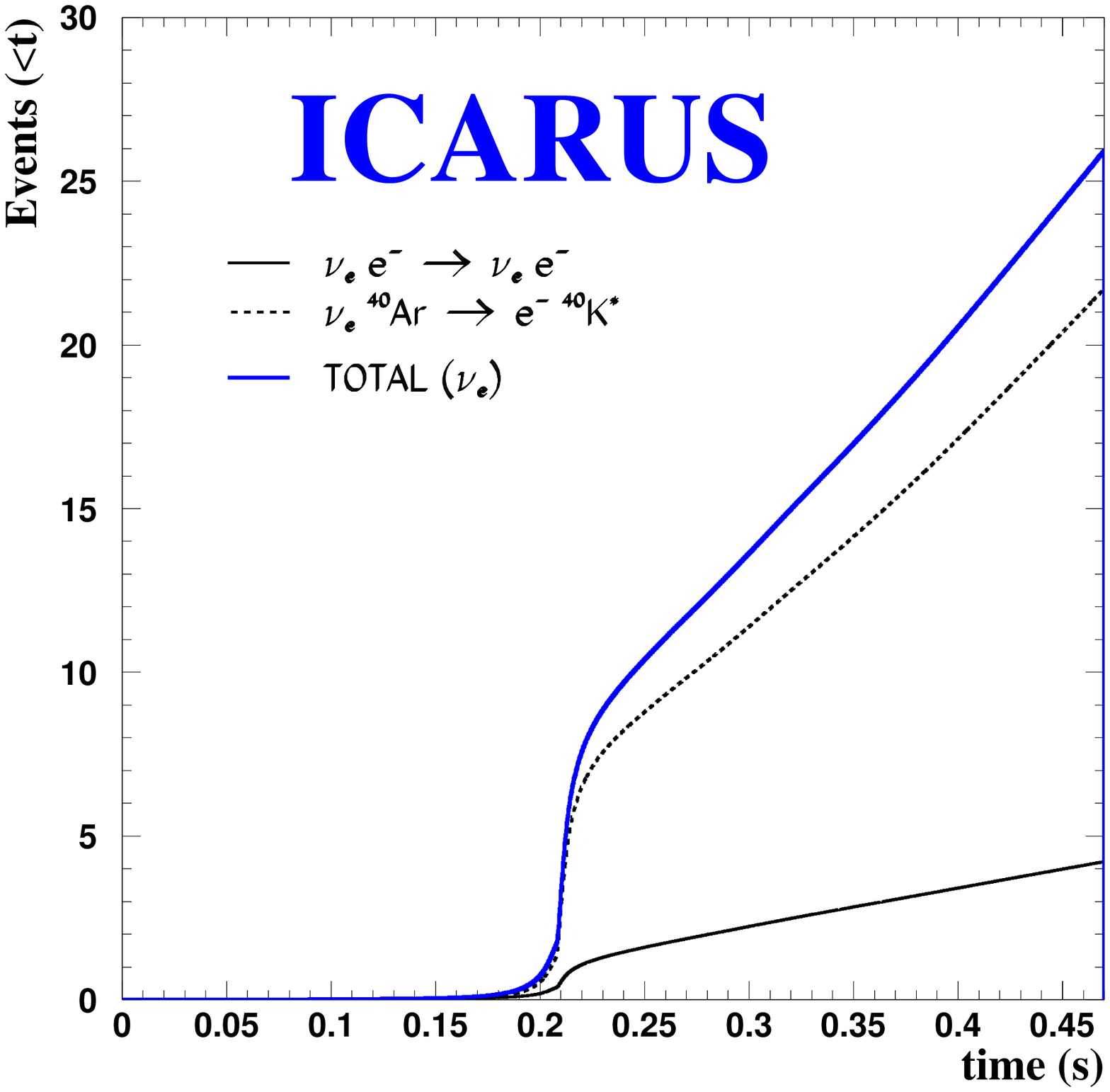,width=12.cm}
\caption{Integrated number of $\nu_e$ events as a function of time
for the elastic and absorption channels. The thick solid line
corresponds to the total number of events for a 3 kton detector. No
oscillation effects are included.} 
\label{fig:timeicarus}
\centering
\begin{tabular}{cc}
\epsfig{file=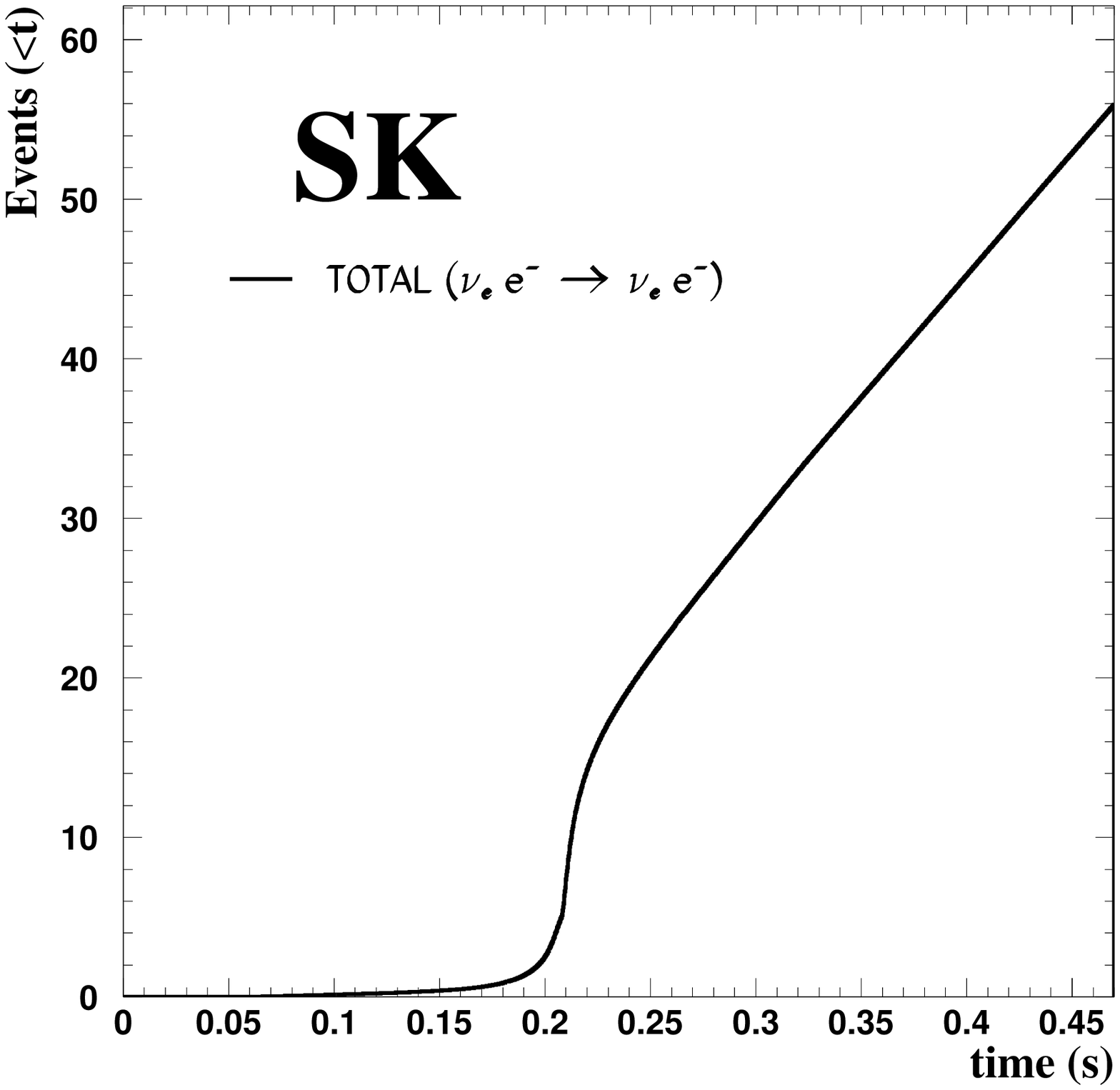,width=8.5cm}
&
\hspace{-1cm}
\epsfig{file=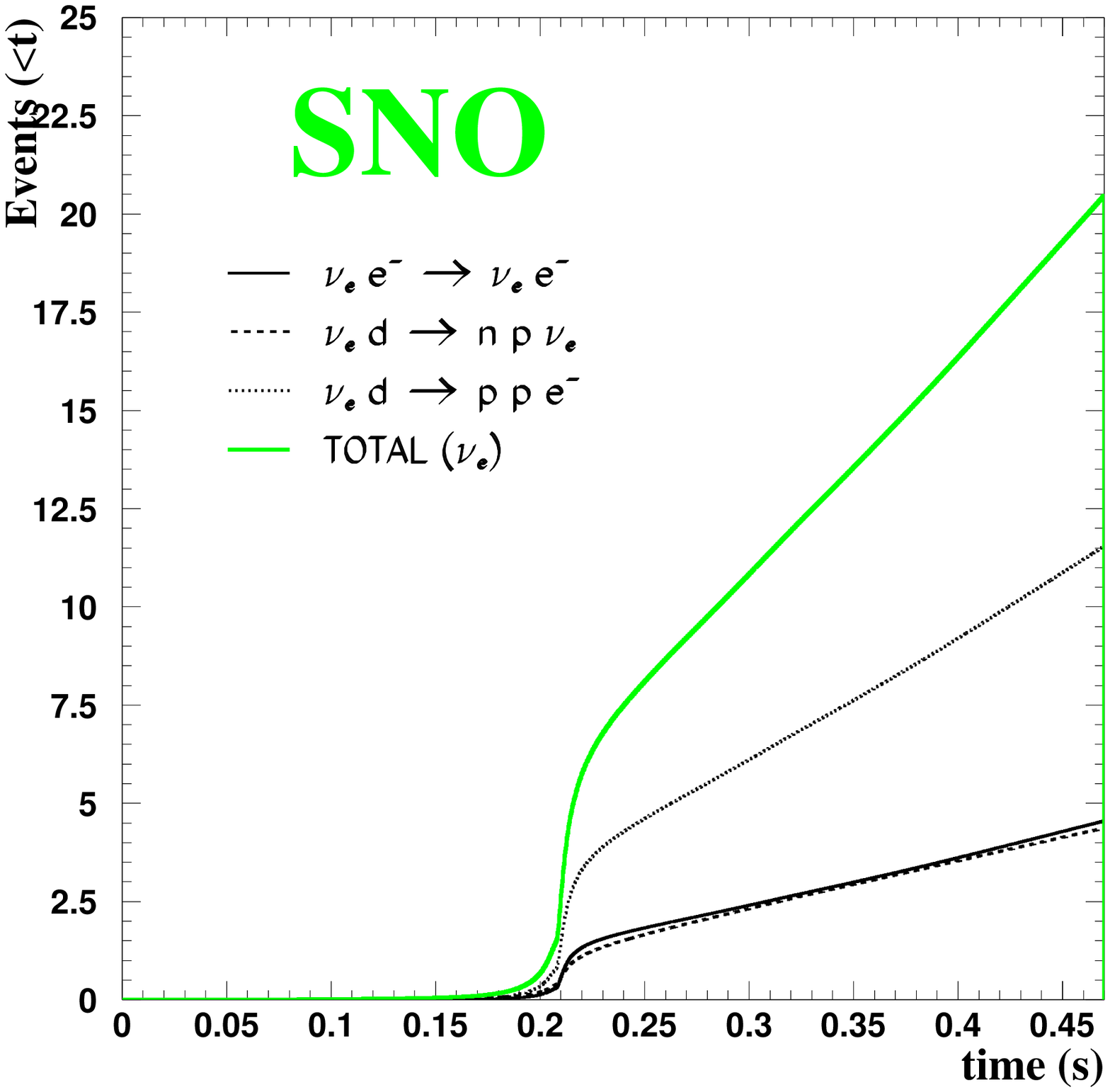,width=8.5cm}
\end{tabular}
\caption{Integrated number of $\nu_e$ events as a function of time
expected in the SK (left) and SNO (right) detectors. No oscillation
effects are included.}
\label{fig:timeothers}
\end{figure}

The integrated number of $\nu_e$ events as a function of time for
elastic and absorption processes without including oscillation effects
are plotted in figure \ref{fig:timeicarus}. The main contribution
comes from the CC channel. Figure \ref{fig:timeothers} shows for comparison similar
plots for SuperKamiokande (SK) and SNO detectors.
 
The SK volume for supernova neutrino detection is 32 kton
of light water. The dominant reaction is the charged--current
absorption of $\anue$ neutrinos on free protons ($\anue p \ra n
e^+$). Neutrino scattering on electrons ($\nu e^- \ra \nu e^-$) also
contributes for all neutrino species. For the study of the shock
breakout we calculate the event rates in SK due to only the $\nue$
elastic scattering. For this process we use the neutrino electron
scattering cross section given in \cite{SKcross} with a threshold of
$\sim$ 5 MeV. If there is large mixing angle between $\nu_e$ and
$\nu_{\mu}$ or $\nu_{\tau}$ neutrinos, the early $\nu_e$ signal will
be reduced significantly in SK.

The SNO detector is a light and heavy water neutrino
detector containing 1 kton of D$_2$O surrounded by a cavity of 1.6
kton of light water. The main reactions for $\nu_e$ neutrinos are the
elastic scattering and the NC and CC reactions on deuterium ($\nu_e d
\ra n p \nu_e$ and $\nu_e d \ra p p e^-$). The cross sections used to
compute these events are taken from \cite{snocross}.

\begin{figure}[htbp]
\centering
\epsfig{file=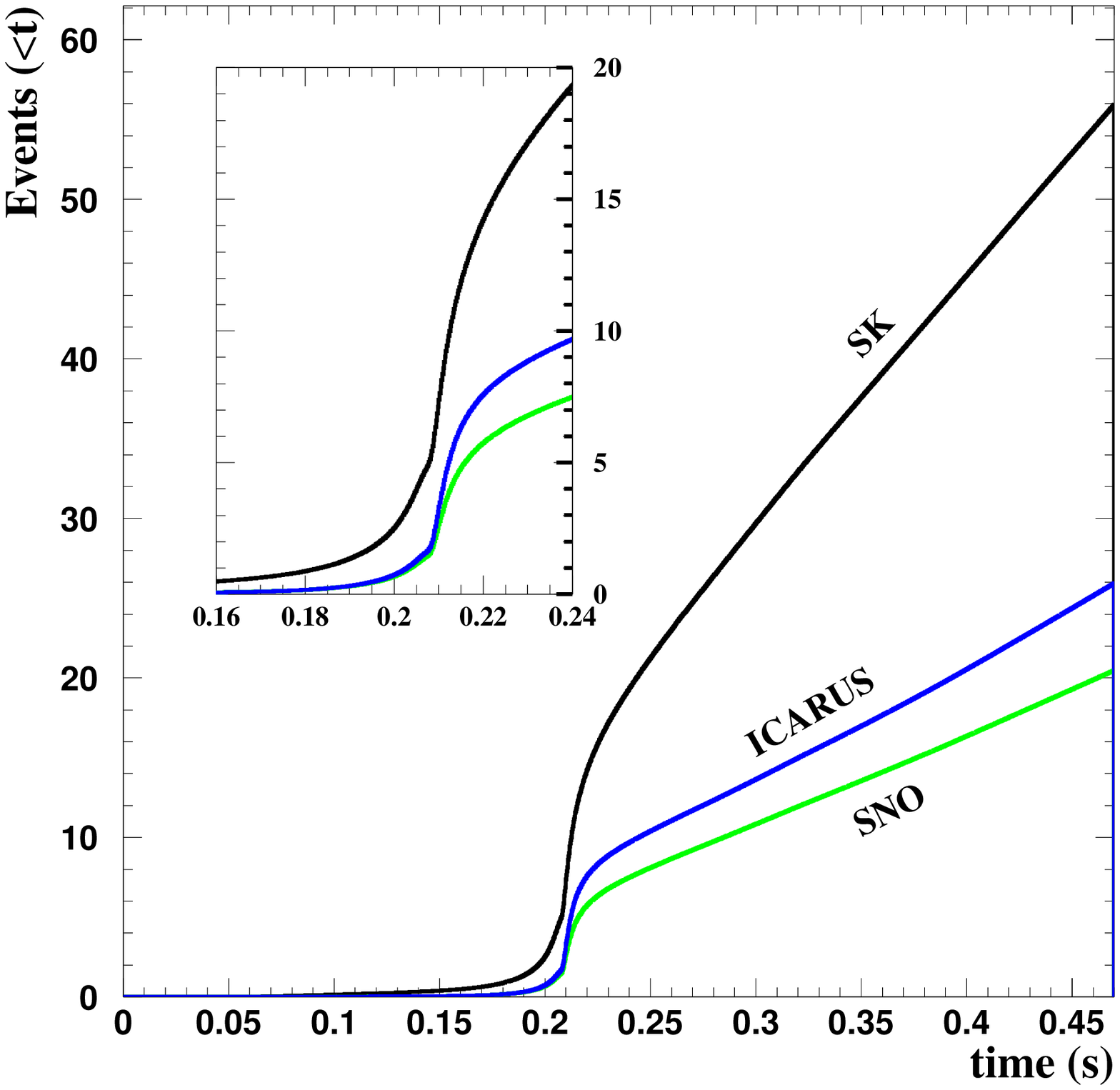,width=14.cm}
\caption{Comparison of the expected number of events from the $\nu_e$
breakout burst in a 3 kton ICARUS, SK and SNO experiments. In the first 40 ms
after bounce a total of 20 events are expected in SK, 10 in ICARUS and
7 in SNO. No oscillation effects are included.} 
\label{fig:timeall}
\end{figure}

Figure \ref{fig:timeall} shows the comparison between the expected
number of events from the $\nu_e$ burst in the three experiments. A
total of 10 events are expected in the 3kton ICARUS in the 40 ms after the
bounce: 8 events from the absorption channel and 2 from the elastic
channel. The contribution of SK and SNO for the same interval of time
is 20 and 7 events, respectively. 

Summarizing, we observe that the event statistics
that would be obtained in SuperKamiokande, SNO and the 3~kton ICARUS
are similar. In this case, SNO and ICARUS benefit largely from
their direct sensitivity to $\nu_e$ via charged current interactions on
resp. deuteron and Argon, while in SuperKamiokande one has
to rely on elastic scattering on electrons, which has a much smaller
cross-section. Hence, smaller detectors like SNO and ICARUS can
compare favorably with the largest SuperKamiokande in the
study of the breakout phase.

We also conclude that given that neutrino oscillations is an established
fact \cite{nuobserv}, it will be difficult to study the neutrino burst within the current round
of experiments given the reduction of statistics following the conversion
of $\nu_e$ into other flavors. In the context of the liquid Argon TPC,
it appears that a 70~kton detector
will be mandatory in order to collect sufficient events to study the
neutrino burst.

\section{Cooling phase}
\label{sec:cool}

In this section we will refer to the main neutrino emission phase in
the core collapse supernovae: the cool-off of the star. 

As seen in previous sections, the energy spectra of the expected
neutrinos at the detector is modified by oscillations among
flavors since the average neutrino energies are flavor dependent and
the interaction cross sections are different for different
flavors. The variations of the neutrino energy will depend on the 
adiabaticity conditions on the H and L resonances in the supernova and
on the neutrino oscillation parameters. Four different cases are
considered in this analysis depending on the type of mass hierarchy
and the value of the $\theta_{13}$ mixing angle.

\subsection{Hierarchical energy distribution -- scenario I}

We first consider the prediction from Ref \cite{Langanke} which
we parameterize as Fermi-Dirac distributions with $\langle E_{\nu_e}
\rangle$ = 11 MeV, $\langle E_{\bar\nu_e} \rangle$ = 16 MeV, $\langle
E_{\nu_{\mu,\tau}} \rangle$ = $\langle E_{\bar\nu_{\mu,\tau}} \rangle$
= 25 MeV and luminosity equipartition among flavors.
In this kind of scenario where the energies are rather different (hierarchical)
between flavors, the effect of the oscillation is the most pronounced.
Indeed, oscillations will as expected harden the $\nu_e$ spectrum,
which has a big incidence of the $\nu_e$ CC channel.

Table \ref{tab:rates} contains the number of neutrino events from a
supernova at 10 kpc expected in the 3 kton detector. Elastic,
CC and NC processes have been considered independently and
oscillation and non oscillation cases have been computed for normal
and inverted hierarchies. However, we recall that four channels are in fact
accessible experimentally: the $\nu_e$ CC, the $\bar\nu_e$ CC,
the NC and the elastic scattering.

We note that:
\begin{itemize}
\item the number of observed  $\nu_e$  CC observed can distinguish
between (a) the non oscillation case, (b) the n.h.-L case and (c) the
three other oscillation scenarios. n.h.-S, i.h.-L and i.h.-S cannot
be distinguished by $\nu_e$  CC.
\item a roughly factor 4 enhancement of observed  $\bar\nu_e$  CC observed  
would be a clear indication for an oscillation matter enhancement only possible
in the inverted mass hierarchy with large mixing angle. Otherwise, a roughly factor 2 enhancement
is expected from the effect of oscillations. 
\end{itemize}

\begin{table}[htbp]
\centering
\begin{tabular}{clccccc} \hline
& &\multicolumn{5}{c}{\bf Scenario I} \\
Reaction & & No & \multicolumn{2}{c}{Oscillation
(n.h.)} & \multicolumn{2}{c}{Oscillation (i.h.)} \\
 & & oscillation & Large $\theta_{13}$ & Small $\theta_{13}$ & Large
$\theta_{13}$ & Small $\theta_{13}$\\ \hline
{\bf Elastic} & & & & & & \\
& $\nu_e \, e^-$ & 20 & 20 & 20 & 20 & 20 \\
& $\bar\nu_e\, e^-$ & 8 & 8 & 8 & 8 & 8 \\
& $(\nu_\mu+\nu_\tau)\, e^-$ & 7 & 7 & 7 & 7 & 7 \\
& $(\bar\nu_\mu +\bar\nu_\tau)\, e^-$ & 6 & 6 & 6 & 6 & 6 \\
& total $\nu\, e^-$ & 41 & 41 & 41 & 41 & 41 \\
\hline
{\bf Absorption} & & & & & & \\
{\bf CC} & $\nu_e$ $^{40}$Ar       & 188 & 962 & 730 & 730 & 730 \\ 
& $\anue$ $^{40}$Ar                  & 15 &  33 &  33 &  75 & 33 \\ 
\hline
{\bf NC} & $\nu$ $^{40}$Ar       & 492 & 492 & 492 & 492 & 492 \\
          & $\bar{\nu}$ $^{40}$Ar & 419 & 419 & 419 & 419 & 419 \\ 
\hline
{\bf Total} & & 1155 & 1947 & 1715 & 1757 & 1715 \\\hline
\end{tabular}
\caption{Expected neutrino events in a 3 kton detector 
including neutrino oscillations with
matter effects inside the supernova. The flux parameters are $\langle E_{\nu_e}
\rangle$ = 11 MeV, $\langle E_{\bar\nu_e} \rangle$ = 16 MeV, $\langle
E_{\nu_{\mu,\tau}} \rangle$ = $\langle E_{\bar\nu_{\mu,\tau}} \rangle$
= 25 MeV and luminosity equipartition among flavors is assumed.}   
\label{tab:rates}
\end{table}

\begin{figure}[htbp]
\centering
\epsfig{file=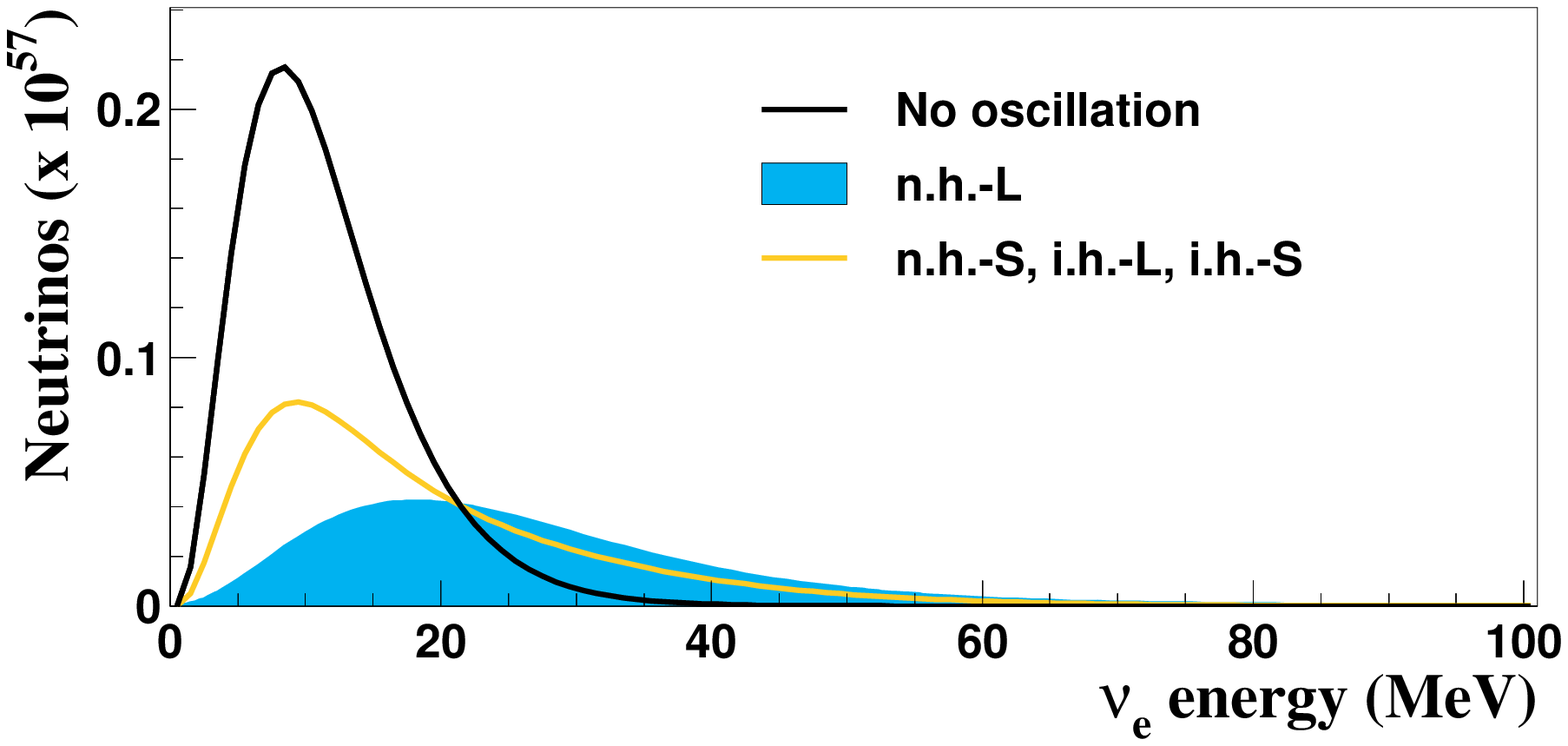,width=8.5cm} \\
\begin{tabular}{cc}
\epsfig{file=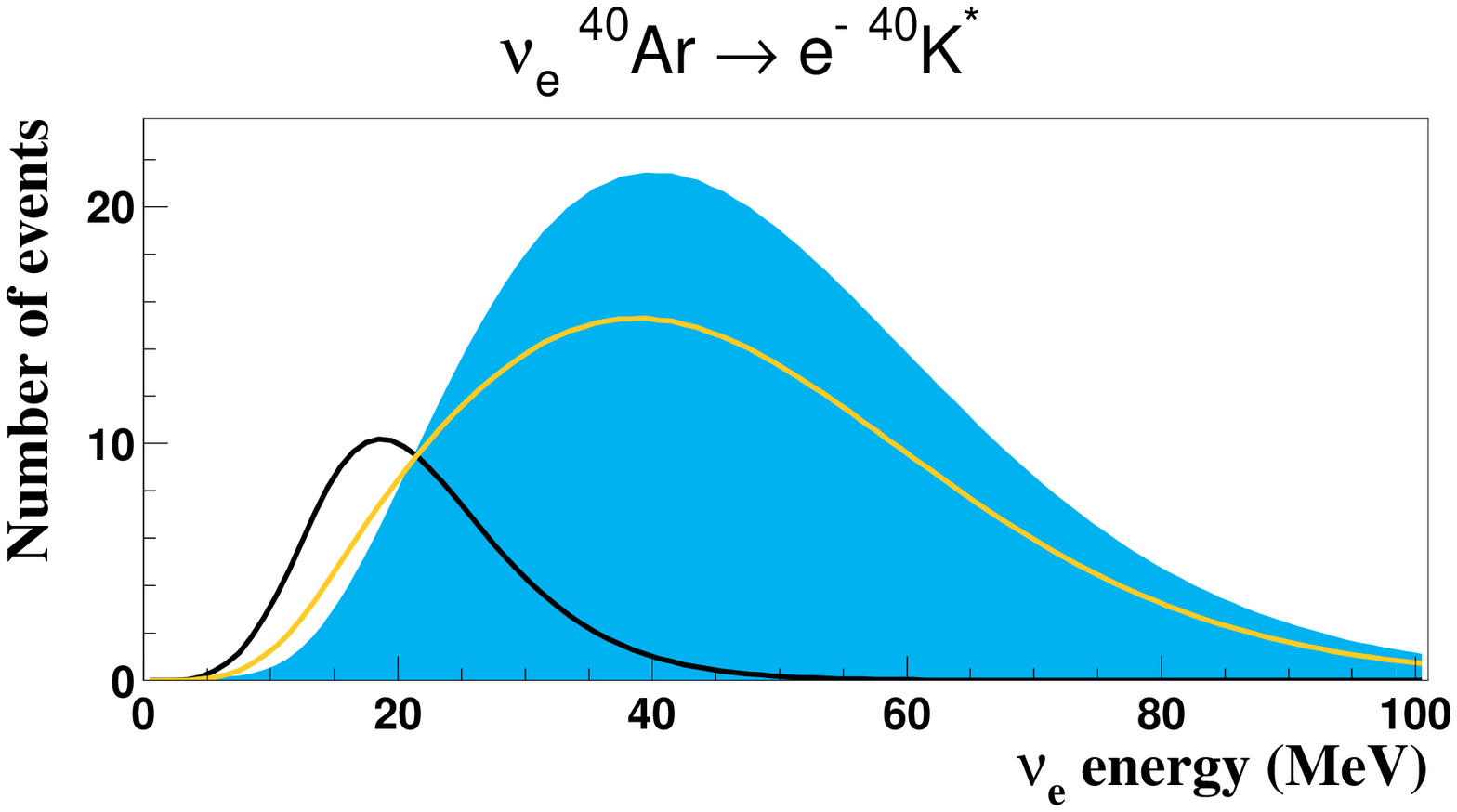,width=8.cm}
&
\hspace{-0.5cm}
\epsfig{file=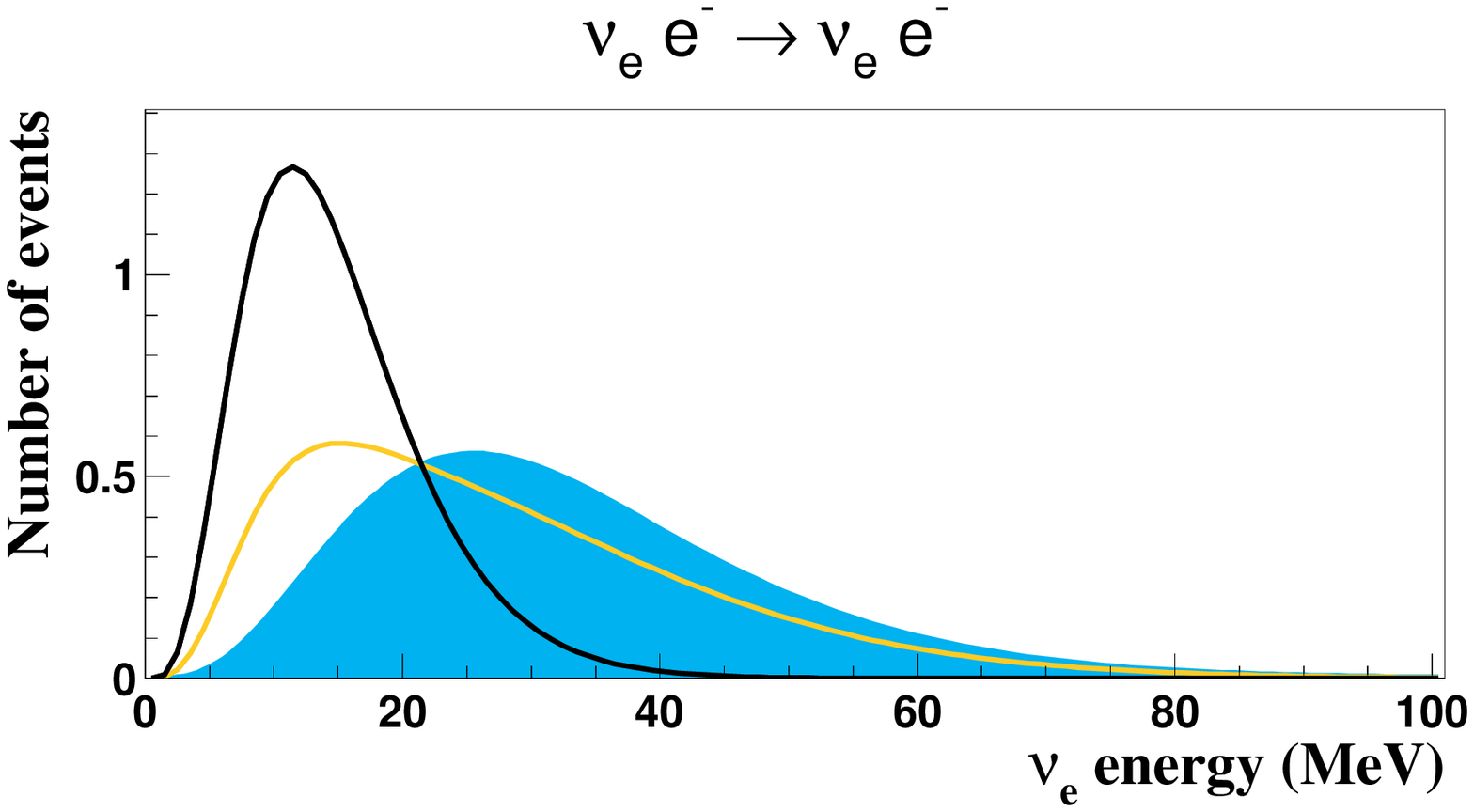,width=8.cm}
\end{tabular}
\caption{Scenario I: Number of $\nue$ neutrinos arriving at Earth (top) and expected
number of events in the 3 kton detector (bottom) for $\nu_e$
neutrinos. The non oscillation and the four
oscillation cases have been taken into account. The CC interaction
process (left) and elastic scattering (right) are split in the figure.}
\label{fig:oschierar_nue}
\end{figure}

\begin{figure}[htbp]
\centering
\epsfig{file=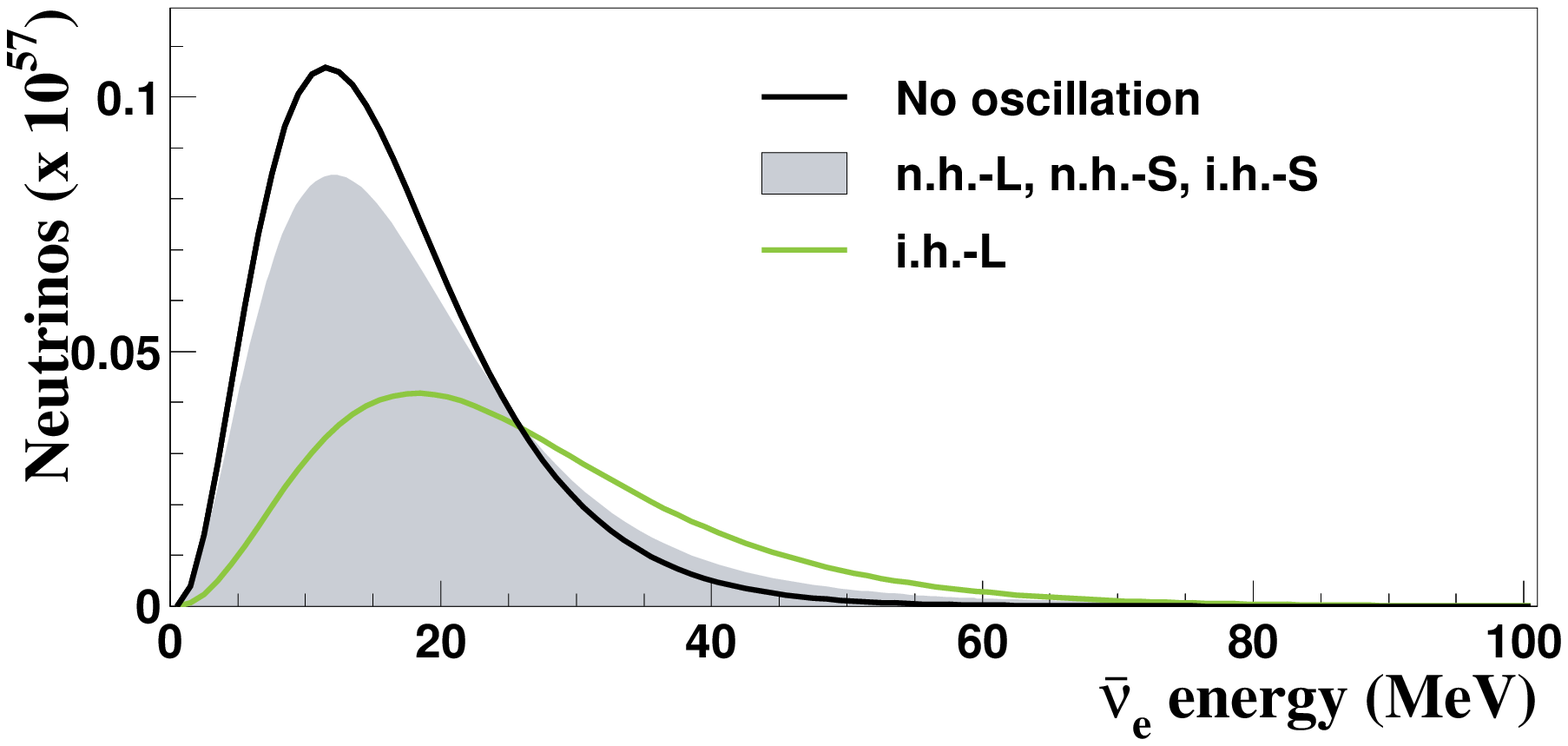,width=8.5cm}
\begin{tabular}{cc}
\epsfig{file=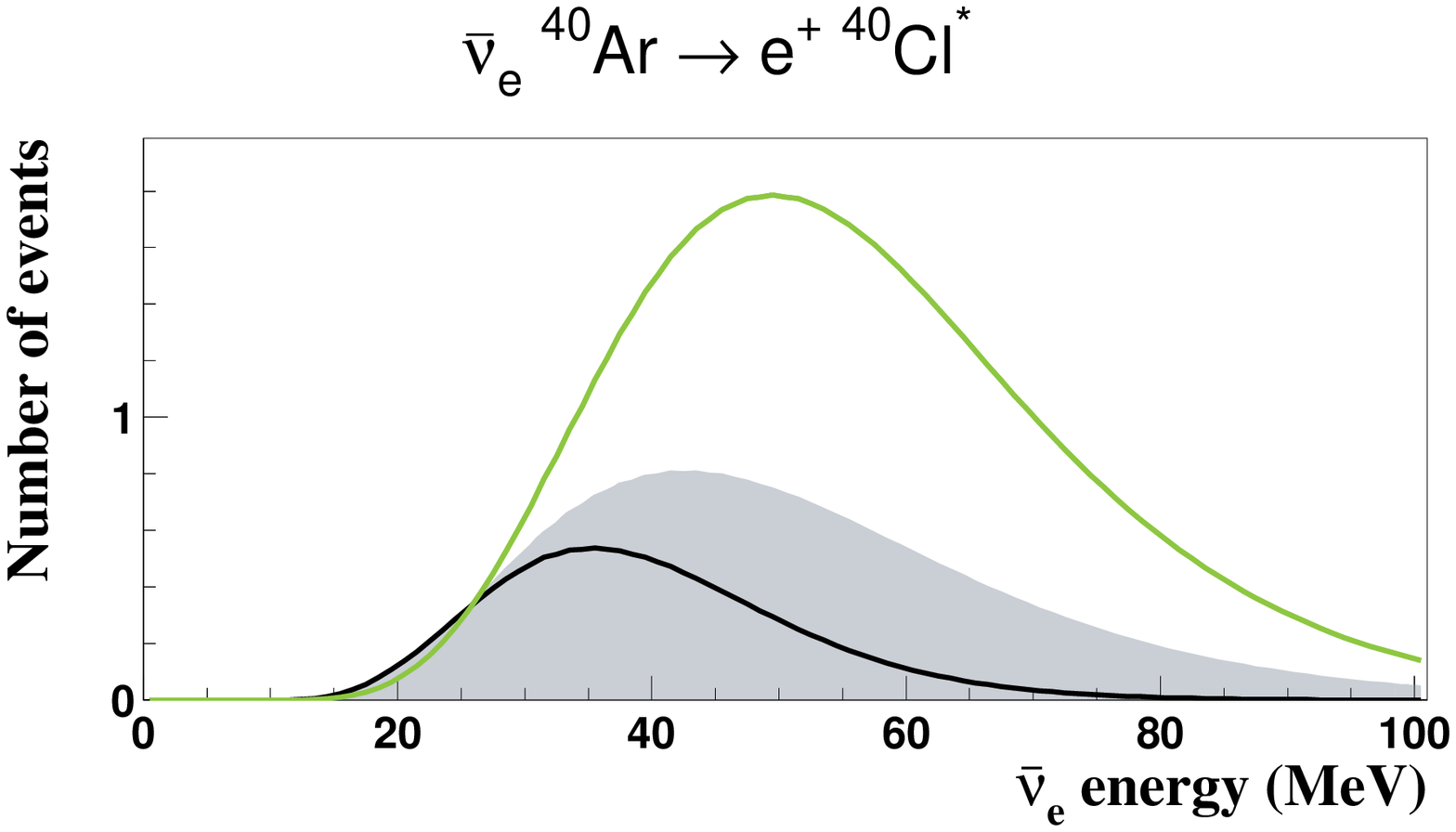,width=8.cm}
&
\hspace{-0.5cm}
\epsfig{file=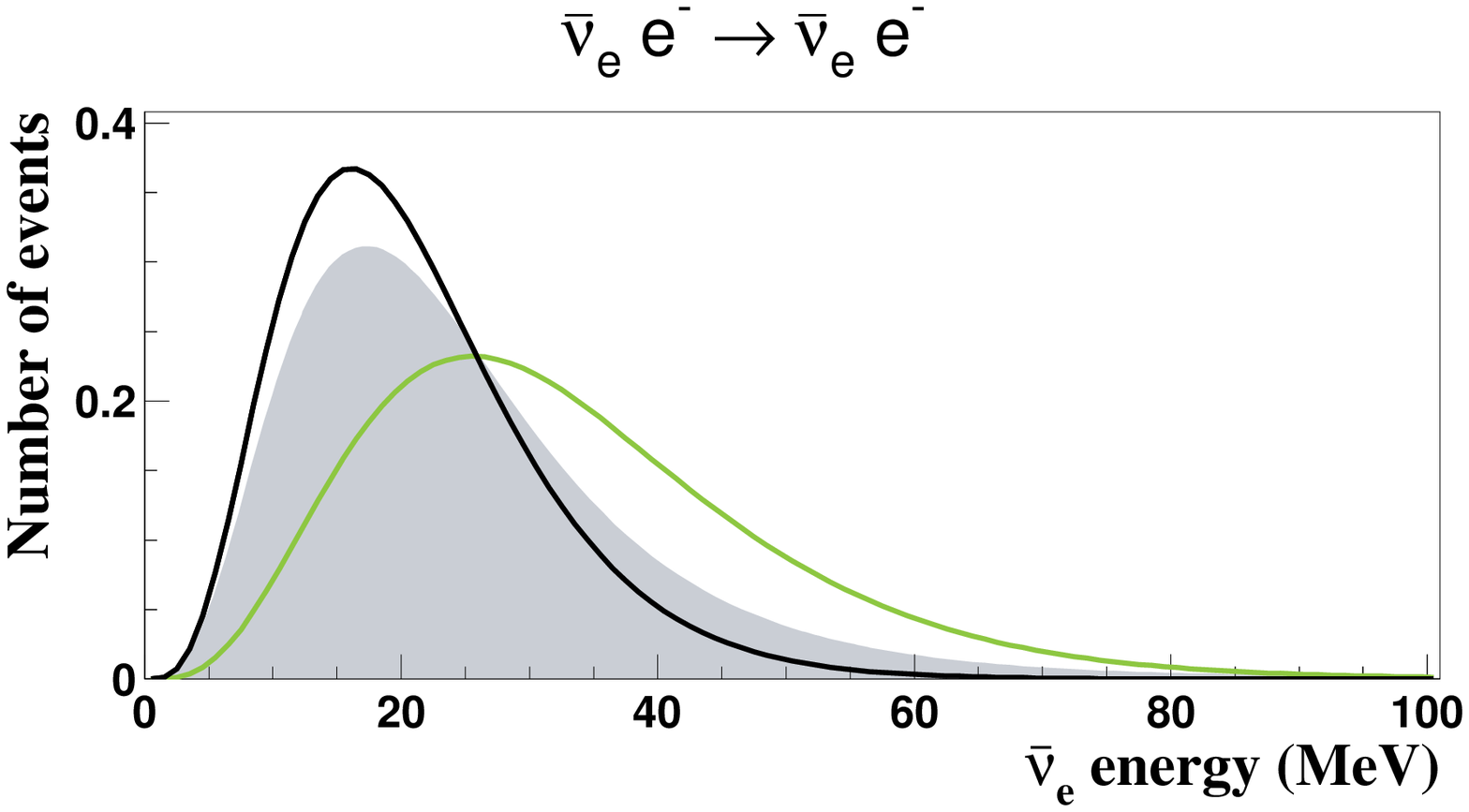,width=8.cm}
\end{tabular}
\caption{Scenario I: Number of $\anue$ neutrinos arriving at Earth (top) and expected
number of events in the 3 kton detector (bottom) for $\bar\nu_e$
neutrinos. The non oscillation and the four
oscillation cases have been taken into account. The CC interaction
process (left) and elastic scattering (right) are split in the figure.}
\label{fig:oschierar_anue}
\end{figure}

\begin{figure}[htbp]
\centering
\begin{tabular}{cc}
\epsfig{file=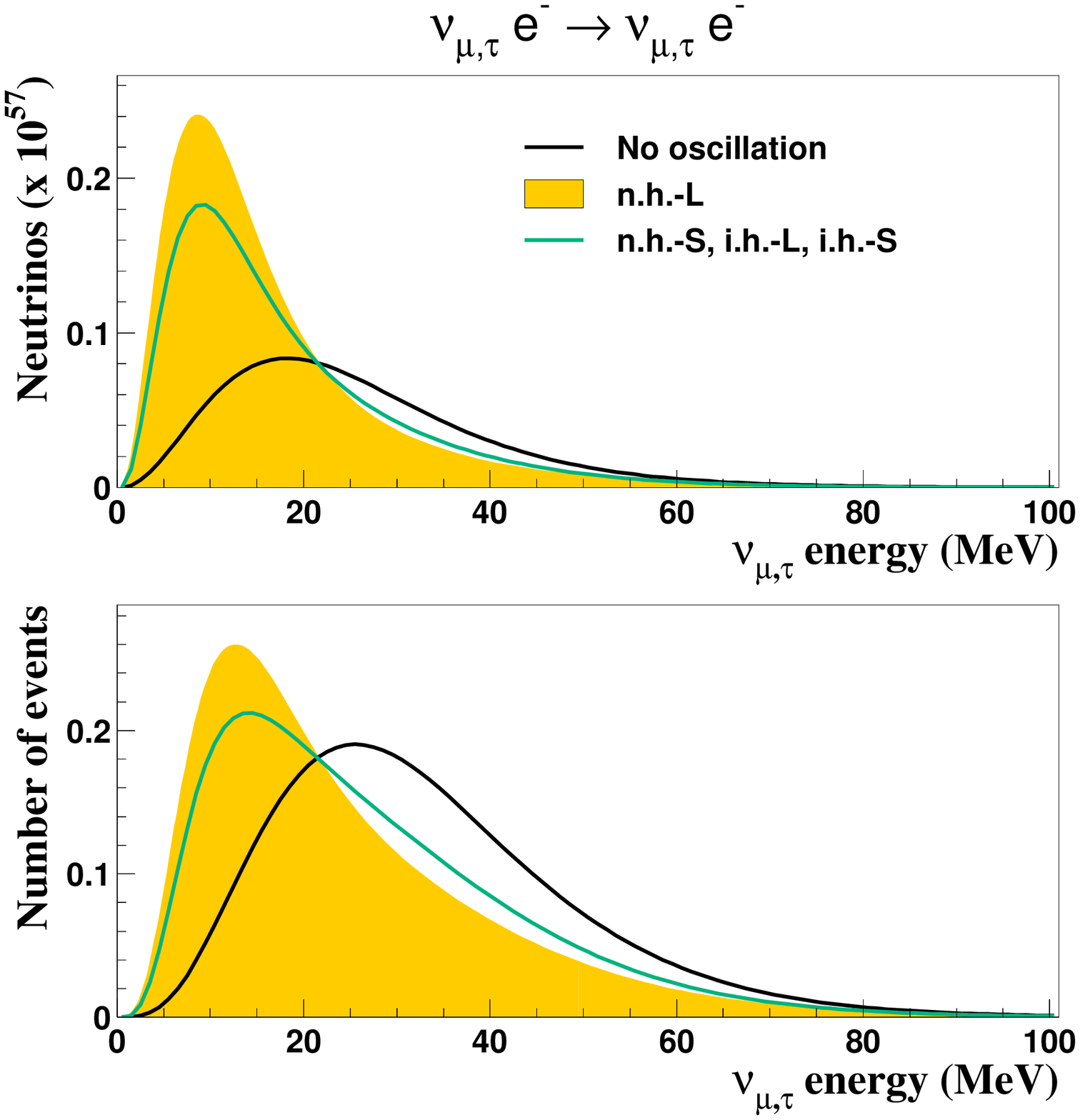,width=8.5cm}
&
\hspace{-1cm}
\epsfig{file=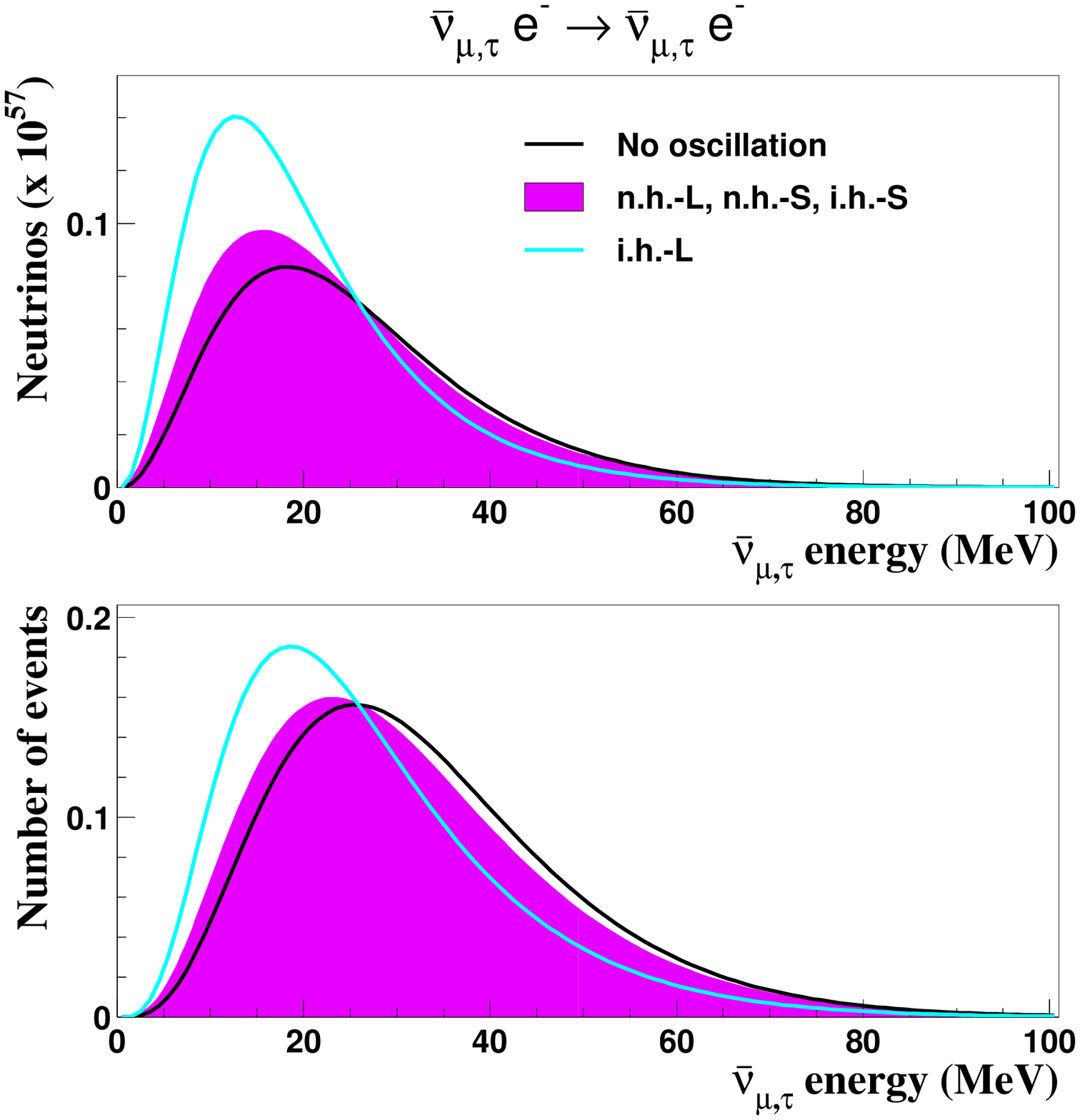,width=8.5cm}
\end{tabular}
\caption{Scenario I: Number of neutrinos arriving at Earth (top) and expected
number of events in the 3 kton detector (bottom) for
$\nu_{\mu,\tau}$ (left) and $\bar\nu_{\mu,\tau}$ (right)
neutrinos. The non oscillation and the four
oscillation cases have been taken into account.}
\label{fig:oschierar_anumu}
\end{figure}

As was mentioned in section \ref{sec:matter}, normal and inverted
hierarchies are degenerated for small \th13 in all the cases. In the
neutrino (antineutrino) channel and i.h. (n.h.) the conversion is
adiabatic and results are independent on \th13. Maximal conversion
occurs for large mixing angle in the neutrino (antineutrino) channel
for n.h. (i.h.).

We stress again that the $\nu_e$ flux and $\bar\nu_e$ flux have
different information about the neutrino oscillation parameters.
The i.h.-L and i.h.-S are distinguishable from $\bar\nu_e$ events
but are not from $\nu_e$ events. Then, a good separation of these
events is important in order to distinguish between models. 

Figures \ref{fig:oschierar_nue}, \ref{fig:oschierar_anue} and
\ref{fig:oschierar_anumu} show these effects in the
neutrino energy spectra. 
The plots on the top correspond to the expected number of
neutrinos arriving at Earth and the bottom plots are the corresponding
number of events detected in a 3 kton detector. The four oscillation cases are
considered in the distributions.

Due to the total conversion $\nu_{\mu,\tau}$ $\to$ $\nue$ for the n.h.-L case,
the $\nue$ energy spectra is harder and this leads a huge increase
of the expected events due to the quadratic dependence of the CC cross
section with energy. The same effect can be seen for $\bar\nu_e$
events and i.h.-L case. The elastic processes are less sensitive to 
the oscillations and smaller number of events is
expected. Nevertheless, the energy spectrum is modified specially for    
$\nu_{\mu,\tau}$ events, moving to lower values of energy due to the
neutrino mixing.

\subsubsection{General results as a function of the mixing angle}

The explicit variation of the expected rates and energy spectra with
the \th13 angle can be computed considering the dependence of the jump
probability with this angle (section \ref{sec:matter}). 



\begin{figure}
\centering
\begin{tabular}{cc}
\epsfig{file=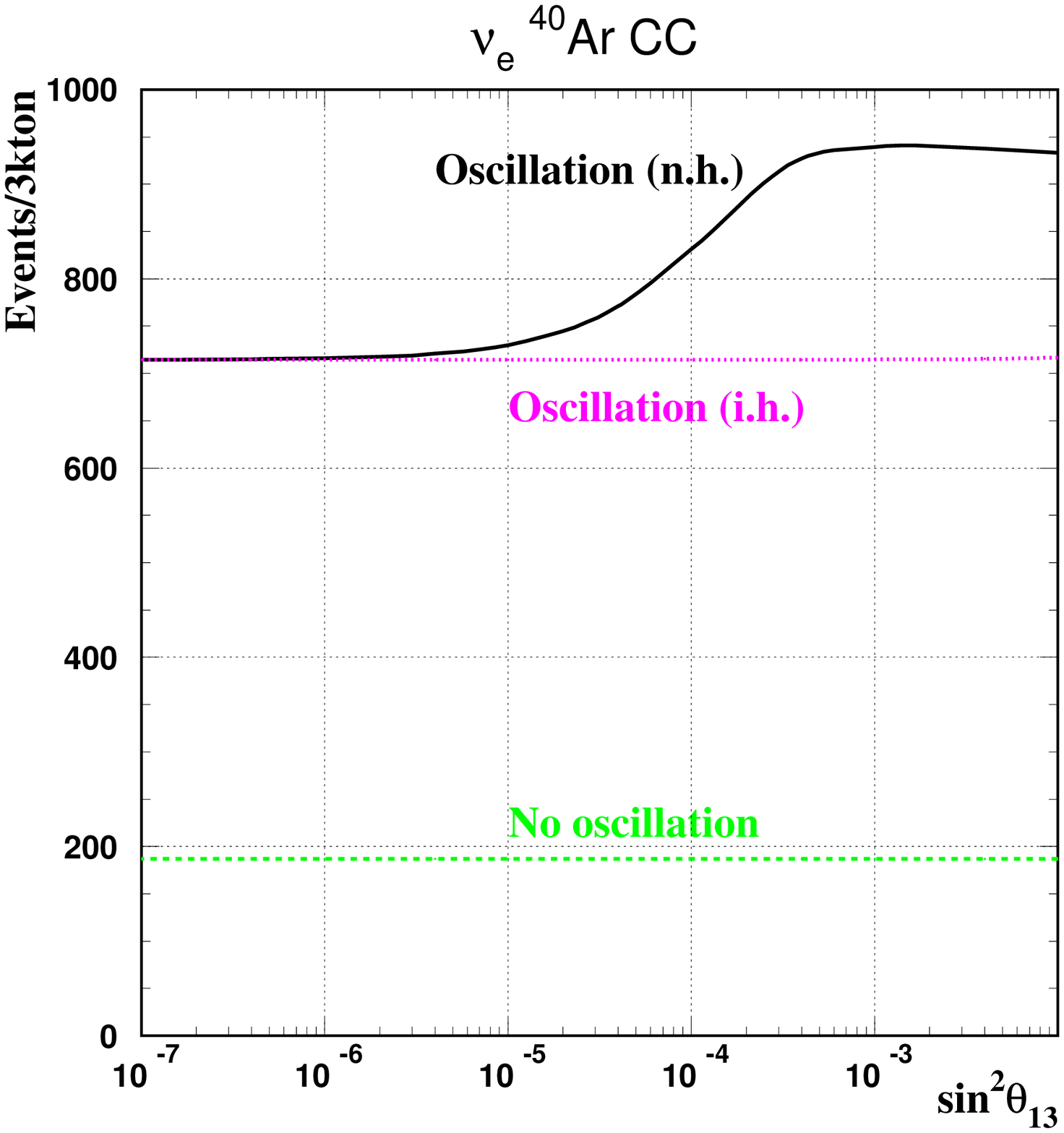,width=8cm}
&
\hspace{-1cm}
\epsfig{file=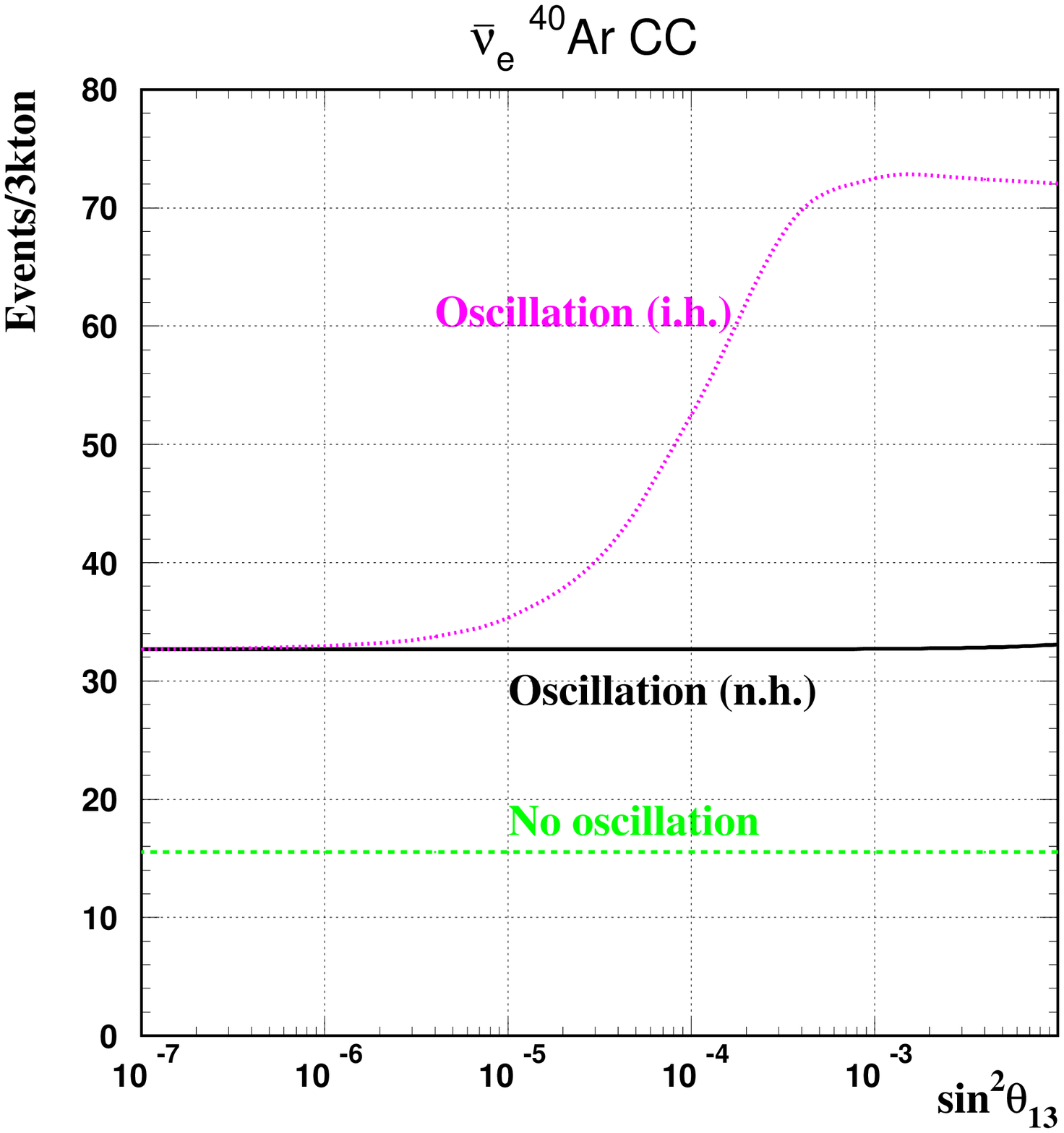,width=8cm}
\end{tabular}
\centering
\begin{tabular}{cc}
\epsfig{file=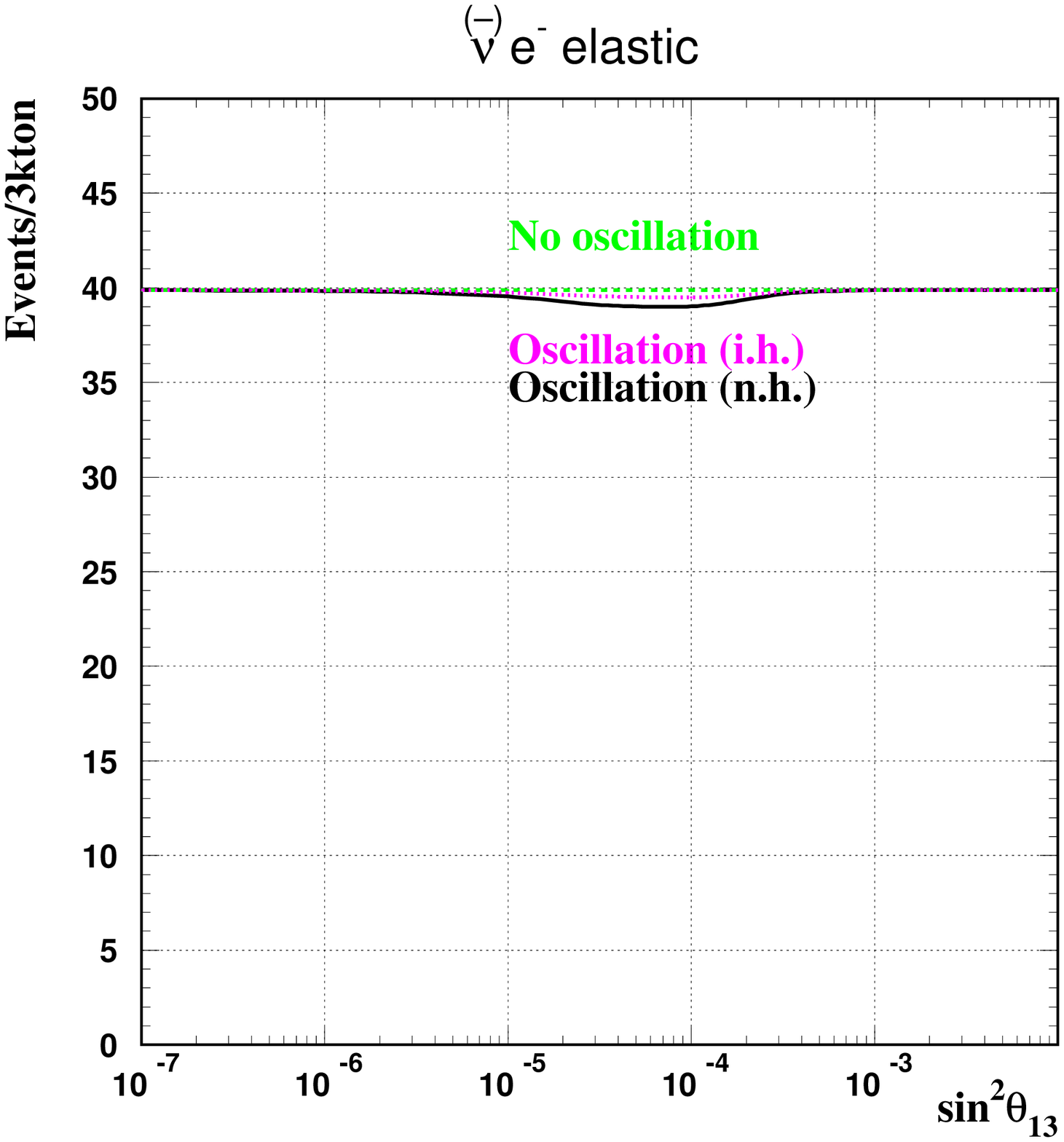,width=8cm}
&
\hspace{-1cm}
\epsfig{file=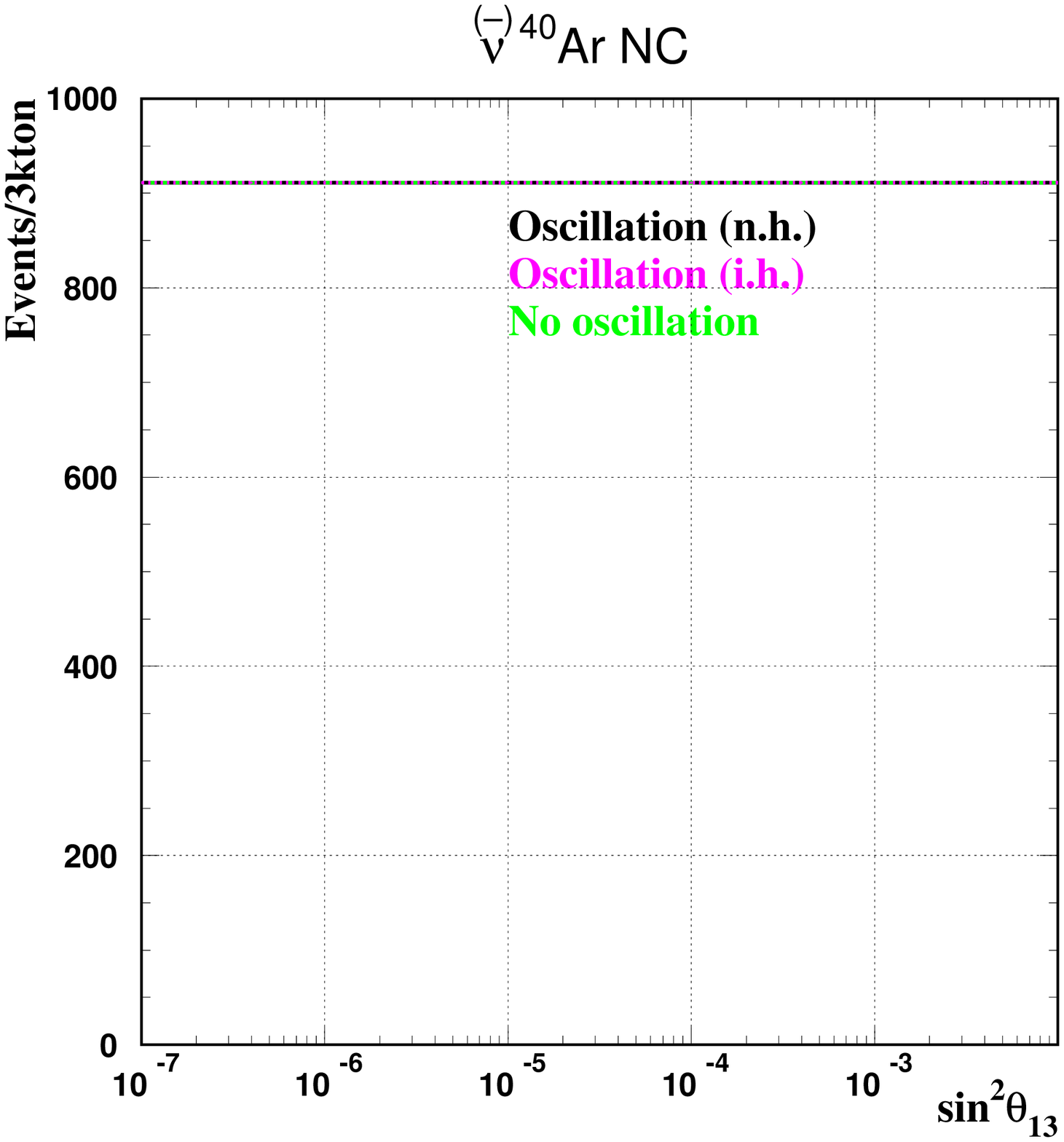,width=8cm}
\end{tabular}
\caption{Expected number of events in a 3 kton detector as a
function of \s2t13. The different neutrino interaction processes
expected are plotted separately. Solid lines correspond to
the n.h. oscillation case, dotted lines to the i.h. case and dashed
lines to the non oscillation case.} 
\label{fig:ratess2t13}
\end{figure}

Figure \ref{fig:ratess2t13} shows the the variation of the neutrino
rates as a function of \s2t13. 
We compare the expected results for the cases of n.h. (solid line),
i.h. (dotted line) and no oscillation (dashed line). CC events are
very sensitive to the change on the energy spectra due to
oscillations. A clear increase on the number of events
is expected, being of a factor (4--5) for $\nu_e$ CC events and a
factor 2 for $\bar\nu_e$ CC interactions. 
The main variations with the angle
\th13 are expected for the $\nu_e$ CC channel in the case of normal
hierarchy and for the $\anue$ CC channel for inverted hierarchy. The
number of $\nue$ CC events increases 30\% from small to large \th13
values and $\anue$ CC events are doubled. Small
variations are expected for the rest of channels.

Although the number of events does not change for some reactions, the energy
spectra are affected by oscillations and in particular by the value of
the \th13 angle. 

\subsection{Other cooling scenarios -- scenarios II--V}

The signature of neutrino oscillations depends on the flavor-dependent
primary supernova neutrino fluxes. Recent SN simulations
\cite{Raffelt:2003en,Raffelt:2002} found that the differences between the
average energies of $\nu_{\mu,\tau}$, $\bar\nu_{\mu,\tau}$ and $\anue$
are in the range 0--20\%, with 10\% being a typical number. The
luminosities of all neutrino flavors are expected to be approximately
equal within a factor two.

In order to consider the range of possibilities for the source fluxes
we compute the expected events for different values of the
neutrino average energies and luminosities, as summarized in table~\ref{tab:sncoolscenario}. The results from the different scenarios
are contained in tables \ref{tab:rates2}--\ref{tab:rates5}. 
Table \ref{tab:rates2} is obtained using the flux parameters found by
the Livermore group \cite{Totani:1997vj}. 

In table \ref{tab:rates3} we
take the results from \cite{Raffelt:2003en,Raffelt:2002} and equipartition
luminosities.  Finally, tables \ref{tab:rates4} and \ref{tab:rates5}
show the expected rates considering the same average energies than
used in table \ref{tab:rates3} but for two extreme cases of luminosity
partition: L$_{\nu_e}$ = L$_{\bar\nu_e}$ = 2 L$_{\nu_x}$ (table
\ref{tab:rates4}) and L$_{\nu_e}$ = L$_{\bar\nu_e}$ = 0.5 L$_{\nu_x}$
(table \ref{tab:rates5}), with $\nu_x$ each of $\nu_\mu$, $\nu_\tau$,
$\bar\nu_\mu$ and $\bar\nu_\tau$ neutrino flavors. 

\begin{figure}[htbp]
\centering
\epsfig{file=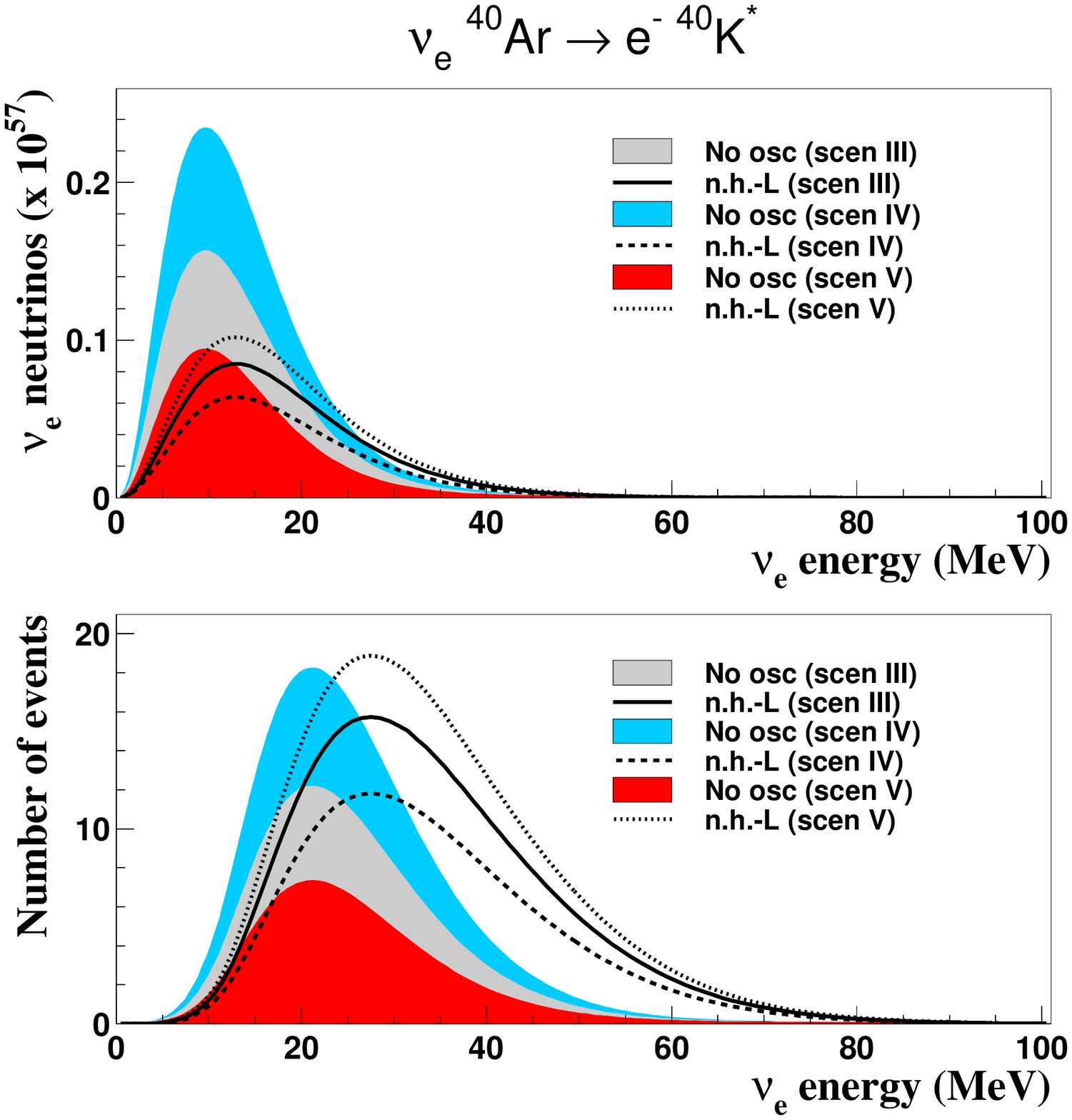 ,width=17cm}
\caption{Scenarios III--V: number of $\nue$ neutrinos arriving at Earth (top) and expected
number of events in the 3 kton detector (bottom) for $\nu_e$
neutrinos. The three scenarios III--V are
considered in the distributions, each time for the non oscillation
and n.h.-L oscillation cases.}
\label{fig:schenother}
\end{figure}

Clearly, scenarios III--V distinguish themselves from the scenarios
I--II in their prediction of a quite flavor degenerate spectra (see
figures \ref{fig:oschierar_nue}, \ref{fig:oschierar_anue} and
\ref{fig:oschierar_anumu}). 
In order to illustrate this, we show in Figure \ref{fig:schenother}
how these affect the neutrino energy spectra. 
The upper plot corresponds to the expected number of
neutrinos arriving at Earth and the bottom plot is the event energy spectrum
normalized to a 3 kton detector. The three scenarios III--V are
considered in the distributions, each time for the non oscillation
and for the oscillation case with a normal mass hierarchy and
large mixing angle.

\begin{table}[tbp]
\centering
\begin{tabular}{clccccc} \hline
& &\multicolumn{5}{c}{\bf Scenario II} \\
Reaction & & No & \multicolumn{2}{c}{Oscillation
(n.h.)} & \multicolumn{2}{c}{Oscillation (i.h.)} \\
 & & oscillation & Large $\theta_{13}$ & Small $\theta_{13}$ & Large
$\theta_{13}$ & Small $\theta_{13}$\\ \hline
{\bf Elastic} & & & & & & \\
& $\nu_e \, e^-$ & 20 & 20 & 20 & 20 & 20 \\
& $\bar\nu_e\, e^-$ & 8 & 8 & 8 & 8 & 8 \\
& $(\nu_\mu+\nu_\tau)\, e^-$ & 7 & 7 & 7 & 7 & 7 \\
& $(\bar\nu_\mu +\bar\nu_\tau)\, e^-$ & 5 & 5 & 5 & 5 & 5 \\
& total $\nu\, e^-$ & 40 & 40 & 40 & 40 & 40 \\
\hline
{\bf Absorption} & & & & & & \\
{\bf CC} & $\nu_e$ $^{40}$Ar       & 264 & 805 & 644 & 644 & 644 \\ 
& $\anue$ $^{40}$Ar                  & 16 &  28 &  28 &  57 & 28 \\ 
\hline
{\bf NC} & $\nu$ $^{40}$Ar       & 415 & 415 & 415 & 415 & 415 \\
          & $\bar{\nu}$ $^{40}$Ar & 355 & 355 & 355 & 355 & 355 \\ 
\hline
{\bf Total} & & 1090 & 1643 & 1482 & 1511 & 1482 \\\hline
\end{tabular}
\caption{Expected neutrino events in a 3 kton detector. 
The flux parameters are $\langle E_{\nu_e}
\rangle$ = 13 MeV, $\langle E_{\bar\nu_e} \rangle$ = 16 MeV, $\langle
E_{\nu_{\mu,\tau}} \rangle$ = $\langle E_{\bar\nu_{\mu,\tau}} \rangle$
= 23 MeV and luminosity equipartition among flavors is assumed.}   
\label{tab:rates2}
\end{table}

\begin{table}[tbp]
\centering
\begin{tabular}{clccccc} \hline
& &\multicolumn{5}{c}{\bf Scenario III} \\
Reaction & & No & \multicolumn{2}{c}{Oscillation
(n.h.)} & \multicolumn{2}{c}{Oscillation (i.h.)} \\
 & & oscillation & Large $\theta_{13}$ & Small $\theta_{13}$ & Large
$\theta_{13}$ & Small $\theta_{13}$\\ \hline
{\bf Elastic} & & & & & & \\
& $\nu_e \, e^-$ & 20 & 20 & 20 & 20 & 20 \\
& $\bar\nu_e\, e^-$ & 8 & 8 & 8 & 8 & 8 \\
& $(\nu_\mu+\nu_\tau)\, e^-$ & 7 & 7 & 7 & 7 & 7 \\
& $(\bar\nu_\mu +\bar\nu_\tau)\, e^-$ & 5 & 5 & 5 & 5 & 5 \\
& total $\nu\, e^-$ & 40 & 40 & 40 & 40 & 40 \\
\hline
{\bf Absorption} & & & & & & \\
{\bf CC} & $\nu_e$ $^{40}$Ar       & 264 & 482 & 417 & 417 & 417 \\ 
& $\anue$ $^{40}$Ar                  & 16 &  18 &  18 &  23 & 18 \\ 
\hline
{\bf NC} & $\nu$ $^{40}$Ar       & 215 & 215 & 215 & 215 & 215 \\
          & $\bar{\nu}$ $^{40}$Ar & 207 & 207 & 207 & 207 & 207 \\ 
\hline
{\bf Total} & & 742 & 962 & 897 & 902 & 897 \\\hline
\end{tabular}
\caption{Expected neutrino events in a 3 kton detector. 
The flux parameters are $\langle E_{\nu_e}
\rangle$ = 13 MeV, $\langle E_{\bar\nu_e} \rangle$ = 16 MeV, $\langle
E_{\nu_{\mu,\tau}} \rangle$ = $\langle E_{\bar\nu_{\mu,\tau}} \rangle$
= 17.6 MeV and luminosity equipartition among flavors is assumed.}   
\label{tab:rates3}
\end{table}

\begin{table}[tbp]
\centering
\begin{tabular}{clccccc} \hline
& &\multicolumn{5}{c}{\bf Scenario IV} \\
Reaction & & No & \multicolumn{2}{c}{Oscillation
(n.h.)} & \multicolumn{2}{c}{Oscillation (i.h.)} \\
 & & oscillation & Large $\theta_{13}$ & Small $\theta_{13}$ & Large
$\theta_{13}$ & Small $\theta_{13}$\\ \hline
{\bf Elastic} & & & & & & \\
& $\nu_e \, e^-$ & 29 & 15 & 19 & 19 & 19 \\
& $\bar\nu_e\, e^-$ & 12 & 10 & 10 & 6 & 10 \\
& $(\nu_\mu+\nu_\tau)\, e^-$ & 5 & 8 & 7 & 7 & 7 \\
& $(\bar\nu_\mu +\bar\nu_\tau)\, e^-$ & 4 & 5 & 5 & 6 & 5 \\
& total $\nu\, e^-$ & 50 & 38 & 41 & 38 & 41 \\
\hline
{\bf Absorption} & & & & & & \\
{\bf CC} & $\nu_e$ $^{40}$Ar       & 397 & 362 & 372 & 372 & 372 \\ 
& $\anue$ $^{40}$Ar                  & 24 &  22 &  22 &  17 & 22 \\ 
\hline
{\bf NC} & $\nu$ $^{40}$Ar       & 188 & 188 & 188 & 188 & 188 \\
          & $\bar{\nu}$ $^{40}$Ar & 199 & 199 & 199 & 199 & 199 \\ 
\hline
{\bf Total} & & 858 & 809 & 822 & 814 & 822 \\\hline
\end{tabular}
\caption{Expected neutrino events in a 3 kton detector. 
The flux parameters are $\langle E_{\nu_e}
\rangle$ = 13 MeV, $\langle E_{\bar\nu_e} \rangle$ = 16 MeV, $\langle
E_{\nu_{\mu,\tau}} \rangle$ = $\langle E_{\bar\nu_{\mu,\tau}} \rangle$
= 17.6 MeV and luminosities L$_{\nu_e}$ = L$_{\bar\nu_e}$ = 2
L$_{\nu_x}$ with $\nu_x$ each of $\nu_\mu$, $\nu_\tau$, $\bar\nu_\mu$
and $\bar\nu_\tau$ neutrino flavors are considered.}   
\label{tab:rates4}
\end{table}

\begin{table}[tbp]
\centering
\begin{tabular}{clccccc} \hline
& &\multicolumn{5}{c}{\bf Scenario V} \\
Reaction & & No & \multicolumn{2}{c}{Oscillation
(n.h.)} & \multicolumn{2}{c}{Oscillation (i.h.)} \\
 & & oscillation & Large $\theta_{13}$ & Small $\theta_{13}$ & Large
$\theta_{13}$ & Small $\theta_{13}$\\ \hline
{\bf Elastic} & & & & & & \\
& $\nu_e \, e^-$ & 12 & 23 & 20 & 20 & 20 \\
& $\bar\nu_e\, e^-$ & 5 & 6 & 6 & 10 & 6 \\
& $(\nu_\mu+\nu_\tau)\, e^-$ & 8 & 6 & 7 & 7 & 7 \\
& $(\bar\nu_\mu +\bar\nu_\tau)\, e^-$ & 7 & 6 & 6 & 5 & 6 \\
& total $\nu\, e^-$ & 32 & 41 & 39 & 42 & 39 \\
\hline
{\bf Absorption} & & & & & & \\
{\bf CC} & $\nu_e$ $^{40}$Ar       & 159 & 578 & 453 & 453 & 453 \\ 
& $\anue$ $^{40}$Ar                  & 10 &  15 &  15 & 28 & 15 \\ 
\hline
{\bf NC} & $\nu$ $^{40}$Ar       & 236 & 236 & 236 & 236 & 236 \\
          & $\bar{\nu}$ $^{40}$Ar & 214 & 214 & 214 & 214 & 214 \\ 
\hline
{\bf Total} & & 651 & 1084 & 957 & 973 & 957 \\\hline
\end{tabular}
\caption{Expected neutrino events in a 3 kton detector. 
The flux parameters are $\langle E_{\nu_e}
\rangle$ = 13 MeV, $\langle E_{\bar\nu_e} \rangle$ = 16 MeV, $\langle
E_{\nu_{\mu,\tau}} \rangle$ = $\langle E_{\bar\nu_{\mu,\tau}} \rangle$
= 17.6 MeV and luminosities L$_{\nu_e}$ = L$_{\bar\nu_e}$ = 0.5
L$_{\nu_x}$ with $\nu_x$ each of $\nu_\mu$, $\nu_\tau$, $\bar\nu_\mu$
and $\bar\nu_\tau$ neutrino flavors are considered.}   
\label{tab:rates5}
\end{table}

\subsection{Discrimination between the various scenarios}

\begin{figure}[htbp]
\vspace{-2cm}
\centering
\epsfig{file=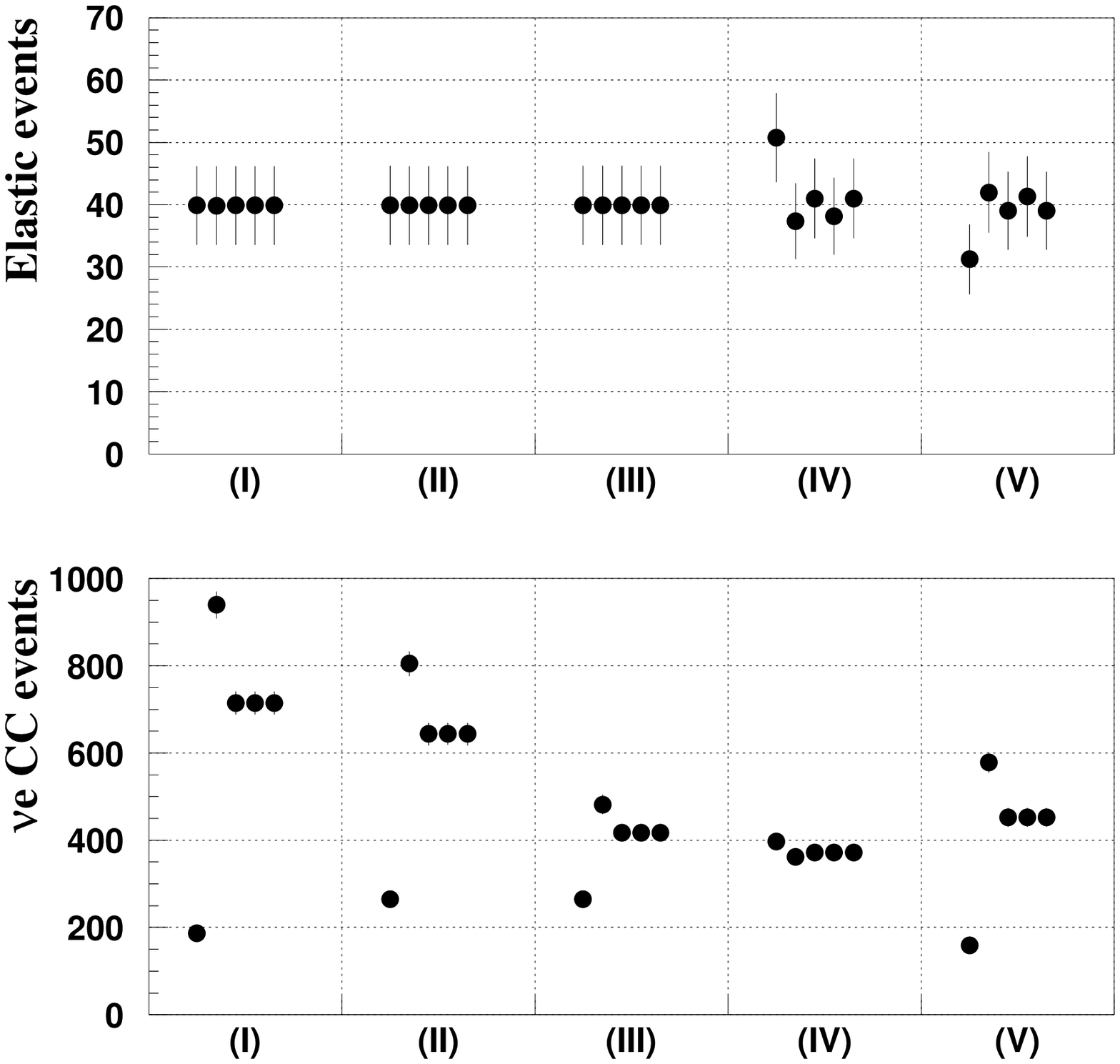,width=14cm,height=11.cm} \\
\vspace{-1.5cm}
\epsfig{file=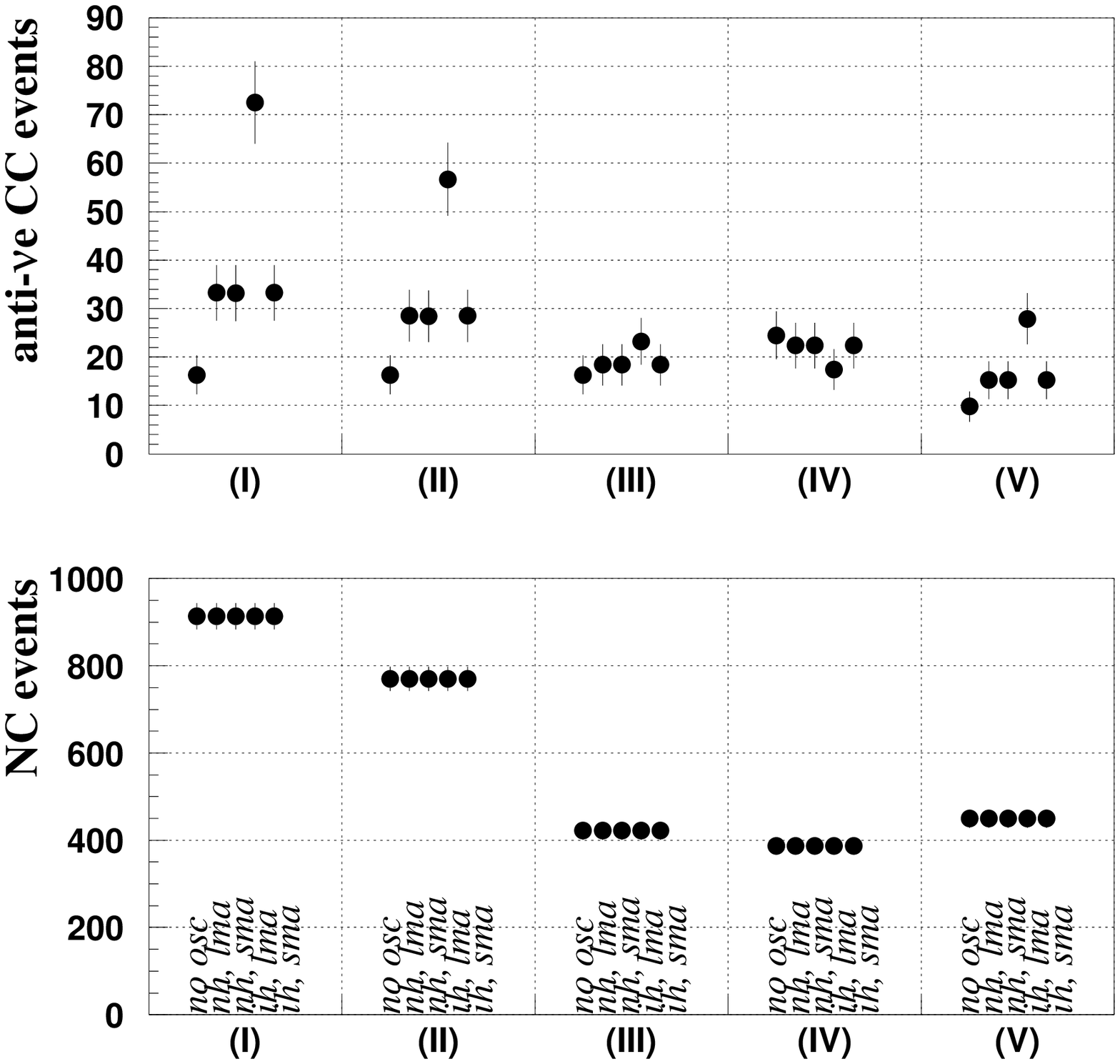,width=14cm,height=11.cm} 
\caption{Expected number of events: the four neutrino processes are considered for
different SN scenarios and oscillation
conditions (see text). The statistical errors are included in the picture.}
\label{fig:ratesall}
\end{figure}

The comparison between the different scenarios is summarized in figure
\ref{fig:ratesall}. The expected number of events 
in the four neutrino detection channels are plotted
for the five scenarios of the SN neutrino flux parameters considered in
this study.  
Moreover, the non oscillation and the four oscillation
cases are plotted for every scenario. The plots are normalized to a supernova at
10 kpc and a 3 kton detector. The errors indicated
in the picture correspond to statistical errors and give a visual feeling
on the ability to separate the various supernova (I--V) and oscillation scenarios
(no osc, n.h.-L, n.h.-S, i.h.-L, i.h.-S).

NC events are sensitive to the primary source parameters and are
independent to the oscillation effects.  They give a direct account
for the total luminosity of the supernova and of the hardness of the
neutrino spectra. Hence, they provide 
the means to discriminate between the various supernova physics
scenarios if the binding energy is assumed to be known. 
This is illustrated in the bottom plot of
Figure~\ref{fig:ratesall}. The rates of the nuclear processes are
such that excellent statistical precision is achieved.
They amount to $911\pm 30$, $770\pm 27$, $422\pm 20$, 
$387\pm 20$, $450\pm 21$ events for
resp. the scenarios I--V, where the error quoted is the statistical
error on the number of events. 

The supernova and neutrino oscillation physics can then be
studied by the simultaneous observation of NC, $\nue$ CC,
$\bar\nue$ CC and elastic channels. We observe:
\begin{itemize}
\item  In the hierarchical scenarios (I--II) the enhancement of $\nue$
due to oscillation is highly significant. As already mentioned,
this is the result of the hardening of the spectrum in all cases
of oscillations. 
\begin{itemize}
\item The effect is of course the strongest in the
case of n.h.-L for the $\nu_e$ and in the case of i.h.-L for $\anue$. 
\item The normal and inverted hierarchy
yield the same number of events in the case of small mixing.
\end{itemize}
\item  In the non-hierarchical scenarios (III--V) the enhancement of $\nue$
due to oscillation is significant compared to the non-oscillation case,
except for scenario (IV) where the quasi-degenerate energies and
the choice of luminosities L$_{\nu_e}$ = L$_{\bar\nu_e}$ = 2
L$_{\nu_x}$ with $\nu_x$ each of $\nu_\mu$, $\nu_\tau$, $\bar\nu_\mu$
and $\bar\nu_\tau$ neutrino flavors just washes out the oscillation
(the disappearing $\nu_e\rightarrow \nu_x$ are compensated
by the appearing $\nu_x\rightarrow \nu_e$). 
\item in the scenario IV this ambiguity in the number of events
can be solved by a study of the event energy distribution
(see Figure~\ref{fig:schenother}). Indeed, the figure clearly
shows that the spectrum is indeed harder for the oscillation
case but that the total number of events is roughly the same. 
\end{itemize}

\section{Conclusions}

An analysis of the neutrino events
expected in a liquid Argon TPC (ICARUS-like) 
from a future galactic supernova has been performed
showing the high capabilities of the detector to observe neutrinos
from stellar collapses and extract information about their
properties. 

Four detection channels have been considered:
elastic scattering on atomic electrons from all
neutrino species, $\nu_e$ charged current absorption on $Ar$ with production
of excited $K$, $\bar\nu_e$ charged current  
absorption on $Ar$ with production of excited $Cl$ and neutral current
interactions on $Ar$ from all neutrino flavors.

We give a detailed account of the dependence with
the mixing angle \th13 and the type of mass hierarchy. 
Neutrino oscillations change significantly the expected supernova neutrino
rates and energy spectra. The matter effects
inside the supernova must be included to predict 
how the spectra of the
neutrino burst is modified by oscillations. 

Information about the supernova explosion mechanism can be obtained
from the study of the neutrino burst. The early phase neutrino signal on Argon
has been computed using the fluxes from a realistic core collapse supernova
model \cite{Burrows}, and including oscillation effects. Thanks to its clean
identification and 
high sensitivity to $\nu_e$ neutrinos, a liquid Argon TPC can provide fundamental
information about the shock breakout mechanism. We find that a 70~kton
detector is best suited to study the SN burst, given its statistics even in case
of the $\nu_e$ suppression expected in case of oscillations.

For the cooling phase of the supernova, we have adopted a simple
Fermi-Dirac parameterization of the neutrino fluxes. We selected
six different scenarios in order to take into account the uncertainties
on the knowledge of the supernova physics and estimate
their effect on our analysis. For the study of the cooling phase, a 3~kton
detector is sufficient to collect hundreds of events which could provide
fundamental information on supernova and neutrino oscillation physics.

We discussed how the various scenarios can be distinguished.
In all cases, we find that the simultaneous observation of the
four channels, in particular nuclear charged and neutral current processes,  is fundamental in order to decouple the supernova
from the neutrino oscillation physics.

The observed event rates and the event energy spectra of the four
classes of events form a set of observables sensitive to the supernova parameters
and to the intrinsic neutrino oscillation parameters (\th13 and mass hierarchy).
We have qualitatively illustrated this with several examples.

In order to understand quantitatively the sensitivity to such parameters,
one should study the classification of actual detected events into the four
channels and perform a global fit to the observed distributions.
Such an analysis will be the subject of a future work~\cite{superfit}.

\section*{Acknowledgments}

We would like to acknowledge T.A. Thompson and A. Burrows for
providing data about the neutrino breakout spectra of core collapse
supernovae and for useful discussions.
We thank G. Martinez-Pinedo, E. Kolbe and
K. Langanke for their calculations on the neutrino cross sections for
Argon nuclei and for very valuable discussions.

%
%

\end{document}